\documentclass[aps,pra,superscriptaddress]{revtex4-1}
\usepackage{preamble}
\usepackage{changes}
\usepackage{gensymb}
\graphicspath{{figs/}}

\begin{document}
%\title{Intercalation kinetics in multiphase layered  materials:
%Modeling lithiation of graphite}
\title{Intercalation kinetics in multiphase layered materials}

\author{Raymond B. Smith}
\affiliation{Department of Chemical Engineering, Massachusetts
Institute of Technology, Cambridge, Massachusetts 02139, USA}

\author{Edwin Khoo}
\affiliation{Department of Chemical Engineering, Massachusetts
Institute of Technology, Cambridge, Massachusetts 02139, USA}

\author{Martin Z. Bazant}
\email[Corresponding author: ]{bazant@mit.edu}
\affiliation{Department of Chemical Engineering, Massachusetts
Institute of Technology, Cambridge, Massachusetts 02139, USA}
\affiliation{Department of Mathematics, Massachusetts Institute of
Technology, Cambridge, Massachusetts 02139, USA}
\date{\today}

\begin{abstract}
Many intercalation compounds possess layered structures or inter-penetrating lattices that enable phase separation into three or more stable phases, or ``stages,'' driven by competing intra-layer  and inter-layer forces.  While these structures are often well characterized in equilibrium, their effects on intercalation kinetics and transport far from equilibrium are typically neglected or approximated by empirical solid solution models.  Here, we formulate a general phase-field model with thermodynamically consistent reaction kinetics and cooperative transport to capture the dynamics of intercalation in layered materials. As an important case for Li-ion batteries, we model single particles of lithium intercalated graphite as having a periodic two-layer structure with three stable phases, corresponding to zero, one, or two layers full of lithium. The electrochemical intercalation reaction is described by a generalized Butler-Volmer equation with thermodynamic factors to account for the flexible structure of the graphene planes. The model naturally captures the ``voltage staircase'' discharge curves as a result of staging dynamics with internal ``checkerboard'' domains,  which cannot be described by solid-solution models based on Fickian diffusion.  On the other hand, the two-layer model is computationally expensive and excludes low-density stable phases with longer-range periodicity, so we also present a reduced model for graphite, which captures the high-density stages while fitting the low-density voltage profile as an effective solid solution. The two models illustrate the general tradeoff between the explicit modeling of periodic layers or lattices and the needs for computational efficiency and accurate fitting of experimental data. 
\end{abstract}

\maketitle

\section{Introduction}
Staging, the phenomenon of pattern formation by intercalants or defects across a layered structure, occurs in a wide range of materials including high temperature superconductors~\cite{wells1996,mohottala2006}, clays and layered double hydroxides~\cite{ijdo1998,iyi2002,pisson2003}, MXenes~\cite{naguib2012,wang2015pseudocapacitance}, metal borocarbides~\cite{joshi2015}, layered metal oxides~\cite{vanderven1998a,chen2002}, and the prototypical case of graphite~\cite{dresselhaus1981}. The properties of these materials can depend strongly on the microscopic staging structure. In high temperature superconductors, the layered segregation of oxygen can lead to regions of insulating behavior~\cite{mohottala2006}.  Staging phase-separation phenomena can also arise in three-dimensional materials with interpenetrating lattices of intercalation sites, as with hydrogen trapping in metals~\cite{dileo2013}. Another example could be lithium intercalation in anatase (TiO$_2$)~\cite{deklerk2016anatase_draft}, which has been suggested to have multiple types of intercalation sites~\cite{lafont2010,shin2011,shen2014} and demonstrates staircase voltage curves~\cite{zachau-christiansen1988}, similar to graphite~\cite{ohzuku1993}.

In graphite, the staging process is common for many intercalants, including lithium which forms LiC$_6$, the most common anode material in Li-ion batteries. The staging of the lithium-graphite system is directly related to the observed voltage in batteries. The dynamics of staging during (de)lithatiation of graphite is important in consideration of Li-ion battery rate capability, because graphite is (de)intercalated with Li during (dis)charging. Li-ion batteries are ubiquitous as energy storage devices for portable electronics, with growing use in the (hybrid) electric vehicle market. In almost all commercial cells, graphite is used as the negative electrode (anode)~\cite{scrosati2010}. Li-ion batteries work by shuttling Li ions back and forth between the two electrodes~\cite{tarascon2001}, a process which involves transport within the solid and electrolyte phases as well as electrochemical reaction to transfer the lithium between these phases. The rates of these processes (and those of competing processes) fundamentally govern the power capabilities of the batteries.

\begin{figure}[h]
    \centering
    (a)
    \includegraphics[width=0.2\textwidth]{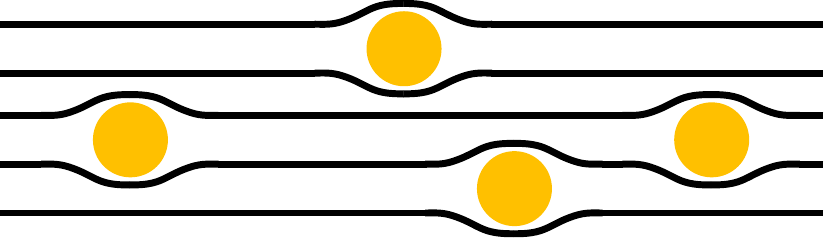}
    \hspace{3mm} (b)
    \includegraphics[width=0.2\textwidth]{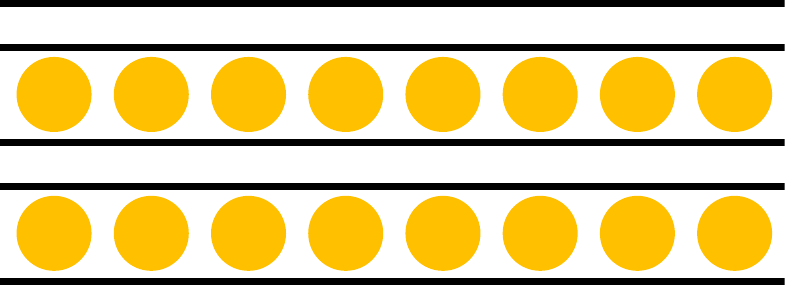}
    \hspace{3mm} (c)
    \includegraphics[width=0.2\textwidth]{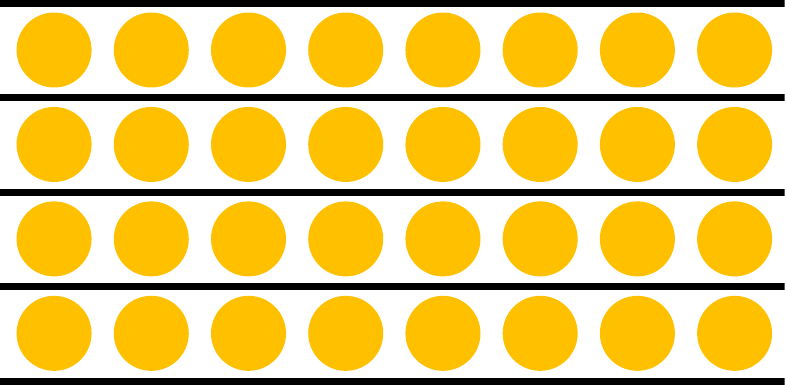}
    \\\vspace{6mm}
    (d)
    \includegraphics[width=0.5\textwidth]{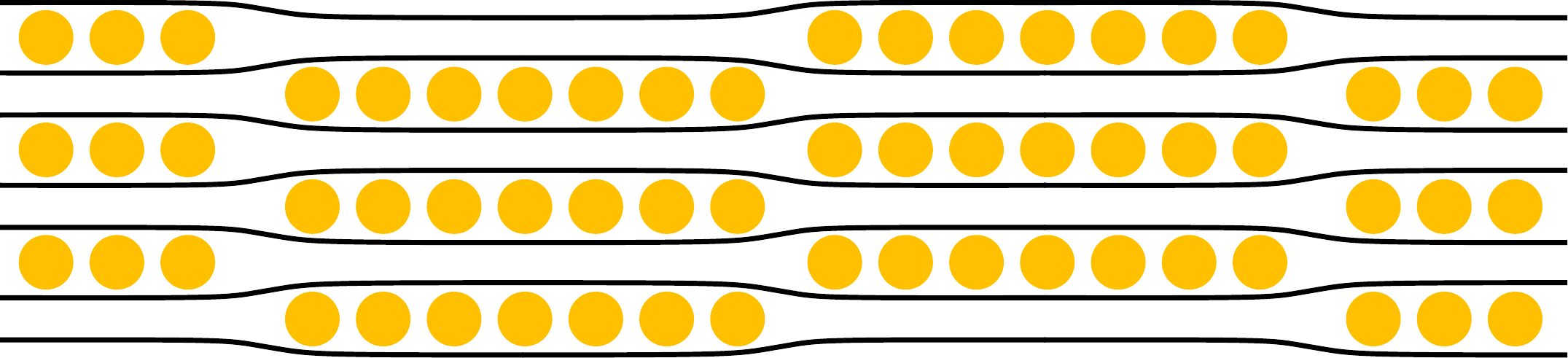}
    \caption{Graphite staging structures. Black lines represent graphene planes, and yellow circles represent intercalated lithium. Dilute stage 1' is represented in (a), half-full stage 2 in (b) and full stage 1 in (c). (d) represents the domain structure that can arise because of intra-layer phase separation of stage $>1$ graphite (stage 2 shown).}
    \label{fig:graphite_struct}
\end{figure}

Complicating our understanding of these processes is the fact that graphite tends to phase separate into Li-rich and Li-poor regions~\cite{dahn1991}, which affects both the reaction and the transport processes~\cite{bazant2013}. As Li is transferred into graphite, it intercalates between graphene planes, tending to form structures with filled layers separated by a number of empty layers. These structures are referred to by a stage number, which refers to the number of planes of graphene between filled layers. For example, stage 1 graphite is nearly full (Figure~\ref{fig:graphite_struct} (c)), whereas high stage number graphite has many empty layers between each full layer (e.g.\ Figure~\ref{fig:graphite_struct} (b) for stage 2).
In the dilute limit, there is no internal structure, which is referred to as stage 1'~\cite{dresselhaus1981} (Figure~\ref{fig:graphite_struct} (a)). Upon formation of the low-stage number structures (stages 1', 2, and 1), graphite undergoes distinct color transformations, which allows direct visual probing of the local state of charge of graphite electrodes~\cite{maire2008,harris2010}. In addition, theoretical~\cite{daumas1969,safran1980,kirczenow1982,krishnan2013} and experimental~\cite{clarke1979,dimiev2013} studies suggest intercalants within graphite tend to phase separate within layered planes such that internal domains are formed (Figure~\ref{fig:graphite_struct} (d)).

Phase separation is common in many battery electrode materials such as lithium iron phosphate~\cite{padhi1997}, spinel manganese oxides~\cite{ariyoshi2004}, and lithium titanate~\cite{ohzuku1995}, and it can dramatically affect the internal dynamics of both single particles~\cite{lim2016origin,tang2010electrochemically,bai2011,cogswell2012,cogswell2013,zeng2014,welland2015miscibility} and porous electrodes~\cite{dreyer2010,chueh2013,li2014,ferguson2012,ferguson2014,orvananos2014,orvananos2015}. As such, in order to develop predictive models of the behavior of such materials with complex phase behavior, it is critical that these models naturally capture the phase separating dynamics. Some models of phase separating electrode materials approximate them as solid solutions described by Fickian diffusion~\cite{fuller1994,verbrugge2003,bernardi2011,safari2011a,srinivasan2004design,baker2012}. Others approximate the phase separation behavior using artificially imposed equilibrium interfaces, which move via Stefan conservation conditions and use the equilibrium voltage curve to determine the functional form of the chemical potential in the solid~\cite{funabiki1999stage,srinivasan2004,baker2012,hess2013}. Others have attempted to describe the volume-averaged phase behavior using nucleation models~\cite{gallagher2012}. However, such approaches cannot capture the formation of interfaces by spinodal decomposition or other processes of non-equilibrium thermodynamics.

Phase-field models, pioneered by Van der Waals~\cite{rowlinson1979translation} and Cahn and Hilliard~\cite{cahn1958,cahn1961}, have been widely applied to phase separating systems to describe the dynamics of their approach to equilibrium states. This approach is based on linear irreversible thermodynamics~\cite{groot1962}, has been extended to electrical systems~\cite{garcia2004}, and has been self-consistently connected to electrochemical reactions to describe phase transformations in chemical and electrochemical systems~\cite{bazant2013}. Here, focusing on staging of lithiated graphite, we implement a Cahn-Hilliard style model in a multi-layer free energy framework to capture the phase behavior of graphite systems, similar to the work of Hawrylak and Subbaswamy~\cite{hawrylak1984}. We also directly relate the phase separation behavior to electrochemical kinetics via a thermodynamically consistent reaction boundary condition~\cite{bazant2013}, which has recently been validated by {\it in operando} scanning X-ray transmission microscopy of individual lithium iron phosphate nanoparticles~\cite{lim2016origin}. A previous work~\cite{guo2016} demonstrates good agreement of the two-layer graphite model with an experiment visually observing the phase separation dynamics during a single particle discharge. In a companion paper~\cite{thomas-alyea2016}, we apply a one-variable reduced model, also presented below, in porous electrode simulations to capture electrochemical and optical measurements on a porous graphite battery electrode. The porous electrode experiments and modeling suggest that the modifications of the basic framework presented here are important to describe the non-equilibrium thermodynamics of Li-ion batteries, as we will discuss in Section~\ref{sec:modSimp}.   In this paper, however, we focus on the development and predictions of the general model, which could be applied to other intercalation materials exhibiting staging associated with layered crystal structures or inter-penetrating lattices.

\section{General Theory}
\label{sec:model}
Layered intercalation materials tend to exhibit stable phases with periodicity in the transverse direction, which could be described by multiple concentration variables.  We assume the layers of the material are stacked and perfectly overlapping, i.e.\@ they look like those in Figure~\ref{fig:graphite_struct} with all layers present throughout the particle and to the edges. Then, we describe the concentration of intercalated species within each layer in only 2 dimensions, effectively depth averaging along the thickness within each individual layer. This allows us to represent the 3-dimensional layered structure as discrete, overlapping, 2-dimensional planes. Thus, we will present a 2-dimensional model with multiple concentration variables corresponding to species in different layers.  As noted above, the same model could also apply to materials with  multiple interpenetrating sub-lattices, represented by separate concentration variables.

\subsection{Multi-Layer Free Energy Model}
Mathematically, we begin by defining the total free energy of the system, $G$, as an integral of the local free energy density, $g$, over the system volume, $V$,
\begin{align}
    &G = \int_{V}^{}g\ud V.
    \label{eq:G1}
\end{align}
The choice for $g$ determines the physical description of the system. We will focus on systems with two-dimensional symmetry in which multiple concentration variables correspond to species occupying different vertically stacked layers. However, the treatment is  analogous to descriptions of three-dimensional materials in which intercalated species can occupy different types of sites, such as hydrogen trapping in metals~\cite{dileo2013} or lithium intercalation in anatase~\cite{zachau-christiansen1988,lafont2010,shin2011,shen2014,deklerk2016anatase_draft}.

For a general system with $n$ layers or site types, each occupying some part of the total system volume, $V_i$,
\begin{align}
    G = \sum_{i=1}^n\left( \int_{V_i}^{}g_i\ud V \right).
    \label{}
\end{align}
For an $n$-layer structure, in which each 2-dimensional layer has area $A_i$ and occupies a fraction, $f_i$, of the height of the unit, $H$, the total free energy is
\begin{align}
    G = H\sum_{i=1}^n f_{i} \left( \int_{A_i}^{}g_i\ud A \right).
    \label{}
\end{align}
Next, we assume individual layer free energy densities are defined by some function, $g_{\ell,i}$ which depends only on the configuration of layer $i$, as well as a series expansion of mixing enthalpies of particle clusters interacting between layers. Thus
\begin{align}
    g_i = g_{\ell,i}
%    + c_\mathrm{ref}c_i\sum_{j\ne i}c_j
    + c_\mathrm{ref}\wt{c}_i\sum_{j\ne i}\wt{c}_j
    \left( h_{ij}
    + \sum_{k\ne i,j}\wt{c}_k
    \left( h_{ijk}
    + \sum_{l\ne i,j,k}\wt{c}_l
    \left( h_{ijkl}
    + \cdots \right)\right)\right).
    \label{}
\end{align}
%with inter-layer interactions and $n$ layers which repeat as a unit. We assume that each layer has a local free energy density, $g_{\ell,i}$, and define the total system free energy density as a sum of the free energy contribution of each layer as well as the coupling energy between the layers. Expanding the free energy density in terms of mixing enthalpies of particle clusters,
%\begin{align}
%%    g = \sum_{i=1}^n g_{\ell,i}(c_i, \bnab c_i)
%%    + c_\mathrm{ref}\sum_{i=1}^n\sum_{j\ne i}\Omega_{2,ij}c_ic_j
%%    + \Omega_{3,ij}c_ic_j\left( 1-c_j \right)
%%    + \Omega_{4,ij}c_i\left( 1-c_i \right)c_j\left( 1-c_j \right),
%    g = \sum_{i=1}^n \left( g_{\ell,i}
%    + c_\mathrm{ref}c_i\sum_{j\ne i}c_j
%    \left( h_{ij}
%    + \sum_{k\ne i,j}c_k
%    \left( h_{ijk}
%    + \sum_{l\ne i,j,k}c_l
%    \left( h_{ijkl}
%    + \cdots \right)\right)\right)\right).
%    \label{}
%\end{align}
%where we have assumed the intra-layer free energy can depend on both the local filling fraction within that layer, $c_i$, and its gradients. The reference concentration, $c_\mathrm{ref}$, can be chosen to be some suitable scale, and the $\Omega$ parameters are energetic interaction parameters representing 2-, 3- and 4-body interactions between particles and particle-hole pairs.
The reference concentration, $c_\mathrm{ref}$, can be chosen to be some suitable scale, the local scaled concentrations are denoted by $\wt{c}_i = c_i/c_\mathrm{ref}$, and the $h$ parameters define the enthalpic interaction energies of multi-body particle clusters.
%When positive, the $\Omega_2$ terms penalize regions of co-filling for layers $i$ and $j$, the $\Omega_3$ terms penalize regions of high filling in layer $i$ with intermediate filling in layer $j$, and the $\Omega_4$ terms penalize regions with intermediate filling fractions in layers $i$ and $j$.

\subsection{Intra-layer Free Energy}
\begin{figure}[h]
    \centering
    \includegraphics[width=0.4\textwidth]{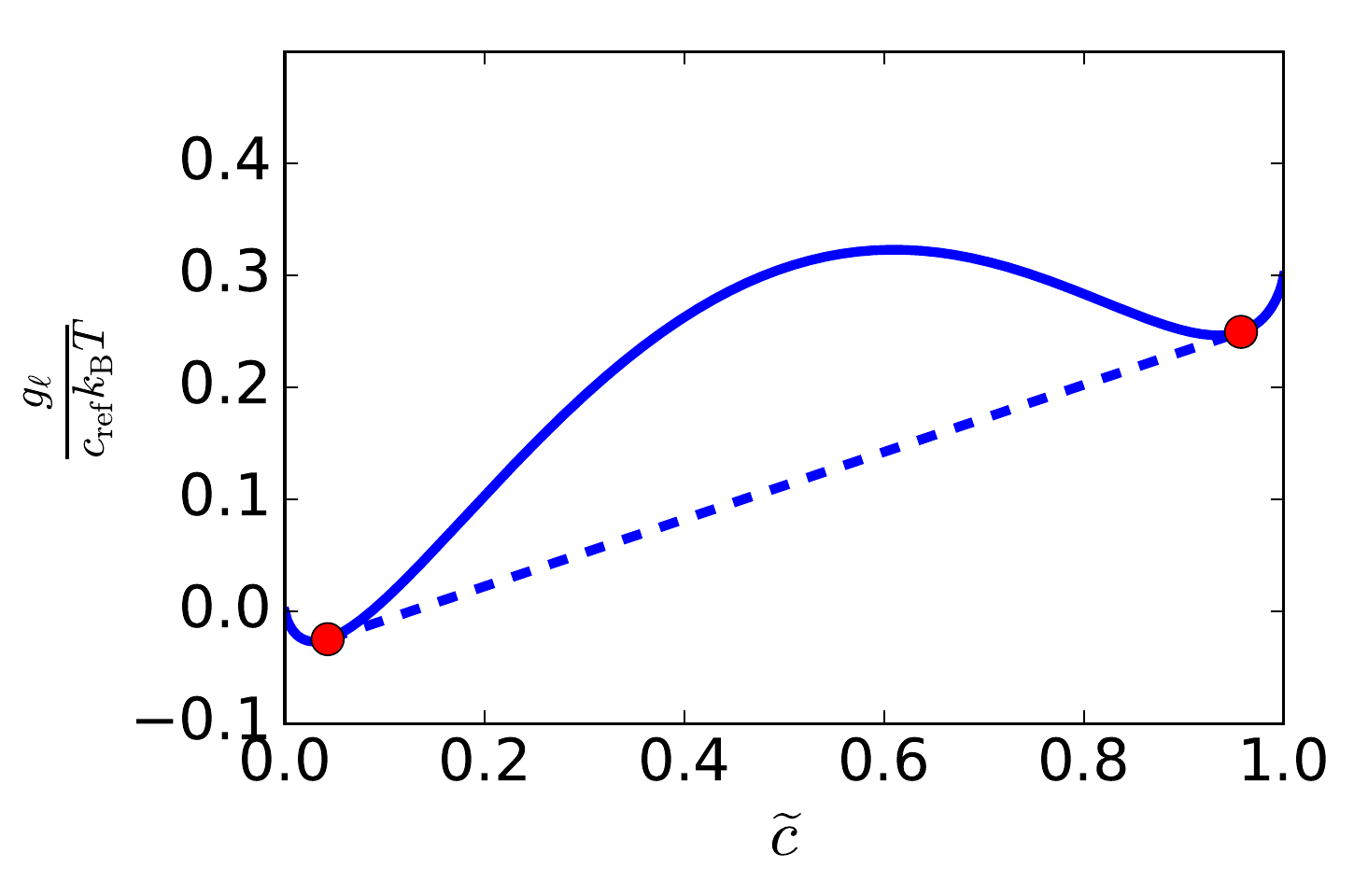}
    \caption{Regular solution free energy density as a function of the local filling fraction, $\wt{c}$. The dashed line represents a common tangent construction intersecting the two energy wells at the red dots. Homogeneous systems with concentration between the two red dots can lower their free energy by phase separating into regions with high and low concentrations given by the red dots and average free energy density lying on the dashed line.}
    \label{fig:gRS}
\end{figure}

We first choose a model for the intra-layer free energy, $g_\ell$, which will not depend on the layer, assuming that each region between adjacent planes is structurally similar. As the simplest model that allows phase separation within layers, we consider a regular solution, a lattice gas model~\cite{derosa1999lattice,ledovskikh2016}, which describes the entropic configurations of particles and vacancies on a lattice along with an enthalpic energy of mixing between particles and vacancies. We add a gradient penalty term to capture the effect of interfacial energy between the intra-layer domains~\cite{cahn1958},
\begin{align}
    g_\ell(\wt{c}) = c_{\textrm{ref}}\left\{k_\mathrm{B}T\left[ \wt{c}\ln(\wt{c}) + (1-\wt{c})\ln(1-\wt{c}) \right] +
    \Omega_a\wt{c}(1-\wt{c}) + \wt{c}\mu^\Theta\right\} + \frac{1}{2}\kappa(\bnab \wt{c})^2
    ,
    \label{eq:layerFED}
\end{align}
where $k_\mathrm{B}$ is Boltzmann's constant, $T$ is the absolute temperature, $\Omega_a$ is the regular solution parameter for interaction of particles and vacancies within a layer, $\mu^\Theta$ is the reference chemical potential, and $\kappa$ is the gradient energy penalty, assumed here to be isotropic within each plane. Of note, we use the gradient, $\bnab$, and divergence, $\dvg$, operators here to indicate their evaluation only within the plane of a single layer, and the non-dimensional concentrations are scaled such that they are filling fractions of individual layers. When $\Omega_a > 2k_\mathrm{B}T$, this model defines a free energy curve with two minima near the extreme filling fractions, leading to a tendency for internal phase separation to regions with high and low concentrations. The function is plotted in Figure~\ref{fig:gRS} using thermodynamic parameters from Section~\ref{sec:implementation}, at $T=298\ \mathrm{K}$, choosing $\mu^\Theta = 0.3\ k_\mathrm{B}T$, and neglecting contributions from the gradient penalty term.

\subsection{Periodic Bilayer Model Applied to Graphite}
The simplest example of the multilayer framework describes a periodic solid with repeated pairs of identical layers for intercalation, which is a natural approximation for lithiated graphite that captures the major staging transitions. A bilayer model allows description of stage 1, 2, and 1' structures, neglecting higher stage number structures in favor of model simplicity and reduced computational cost. Each modeled layer represents a plane of Li sites between graphene planes in (de)intercalated graphite, and the two modeled layers repeat in the direction normal to the modeled planes. Thus, we have a discretely ``stacked'' two-dimensional continuum description of the intercalant within the graphite. We do not assign any clear physical separation between layers in the vertical direction other than through energetic interactions, which act as local approximations of the misfit strain energy in graphite~\cite{schiffer2016strain}, without accounting for nonlocal coherency strain~\cite{cogswell2012}. From the choice of $n=2$ structurally similar layers, each of which occupies half the total volume of the system, the reference concentration, $c_\mathrm{ref} = 0.5c_\mathrm{max}$, where $c_\mathrm{max}$ is the concentration of Li in LiC$_6$.

\begin{figure}[h]
    \centering
    (a)
    \includegraphics[width=0.9\textwidth]{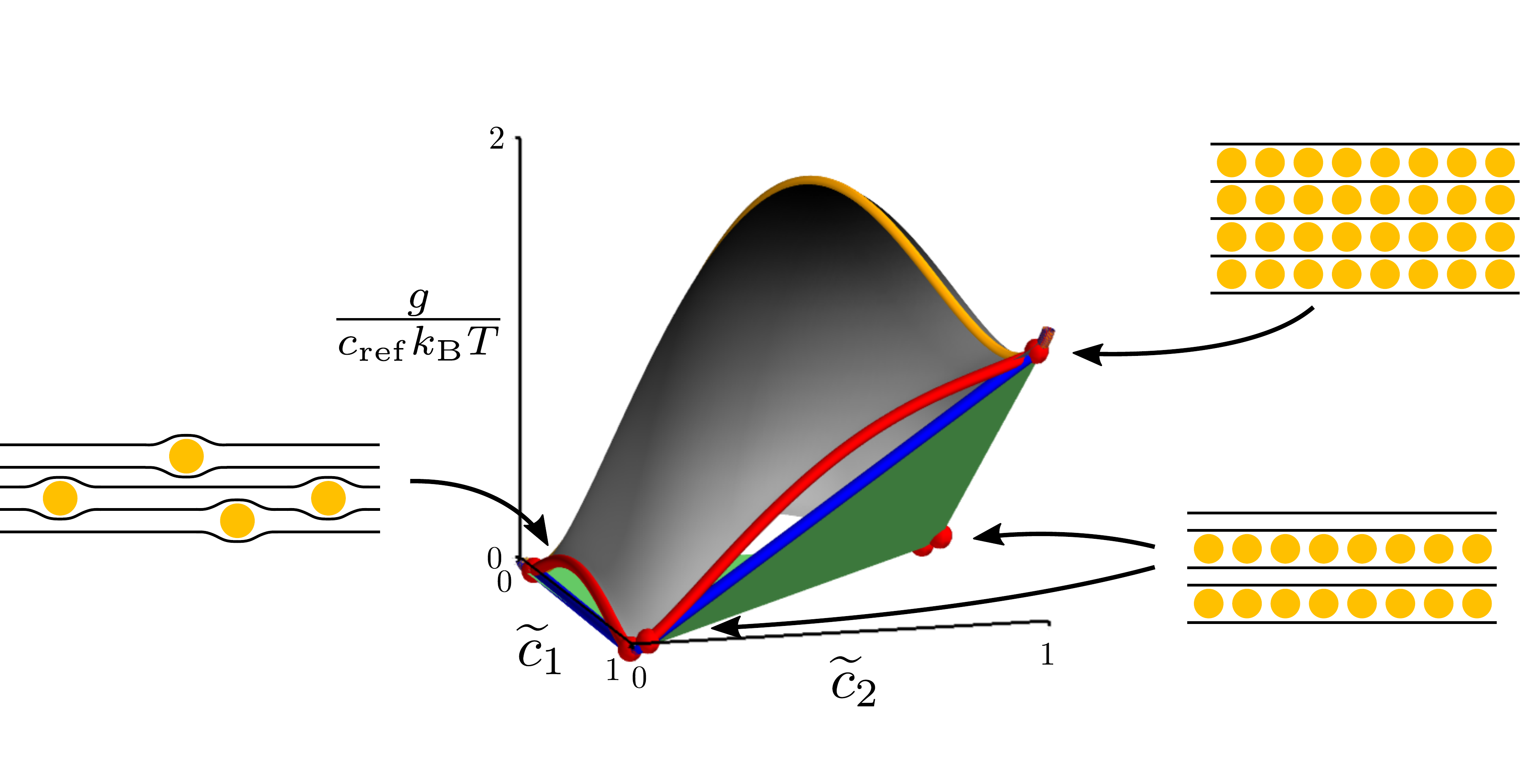}
    \\
    (b)
    \includegraphics[width=0.4\textwidth]{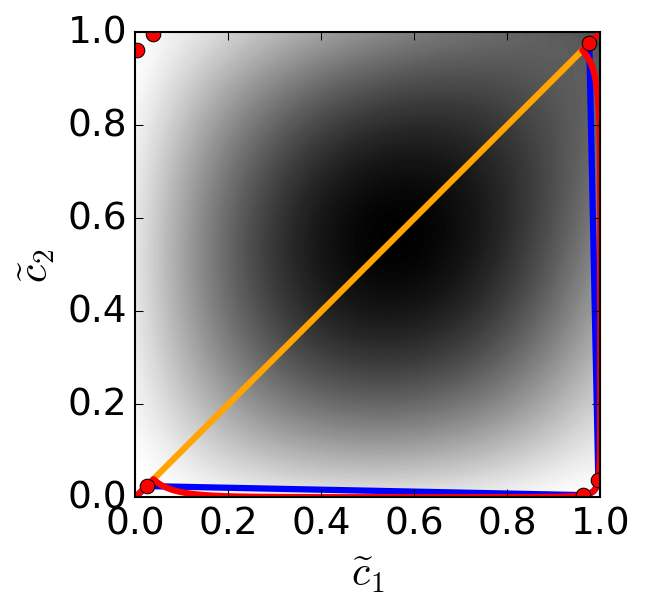}
    \ \
    (c)
    \includegraphics[width=0.4\textwidth]{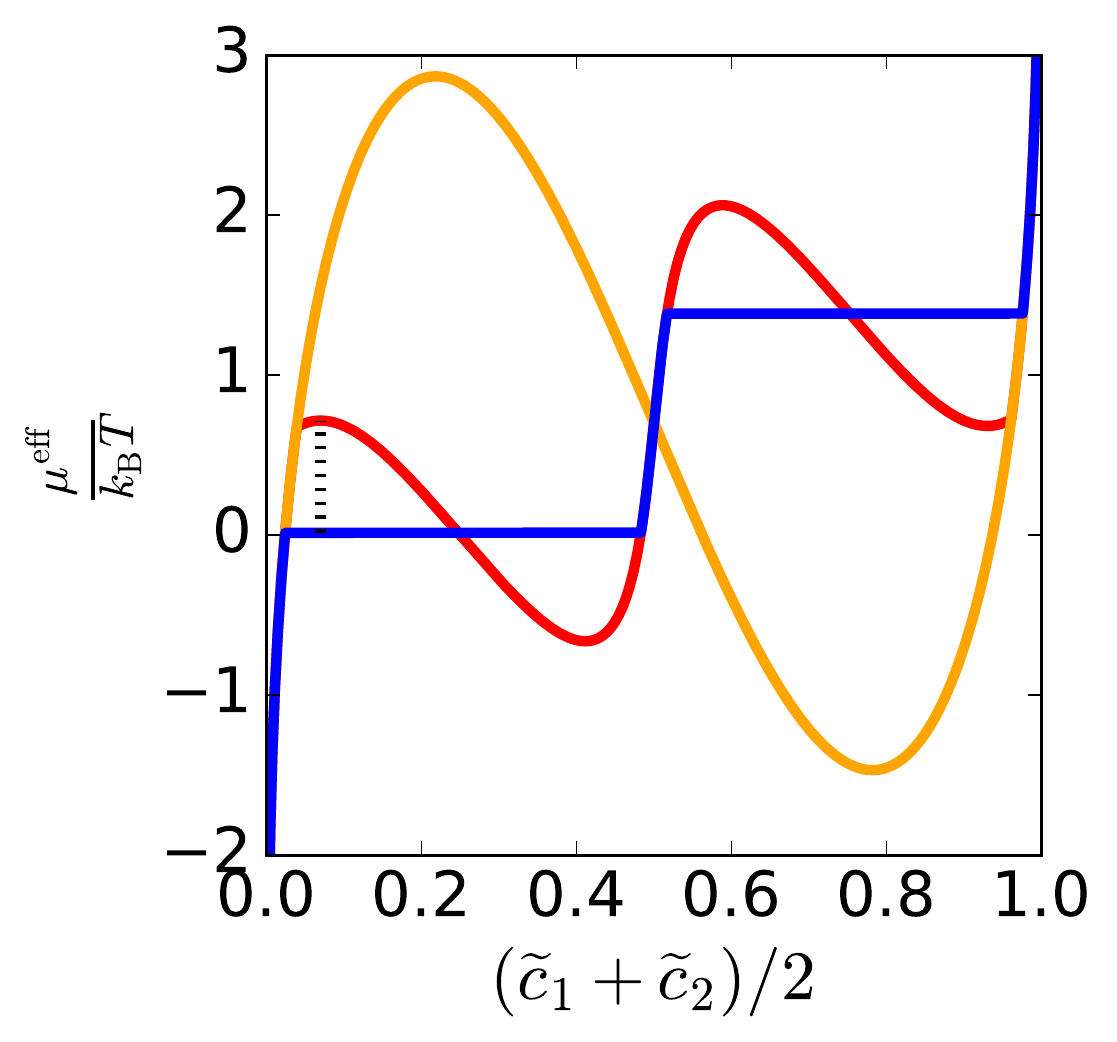}
    \caption{Concentration dependence of the local homogeneous free energy density, $g$ in (a) and (b), and the effective chemical potential, $\mu^\mathrm{eff}$ in (c) assuming minimal contribution from intra-layer concentration gradients. The plot in (b) is shaded to correspond to the surface in (a) to more clearly show the colored paths which represent selected filling/emptying paths which are explained in the text.}
    \label{fig:NRGD}
\end{figure}

For the inter-layer interactions, we propose a model incorporating 2- and 4-body interactions,
\begin{align}
    g = g_\ell(\wt{c}_1, \bnab \wt{c}_1) + g_\ell(\wt{c}_2, \bnab \wt{c}_2) +
    c_{\textrm{ref}}\left[\Omega_b\wt{c}_1\wt{c}_2 + \Omega_c\wt{c}_1(1-\wt{c}_1)\wt{c}_2(1-\wt{c}_2)\right],
    \label{eq:totalFED}
\end{align}
where subscripts 1 and 2 represent the repeated layers 1 and 2 respectively. For graphite, the $\Omega_b$ parameter is related to a repulsive interaction between cross-plane lithium~\cite{safran1979}, and it allows for prediction of the ``staircase'' voltage plateau measured in near-equilibrium (dis)charge curves for graphite electrodes~\cite{ohzuku1993,ferguson2014} by increasing the free energy of the full stage 1 state compared to that of the stage 1' and stage 2 states (see Figure~\ref{fig:NRGD}). The $\Omega_c$ term can be interpreted as a particle-vacancy cross-plane mixing enthalpy, which Ferguson and Bazant found to be necessary to push particles toward stage 2 in the filling process rather than along the $\wt{c}_1=\wt{c}_2$ direction, although we will reexamine this in Sections~\ref{sec:spinodal} and~\ref{sec:CC}. The primary change of this free energy model beyond that of Ferguson and Bazant~\cite{ferguson2014} is the gradient penalty in Eq.~\ref{eq:layerFED}, which allows us to predict intra-layer phase separation. The thermodynamic treatment is similar to that in previous work describing staging in graphite~\cite{hawrylak1984,safran1980b,kirczenow1984}. For a review of other theoretical work on graphite staging, the reader is directed to the informative ref.~\cite{kirczenow1990}.

Here, it is helpful to examine the basic thermodynamic structure of the model, which is presented in Figures~\ref{fig:gRS} and~\ref{fig:NRGD}. Using the thermodynamic parameters from Section~\ref{sec:implementation} (and choosing $\mu^\Theta = 0$), the free energy density of the system, $g$, is plotted in Figure~\ref{fig:NRGD}~(a) and~(b) as a function of the filling fractions in each layer, $\wt{c}_1$ and $\wt{c}_2$, assuming minimal energetic contribution from concentration gradients. Because the predicted free energy density is not a convex function, the system will tend to phase separate. For a simpler, single-variable free energy model such as the regular solution model applied to individual layers (Figure~\ref{fig:gRS}), the equilibrium phase separating behavior can be described by creating a common tangent line between the energy wells. When the concentration of a homogeneous system lies between the two intersection points, the system can lower its free energy to the common tangent by separating into two phases, each of which has equilibrium concentration set by the intersection points. The overall ratio of phase amounts is set by the lever rule. For the two-variable free energy model developed here, we instead construct common tangent planes, pictured in green in Figure~\ref{fig:NRGD}~(a). Similar to the simpler case, the intersections of these planes with the homogeneous free energy surface, depicted as red dots, correspond to the stable equilibrium concentrations of the phases that will develop from overall system concentrations lying within the triangles formed by the three red dots of a given plane. This predicts that the equilibrium concentrations of stage 2 phases depend on whether the system is all in stage 2, in equilibrium with state 1' (lower plane), or in equilibrium with stage 1 (upper plane).

In Figure~\ref{fig:NRGD}~(a) and~(b), three paths are plotted, which correspond to selected filling/emptying paths as the graphite is (de)intercalated. The orange curve represents a fully homogeneous (and energetically unfavorable) filling path in which $\wt{c}_1 = \wt{c}_2$. The red curve represents the minimum energy path for filling with individual layers remaining fully homogeneous (no intra-layer phase separation). The blue curve represents an equilibrium path allowing intra-layer phase separation. Following these paths motivates the definition of a quantity which we will refer to as an effective chemical potential,
\begin{align}
    \mu^\mathrm{eff} = \frac{\partial g^*}{\partial\left( c_1 + c_2 \right)}.
    \label{eq:muEff}
\end{align}
This quantity represents the change in free energy along the  given filling path, $g^*$, to insert a particle in either of the layers; it is dependent on the state of the system. $\mu^\mathrm{eff}$ is related to predicted equilibrium voltage curves~\cite{ferguson2014}, as we will see in Section~\ref{sec:OCV}. Examples of its values over the three selected filling paths are presented in Figure~\ref{fig:NRGD}~(c).

Returning to the formulation of the model, we now focus on the kinetic processes. The interstitial lithium flux, $\textbf{F}_i$, occurs within each modeled plane of sites and is driven by gradients in the diffusional chemical potential~\cite{groot1962}, $\mu_i$, defined as the variational derivative of the free energy with respect to the concentration within the layer~\cite{cahn1962},
\begin{align}
    &\mu_i = \frac{1}{c_{\textrm{ref}}}\frac{\delta G}{\delta \wt{c}_i}
    = \frac{1}{c_{\textrm{ref}}}\left( \frac{\partial g}{\partial \wt{c}_i} -
    \dvg\frac{\partial g}{\partial\bnab \wt{c}_i}\right).
    \label{}
\end{align}
Using Eqs.~\ref{eq:layerFED} and~\ref{eq:totalFED},
\begin{align}
    &\mu_i = k_\mathrm{B}T\ln\left(
    \frac{\wt{c}_i}{1-\wt{c}_i} \right) + \Omega_a(1-2\wt{c}_i) -
    \frac{\kappa}{c_{\textrm{ref}}}\nabla^2\wt{c}_i
    + \Omega_b\wt{c}_j + \Omega_c(1-2\wt{c}_i)\wt{c}_j(1-\wt{c}_j) + \mu^\Theta,
    \label{eq:chemPot}
\end{align}
where $j \ne i$.

\subsection{Concentrated Solid Diffusion}
From the assumptions of linear irreversible thermodynamics~\cite{groot1962}, the flux can be expressed as
\begin{align}
    &\textbf{F}_i = -\frac{D_i}{k_\mathrm{B}T}c_i\bnab \mu_i.
    \label{eq:flux_base}
\end{align}
where $D_i$ is the tracer diffusivity, related to the mobility by the Einstein relation, $D_i = M_{i}k_\mathrm{B}T$. Whereas Hawrylak and Subbaswamy chose a constant flux prefactor investigating a similar model~\cite{hawrylak1984}, we propose a flux relationship describing diffusion on a lattice~\cite{bazant2013}. First, we begin with tracer diffusivity given by
\begin{align}
    D_i = D_0\frac{\gamma_i}{\gamma_{\ddagger,i}^d},
    \label{eq:flux_prefactor_gammas}
\end{align}
where $D_{0}$ is the tracer diffusivity in the dilute limit, $\gamma_i$ is the activity coefficient of the diffusing species, and $\gamma_{\ddagger,i}^d$ is that of the diffusing transition state. Following the assumptions outlined in Appendix~\ref{sec:appTransCoeff}, we find
\begin{align}
    &\textbf{F}_i = -\frac{D_{0}}{k_\mathrm{B}T}(1-\wt{c}_i)c_i\bnab \mu_i.
    \label{eq:flux_with_prefactor}
\end{align}
Here, we have assumed that the transition state only differs thermodynamically from lithium in a stable lattice state in the entropic contribution. However, the enthalpic terms could differ, and extra, local stresses could also affect the energetics of the diffusing transition state via an activation volume\cite{aziz1991,bazant2013}.

Using these flux relations, we can solve for the time and space evolution of concentration profiles within the layers using species conservation,
\begin{align}
    \frac{\partial c_i}{\partial t} = -\dvg\mathbf{F}_i + R_i
    \label{eq:massCons}
\end{align}
where $R_i$ is a volumetric source/sink term, which could describe, e.g., inter-layer reaction in which Li can move from layer 1 to layer 2 in the bulk via defects (Li hopping through pristine graphene is unlikely because of large energetic barriers~\cite{meunier2002,yao2012diffusion}). We will set $R_i = 0$ here except as noted in Section~\ref{sec:CCRxn}.

\subsection{Thermodynamically Consistent Reaction Boundary Conditions}
The system is closed by specifying boundary conditions on the spatially 4th order partial differential equation for each layer, according to the Cahn-Hilliard reaction model for heterogenous reactions~\cite{bazant2013,zeng2014}. The concentration at the surface of the particle is governed by the natural boundary condition~\cite{cahn1977,bazant2013}, $\widehat{\textbf{n}}\cdot\left(\kappa\bnab \wt{c}_i\right)_S = \frac{\partial\gamma_S}{\partial \wt{c}_i}$, where $\widehat{\textbf{n}}$ is an outward facing unit normal vector, the $S$ subscript indicates evaluation at the surface, and $\gamma_S$ is the surface energy. Although surface ``wetting'' has been shown to be important in nanoparticle dynamics~\cite{cogswell2013}, we set $\widehat{\textbf{n}}\cdot\left(\kappa\bnab \wt{c}_i\right)_S = 0$ here. The remaining boundary condition is a flux specification at the surface, which is determined by the dynamic process being examined. For example, in order to study spinodal decomposition, no flux boundary conditions can be imposed~\cite{cahn1962}. Here, because we are interested in examining battery operation, we specify the total current, $I$, into the particle as the sum of integrated current densities, $J_i$, into the perimeter of the particle $S$,
\begin{align}
    I = H\sum_{i=1}^2\int_{S}^{}f_{i}J_{i}\ud S
%    = H\int_{S}^{}\left(f_1J_1 + f_2J_2\right)\ud S
    \label{}
\end{align}
where $H$ is the height of the particle and $f_1 = f_2 = 0.5$ are the fractions of the surface area occupied by each layer.
The current density into each layer is related to the flux from the electrochemical reaction associated with intercalation into each of the repeating layers,
\begin{align}
%    \textrm{Li}^\textrm{+} + \mathrm{C}_6 + \textrm{e}^\textrm{-} \to \textrm{LiC}_6.
    \textrm{Li}^\textrm{+} + \textrm{e}^\textrm{-} \to \textrm{Li}.
    \label{}
\end{align}
We describe the reaction using a thermodynamically consistent modification~\cite{bazant2013} to the Butler-Volmer equation~\cite{bard2001}, which depends on the diffusional electrochemical potentials of the oxidized ($\textrm{Li}^\textrm{+} + \textrm{e}^\textrm{-}$) and reduced ($\textrm{Li}$) states as well as the activity coefficient of the reaction transition state,
\begin{align}
    \frac{J_i}{e} =
    -\widehat{\textbf{n}}\cdot\left(\textbf{F}_i\right)_S =
    \frac{f_ik_0a_O^{1-\alpha}a_{i}^{\alpha}}{e\gamma_{\ddagger,i}}
    \left[ \exp\left(-\alpha e\eta_i/k_\mathrm{B}T\right) -
    \exp\left(\left(1-\alpha\right)e\eta_i/k_\mathrm{B}T\right) \right]
    ,
    \label{eq:rxn}
\end{align}
where $i$ subscripts indicate variables corresponding to layer $i$, $k_0$ is the reaction rate constant defined per unit reaction area of the particle, $e$ is the elementary charge, $a_O$ is the activity of the oxidized state (assumed to be unity here), $a_{i}$ is the activity of the reduced state in layer $i$ ($k_\mathrm{B}T\ln{a_{i}} = \mu_i-\mu^\Theta$), $\alpha$ is the reaction symmetry factor, $\gamma_{\ddagger,i}$ is the activity coefficient of the transition state for reaction into/out of layer $i$, and $\eta_i$ is the activation overpotential, or reaction driving force, for layer $i$ ($e\eta_i = \mu_{i} - \mu_O$). $\mu_O$ is the diffusional electrochemical potential of the oxidized state,
\begin{align}
    \mu_O = \mu_O^\Theta + e\left(\phi_\mathrm{elyte} - \phi_\mathrm{sld}\right) = \mu_O^\Theta - e\Delta\phi
    \label{}
\end{align}
where $\mu_O^\Theta$ is the reference chemical potential in the oxidized state, $\phi_\mathrm{elyte}$ is the electric potential in the electrolyte, and $\phi_\mathrm{sld}$ is the electric potential in the graphite. Thus, we see that the total integrated current will be directly related to the applied interfacial potential, $\Delta\phi = \phi_\mathrm{sld} - \phi_\mathrm{elyte}$, which will allow simulations with either specified current or specified voltage. We note that use of the Butler-Volmer equation ignores recent evidence of departures from its predictions, especially at large driving forces~\cite{chidsey1991,bai2014,laborda2013}. Here, it is convenient to group the standard potentials, $eE^\Theta = \mu^\Theta - \mu_O^\Theta$, to be the standard half-cell potential relative to the Li/Li$^+$ metal standard half-cell potential.

\begin{figure}[h]
    \centering
    \includegraphics[width=0.4\textwidth]{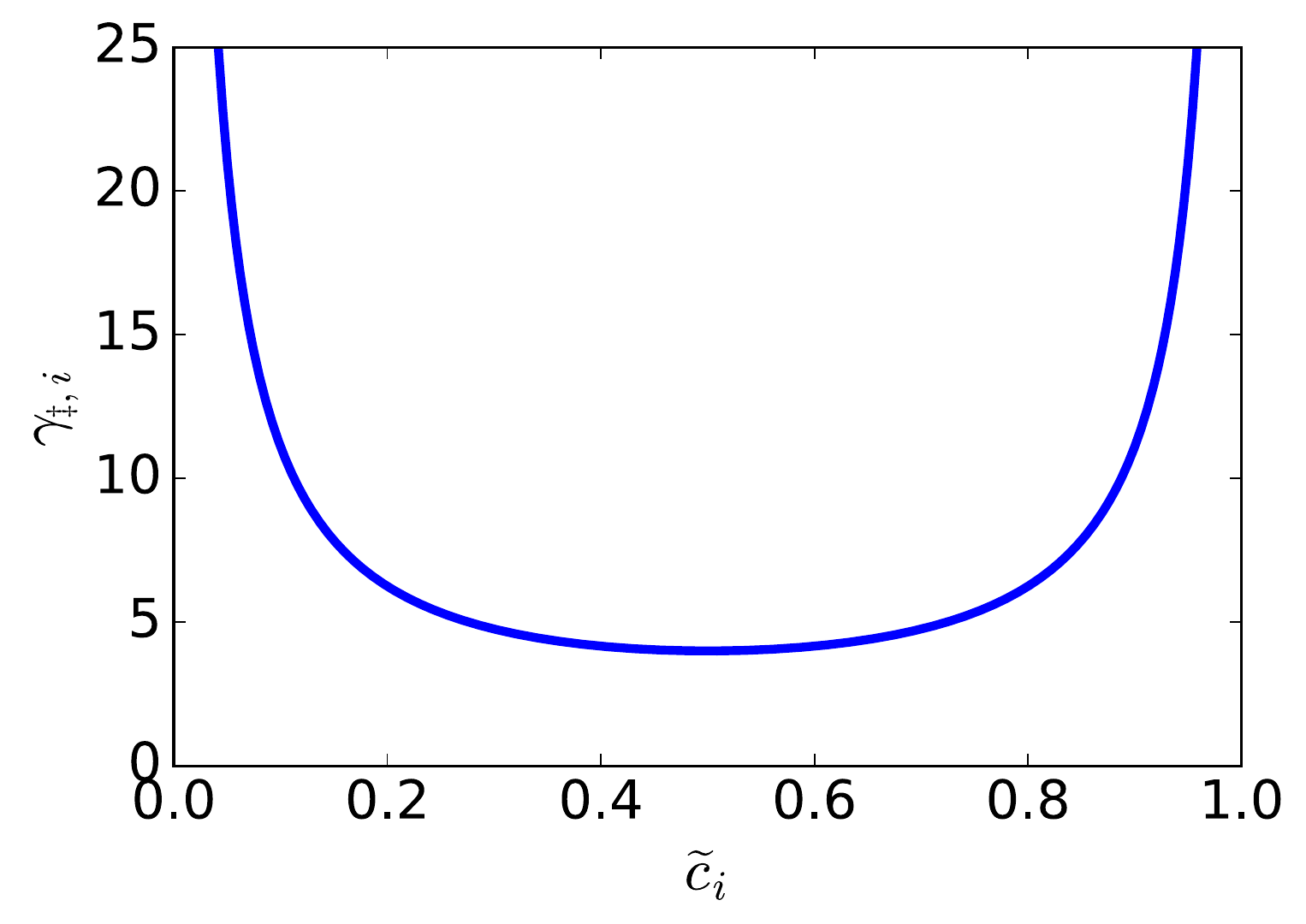}
    \caption{Reaction transition state contribution to (de)intercalation rate. The exchange current density for a layer scales with $\gamma_{\ddagger,i}^{-1}$, so the resistance scales with $\gamma_{\ddagger,i}$.}
    \label{fig:gamma_TS}
\end{figure}

The reaction transition state activity coefficient is used to describe physics of the reaction process. For example, for Li intercalation into the rigid ion channels of the positive electrode material FePO$_4$ (iron phosphate), Bai et al.\@ assumed the reaction transition state requires (excludes) a single empty lattice site, such that $\gamma_\ddagger = 1/(1-\wt{c})$~\cite{bai2011}. To describe reaction at the surface single crystal of graphite, ref.~\cite{guo2016} used
\begin{align}
    \gamma_{\ddagger,i} = \frac{1}{\wt{c}_i\left(1-\wt{c}_i\right)},
    \label{eq:gamma_ts}
\end{align}
which is based on the assumption that a reaction event in layer $i$ requires both a vacant site and a nearby filled site for the reaction to occur. This requires that an empty site be present near the surface for intercalation. In addition, when there is no nearby filled site, the graphene layers can collapse, making the reaction unlikely. Thus Eq.~\ref{eq:gamma_ts} predicts that the reaction resistance for the layer, which scales with $\gamma_{\ddagger,i}$, diverges at empty and full filling fractions, as presented in Figure~\ref{fig:gamma_TS}.
This implicitly assumes that the reaction transition state behaves like an ideal solution except for entropic exclusion effects, unlike the diffusion transition state (Eq.~\ref{eq:TS_d}), which we assume also depends on all the enthalpic terms of the diffusional chemical potential because it occurs in the bulk of the system rather than at the interface between one phase and another. We use this reaction model here except where noted to demonstrate some of its predictions. However, because of an inability to carefully experimentally control for other kinetic limitations in ref.~\cite{guo2016}, more detailed consideration of this reaction model would be worthwhile. For example, Thomas-Alyea et al.~\cite{thomas-alyea2016} find when describing secondary graphite particles with a variant of this model (developed in Section~\ref{sec:modSimp}) that a reaction model dominated by film resistance enabled much better agreement with the experimental results. Thus, we vary this model where noted below.

\section{Model Implementation}
\label{sec:implementation}
The energetic parameters in the free energy are taken as those fit to thermodynamic data in previous work~\cite{ferguson2014}: $\Omega_a = 3.4\ k_\mathrm{B}T_\mathrm{ref}$, $\Omega_b = 1.4\ k_\mathrm{B}T_\mathrm{ref}$, and $\Omega_c = 20\ k_\mathrm{B}T_\mathrm{ref}$ with $T_\mathrm{ref} = 298\ \mathrm{K}$, and the system is simulated at room temperature, $T=T_\mathrm{ref}$. Here, we have adjusted their fit value of $E^\Theta$ from 0.1366 V to 0.12 V in order to account for the fact that our particles can phase separate within the layers, which changes the observed voltage plateau. The difference corresponds to the dimensional length of the black dashed line in Figure~\ref{fig:NRGD} (c) as explained previously~\cite{levi1997simultaneous,dreyer2010} and demonstrated for a different material model in the same paper~\cite{ferguson2014}. With the intra-layer phase separating particles, the value of $E^\Theta$ corresponds to the voltage of the low-filling-fraction plateau, which we find to be 0.12 V from near-open-circuit experimental data~\cite{ohzuku1993}. We choose $c_\mathrm{max} = 28.2\ \mathrm{M}$ to match the experimental and theoretically calculated density of lithiated graphite~\cite{wissler2006,kganyago2003}.

The values of $D_0$ and $\kappa$ are taken from ref.~\cite{guo2016} in which we used this model and found excellent agreement with optical measurements of the concentration profiles of a single graphite crystal in time and space. From this, $D_0 = 1.25\times10^{-12}\ \textrm{m}^2/\textrm{s}$ and $\kappa = 4\times10^{-7}$ J/m, although the fit was relatively insensitive to the choice of $\kappa$, and the value of $D_0$ agrees well with \emph{ab initio} calculations~\cite{persson2010}. The reaction symmetry factor, $\alpha$, is set to 0.5, which implies a symmetric electrochemical reaction~\cite{bard2001}. Because the single particle electrode experiment did not allow unique determination of the reaction rate constant, $k_0$, we do not take the value used in ref.~\cite{guo2016} and instead vary this parameter here as stated.

Inspired by our previous work~\cite{guo2016}, we simulate a cylindrical particle in which intercalation occurs around the circumference of the particle into the circular layers. For simplicity, we assume axial symmetry about an axis normal to the simulated plane in the center of the particle, allowing us to reduce the model to solving in time, $t$, and the radial coordinate, $r$. We use a mass-conserving variant of finite volume discretization in space following ref.~\cite{zeng2014}, ensuring that the grid spacing is less than the scale for the interfacial width, $\lambda_b\sim\left( \kappa/\left( c_\mathrm{ref}\Omega_a \right) \right)^{1/2}=5.8\times10^{-8}\ \mathrm{m}$. Time integration is carried out using DAE Tools~\cite{nikolic2016}, which wraps the SUNDIALS integration suite~\cite{hindmarsh2005} with the ADOL-C automatic differentiation library~\cite{griewank1996} to facilitate the non-linear solver involved in the implicit time stepping. Single particle simulations took between a few minutes and a few hours on one CPU\@. In the following two sections, we validate the simulation by implementing the same model in COMSOL Multiphysics and comparing outputs, then using the COMSOL Multiphysics implementation in 2D to examine the validity of the assumption of axial symmetry in context of the experiment simulations from ref.~\cite{guo2016}.

\subsection{Simulation Validation}
\label{sec:validation}
\begin{figure}[h]
    \centering
    (a)
    \includegraphics[width=0.4\textwidth]{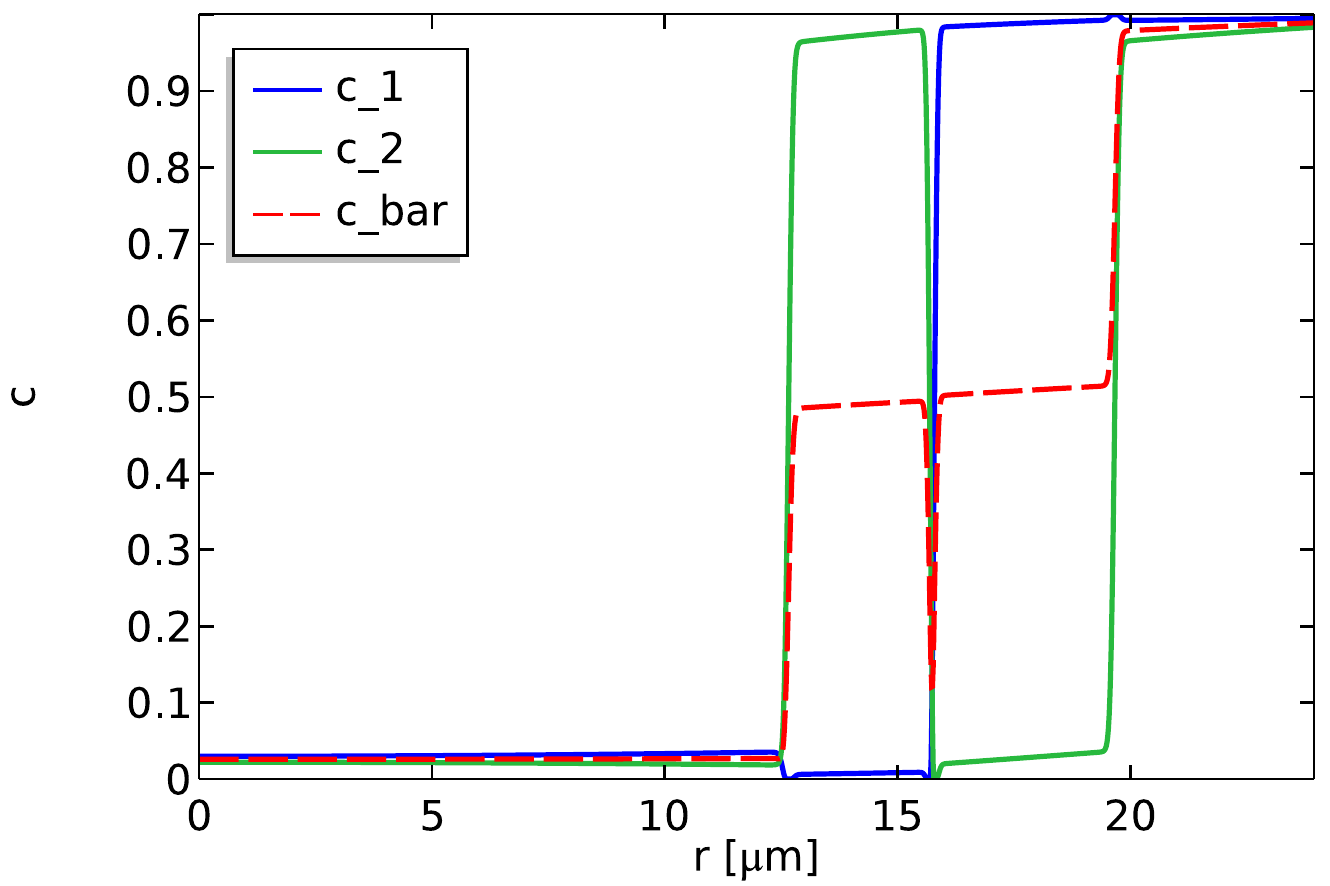}
    \ \
    (b)
    \includegraphics[width=0.4\textwidth]{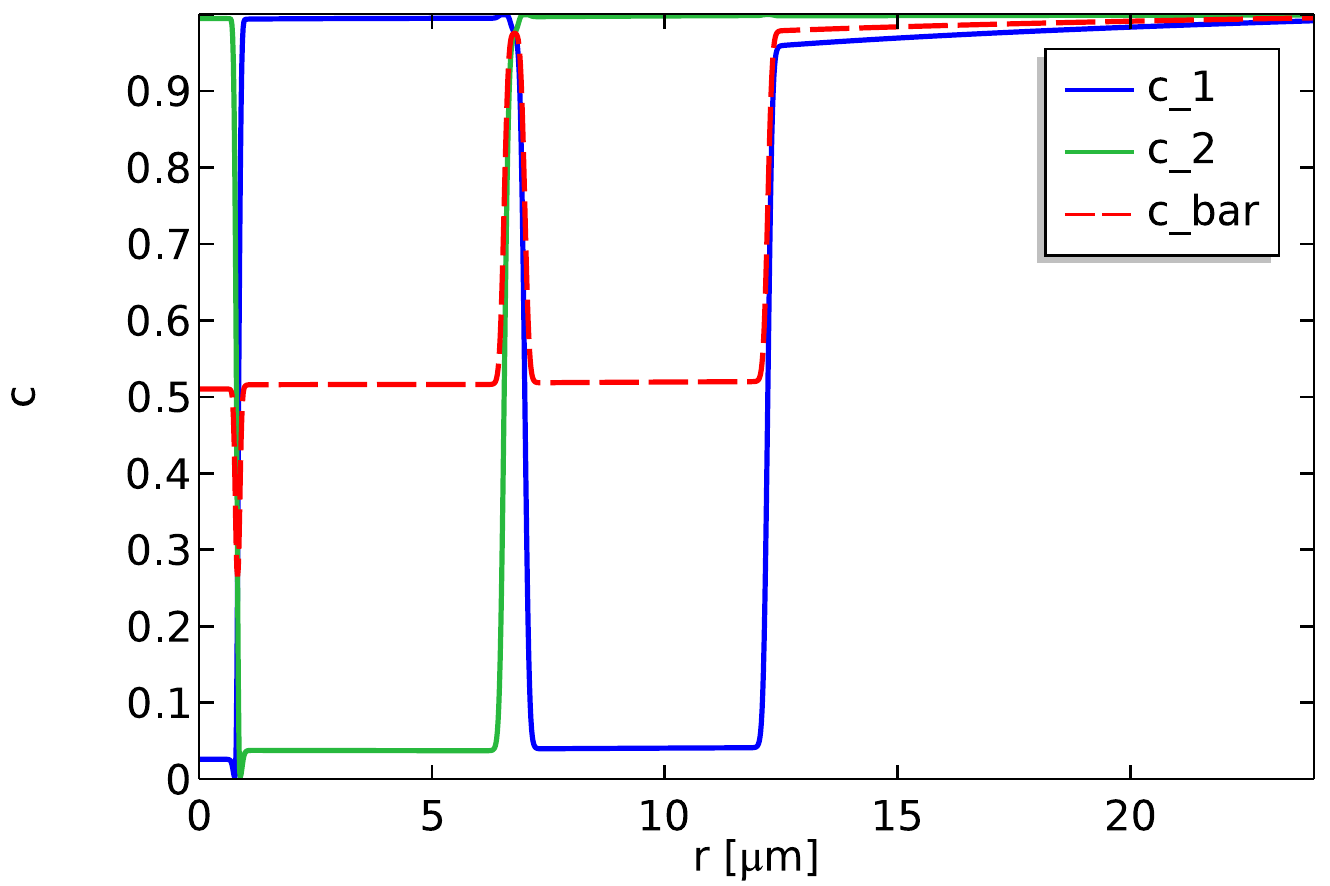}
    \\
    (c)
    \includegraphics[width=0.4\textwidth]{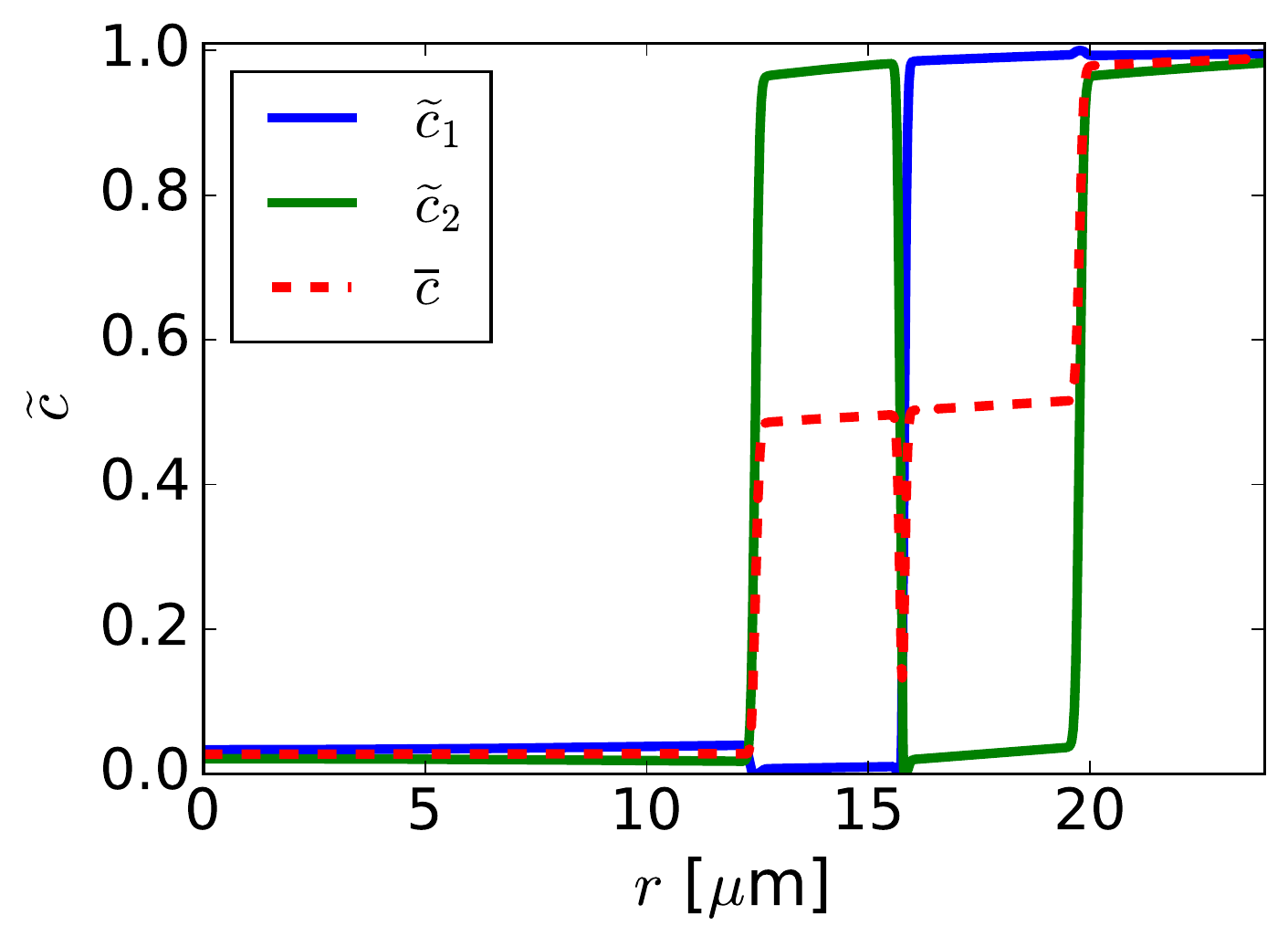}
    \ \
    (d)
    \includegraphics[width=0.4\textwidth]{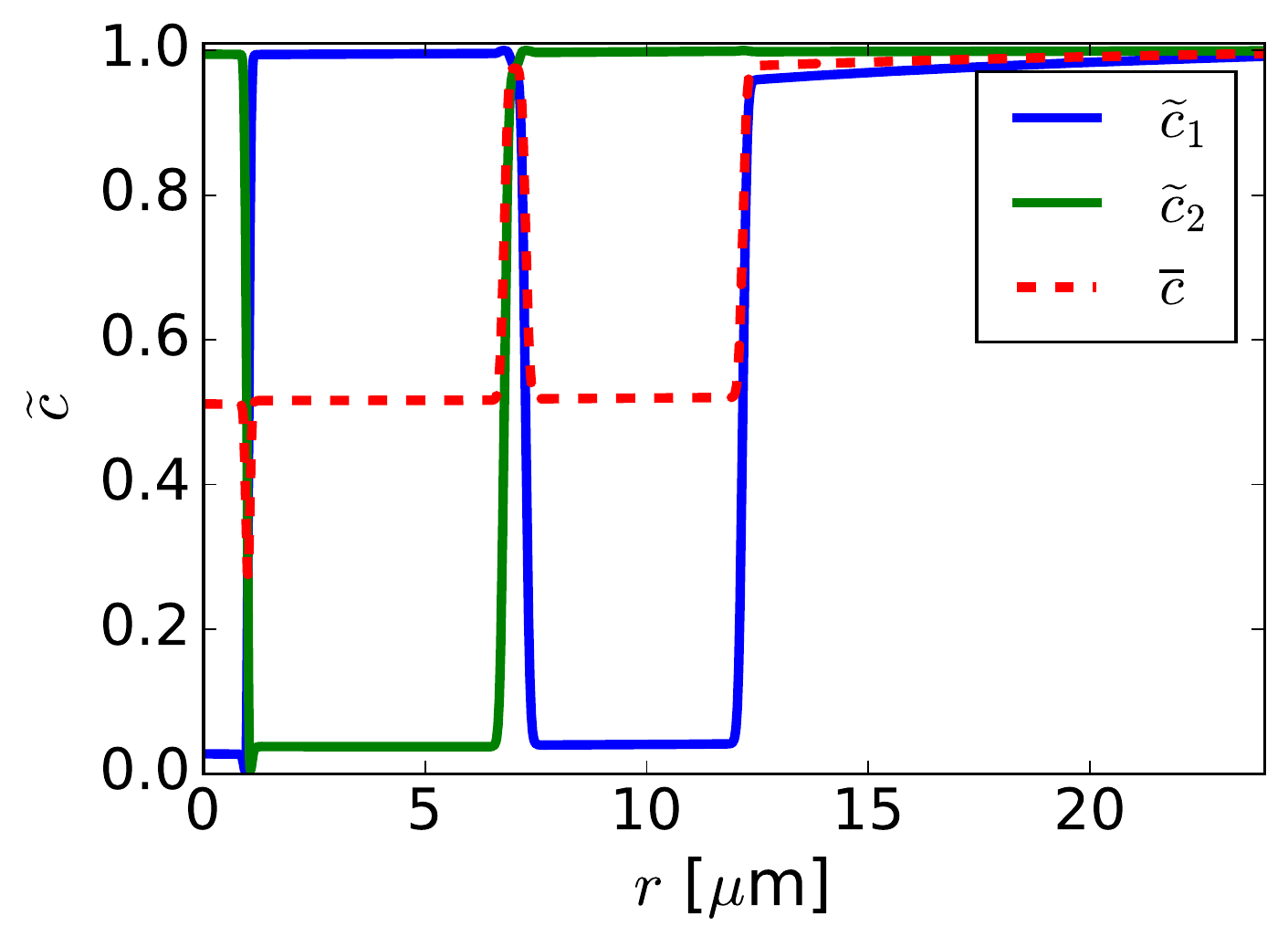}
    \caption{Simulations from ref.~\cite{guo2016} performed using COMSOL Multiphysics (finite element discretization) and finite volume discretization. Snapshots in (a) and (b) are obtained from COMSOL Multiphysics and those in (c) and (d) are obtained from finite volume discretization. Snapshots in (a) and (c) are taken at $t = 2070\ \textrm{s}$ and those in (b) and (d) are taken at $t = 4140\ \textrm{s}$. $\overline{c}$ is defined in Eq.~\ref{eq:cbar}.}
    \label{fig:1d_comparison}
\end{figure}

Beginning from the simulations in ref.~\cite{guo2016}, we first validate the 1D finite volume simulation by implementing the same model in COMSOL Multiphysics, which uses the finite element method, to verify that both methods give similar results. In each case, we use the parameters given in ref.~\cite{guo2016}: $R=23.9\ \mu\textrm{m}$, $c_\textrm{ref} = 9.3\ \textrm{M}$, $k_0 = 0.03\ \textrm{A/m}^2$, $V_\textrm{set} =-0.38\ \textrm{V}$ with respect to Li/Li$^+$. The low value of $c_\textrm{ref}$ could be thought of as adjusting the value of $\kappa$, as the two compensate for each other (changing $c_\textrm{ref}$, $k_0$, and $\kappa$ by the same factor leads to the same simulation outputs). Because the results in ref.~\cite{guo2016} were not sensitive to the value of $\kappa$, we simply use their reported value of $\kappa$ directly here with the value for $c_\textrm{ref}$ in Section~\ref{sec:implementation} for the remainder of this work. Using a grid spacing of $1.5 \times 10^{-8}\ \textrm{m}$, both simulations reproduce the previous results and agree with each other in Figure~\ref{fig:1d_comparison}. We note that the model is symmetric in $\wt{c}_1$ and $\wt{c}_2$. Therefore, in both the COMSOL and finite volume simulations, $\wt{c}_1$ and $\wt{c}_2$ profiles can be flipped, and the occurrence of this flipping is random and depends on numerical noise. Full movies are provided in the supplement.

\subsection{Axial Symmetry Assumption}
\label{sec:axsym}
\begin{figure}[!h]
    \centering
    (a)
    \includegraphics[width=0.4\textwidth]{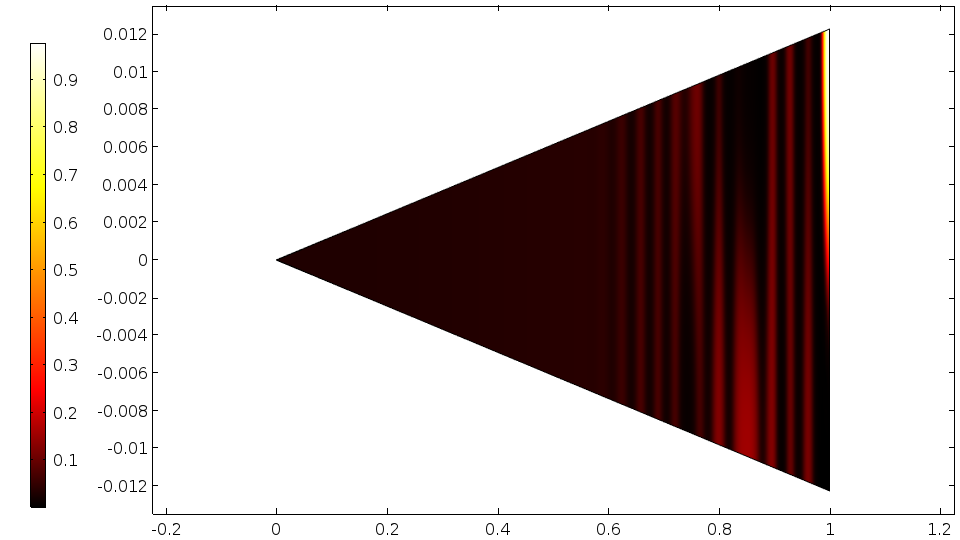}
    \ \
    (b)
    \includegraphics[width=0.4\textwidth]{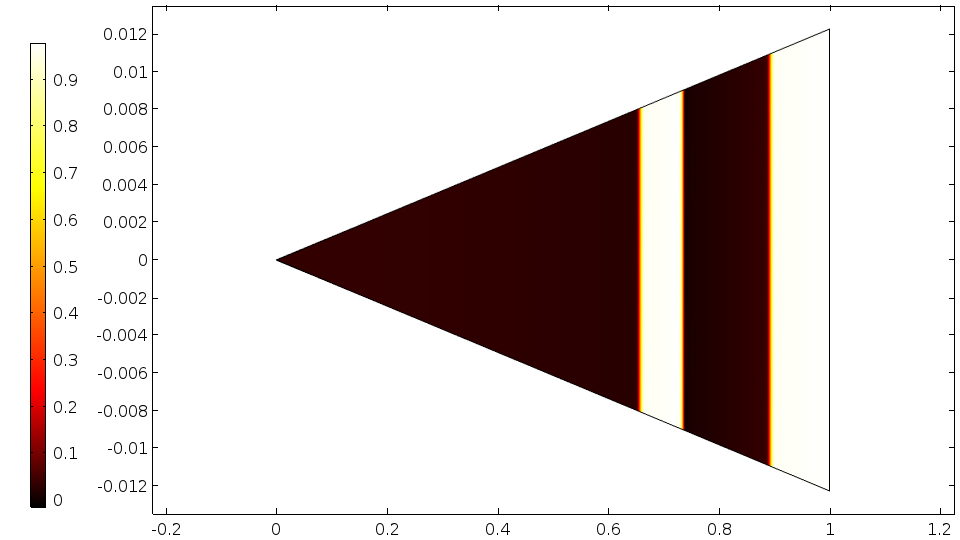}
    \caption{2D versions of simulations from ref.~\cite{guo2016} performed using COMSOL Multiphysics (finite element discretization). The horizontal and vertical axes in the contour plots (a) and (b) are nondimensionlized by the particle radius $R = 23.9\ \mu\textrm{m}$. The plots represent $\wt{c}_1$ at (a) $t = 670\ \mathrm{s}$ and (b) $t = 1550\ \mathrm{s}$ showing axial symmetry initially broken and recovered at later times.}
    \label{fig:2d_sim}
\end{figure}

We examine the assumption of axial symmetry by relaxing it and implementing the resulting 2D model in COMSOL Multiphysics, again using parameters from ref.~\cite{guo2016}. We choose the simulation domain to be the largest possible 2D slice of a circular disk such that the simulation finishes within a few hours on one CPU\@. This periodic simulation domain is replicated as many times as necessary to give the full circular disk. With this constraint, we use a sector with a central angle of $2^{-8} \times 360\degree \approx 1.41\degree$ and a grid spacing of $1.43 \times 10^{-7}\ \textrm{m}$. Taking a slice imposes an extra non-physical length scale on the simulation, related to the arc length of the simulated slice, and small enough slices may prevent formation of phase interfaces with normal components in the $\theta$-direction and artificially encourage axial symmetry. Our simulated arc length of $5.9\times10^{-7}\ \mathrm{m}$ is several times the interface width for this simulation, so we expect this not to prevent breaking of axial symmetry at least near the particle surface. We see in Figure~\ref{fig:2d_sim} that the axial symmetry assumption is broken at early times before complete phase separation occurs, but as the concentration front propagates inward from the particle surface, the particle regains axial symmetry and profiles become nearly identical to that predicted by the 1D model. In Appendix~\ref{sec:appAxsym}, we consider a more detailed comparison of the 2D concentration profiles with a simulation of the same system in 1D. Movies are also provided in the supplement. Overall, at early times, the 1D and 2D cases show moderate differences, but at later times, the simulations become quite similar, so these simulations support the validity of the axial symmetry assumption.

\section{Results}
Because the general intercalation framework is consistently connected to electrochemical reactions, we are able to investigate typical experimental procedures such as constant current and constant voltage (dis)charge. We present here a few example predictions made by the model to highlight some of its key features, with focus on those which deviate strongly from models based on Fickian diffusion. Given initial conditions and a specified current (voltage) profile, the model predicts the concentration profiles $\wt{c}_1(r, t)$ and $\wt{c}_2(r, t)$ as well as the output voltage (current) profile. Here, we also define the local average filling fraction,
\begin{align}
    \overline{c} = \frac{\wt{c}_1 + \wt{c}_2}{2},
    \label{eq:cbar}
\end{align}
which we will relate to the stage number.

In the following sections, we explore a number of model predictions. In Section~\ref{sec:equilibrium}, we focus on (near-)equilibrium properties of the model and relate them to the free energy model by examining the open circuit voltage of a single particle, the equilibrium structures and concentrations, and the phase diagram predicted by the thermodynamic model. In Section~\ref{sec:dynamics}, we examine various predictions the model makes about transient processes. First, in~\ref{sec:CV}, we examine a constant voltage intercalation process, which highlights the overall phase behavior the model predicts, including propagating phase boundaries and diffusive profiles in the stable phases between phase boundaries, as well as the formation of checkerboard patterns during the filling process. In Section~\ref{sec:spinodal}, we study the process of spinodal decomposition of the particle from an initially homogeneous state to stable phases. This again highlights the checkerboard patterns that emerge because of the free energy structure of this model which could not be captured by some other models which are commonly applied to phase separating systems. In Section~\ref{sec:CC}, we examine how the single particle behaves under a constant current discharge with particular attention given to the impact of the reaction rate model on the predicted system voltage. We look at discharge voltage curves both with the reaction model presented in Eq.~\ref{eq:rxn} and an alternate version to highlight the importance of this component of the model in the predictions it makes. We show that the model in Eq.~\ref{eq:rxn} (used in ref.~\cite{guo2016}) does not give predictions qualitatively matching experimental graphite discharge curves, whereas a simple alternative choice works better. We also give some discussion about the significance of this observation in context of applying the model to battery electrodes as in ref.~\cite{thomas-alyea2016}. In Section~\ref{sec:CCRxn}, we repeat the constant current discharge but allow for bulk inter-layer exchange of lithium and note the differences. This exchange is not relevant for pristine graphite, but demonstrates qualitative differences in predictions for systems in which exchange between lattices is possible, including graphite with defects~\cite{yao2012diffusion}.
We conclude the results with Section~\ref{sec:fickCompar}, in which we make some comparisons of this model with a model based on Fickian diffusion (solid solution) with a voltage curve fit to the predicted open circuit voltage (OCV) of the layered phase separating model. We choose to compare the model predictions of the solid solution and phase separating models when exposed to a current pulse within a two-phase plateau, where the models give particularly different predictions of both concentration profiles and predicted voltage.
%In section~\ref{sec:modSimp}, we develop a simplified version of the thermodynamic model, which we apply in ref.~\cite{thomas-alyea2016}, to porous battery electrodes. We comment here about certain changes to the originally proposed flux and reaction models which better describe the behavior of secondary battery particles in battery electrodes.

\subsection{Equilibrium Behavior}
\label{sec:equilibrium}
\subsubsection{Open Circuit Voltage}
\label{sec:OCV}
\begin{figure}[h]
    \centering
    \includegraphics[width=0.5\textwidth]{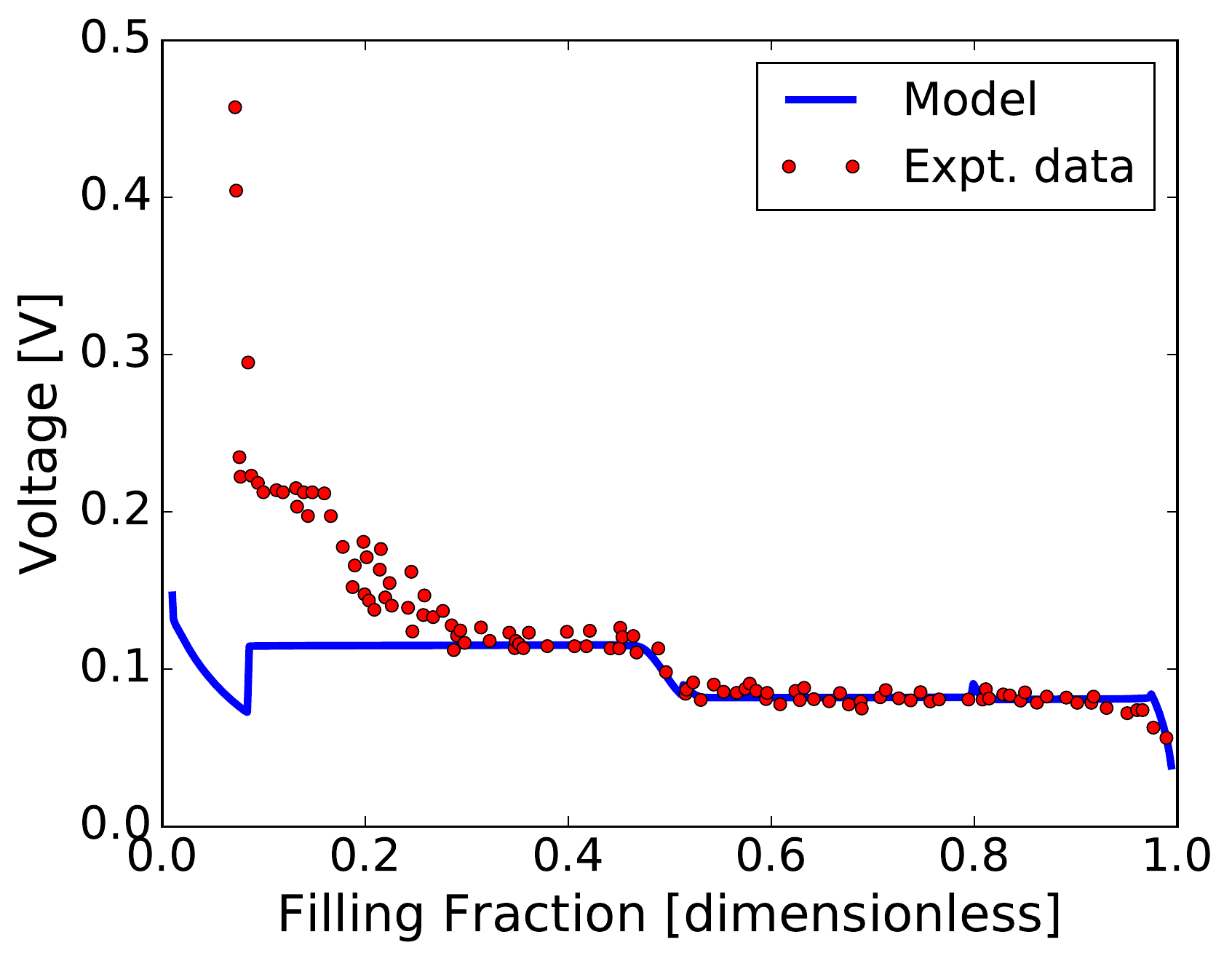}
    \caption{Graphite simulated open circuit voltage of a single particle (vs.\ Li/Li$^+$). This voltage curve represents an electrochemical phase diagram and characterizes equilibrium system behavior. The initial overshoot of the plateau at low filling fraction is caused by the system entering the metastable region before phase separating to the equilibrium structure. Experimental data are from~\cite{ohzuku1993}.}
    \label{fig:graphiteOCV}
\end{figure}

First, we simulate a single particle with radius $R = 10\ \mu\mathrm{m}$ under a constant current discharge at a rate of C/10,000, (where $n$ C corresponds to the rate to (dis)charge the particle in $1/n$ hours). With a rate constant of $k_0 = 0.1\ \mathrm{A/m}^2$, the characteristic reaction time is given by
\begin{align}
    \tau_R \sim \frac{ec_\mathrm{ref}R}{k_0} = 38\ \mathrm{h},
    \label{}
\end{align}
and the characteristic time for species transport is
\begin{align}
    \tau_D \sim \frac{R^2}{D_0} = 80\ \mathrm{s}.
    \label{}
\end{align}
We note that another time scale could be formed for transport over the length of a phase boundary, $\lambda_b$, but that is a significantly shorter time scale. Thus, because the imposed discharge process time scale (10,000 h) is significantly longer than any inherent time scale in the system, we expect the system to be near equilibrium during the entire process. Because the system remains near equilibrium, the system voltage should be determined by a Nernstian relationship,
\begin{align}
    V^\mathrm{eq} = E^\Theta - \frac{\mu^\mathrm{eff}}{e}
    \label{}
\end{align}
where $\mu^\mathrm{eff}$ is defined in Eq.~\ref{eq:muEff} and corresponds to the blue paths in Figure~\ref{fig:NRGD} in which intra-layer phase separation is allowed. This is what we see in Figure~\ref{fig:graphiteOCV}, which also presents experimental data of a very slow graphite electrode discharge~\cite{ohzuku1993}. As the simulation proceeds, lithium initially fills either layer 1 or layer 2. Each layer phase separates internally into high/low concentration regions, until the system is at an overall filling fraction of 0.5 and in a stage 2 structure everywhere. This process corresponds to moving along the lower common tangent plane in Figure~\ref{fig:NRGD}~(a). Then, lithium proceeds to fill regions into a full, stage 1 structure, which corresponds to moving along the upper common tangent plane in Figure~\ref{fig:NRGD}~(a). Thus, in this simulation, the system follows along common tangent planes, similar to the blue paths in Figure~\ref{fig:NRGD}, which predicts the ``staircase'' behavior of the voltage curve. Of note, the initial ``overshoot'' until a filling fraction near $0.1$ is caused by the system entering a metastable region before falling to the stable-equilibrium plateau. A movie of the simulated concentration profiles is included in the supplement.

The disagreement between model and data in Figure~\ref{fig:graphiteOCV} at low filling fraction is related to the model assumption of only having two repeating, structurally similar layers, whereas real graphite can form high stage number structures with different energies~\cite{dahn1991}. Thus, because each layer is treated identically for physical reasons, the first half and second half of the equilibrium discharge voltage curve should look translationally similar, and we are unable to capture the details of the low-filling voltage curve with a 2-layer model. As we have only implemented the model with two layers here, we can only conjecture that adding more layers to the model may enable us to capture the higher stage structures and also the shape of the low-filling open circuit voltage curve.

\subsubsection{Equilibrium Structures}
\label{sec:eqmStruct}
\begin{figure}[h]
    \centering
    (a)
    \includegraphics[width=0.4\textwidth]{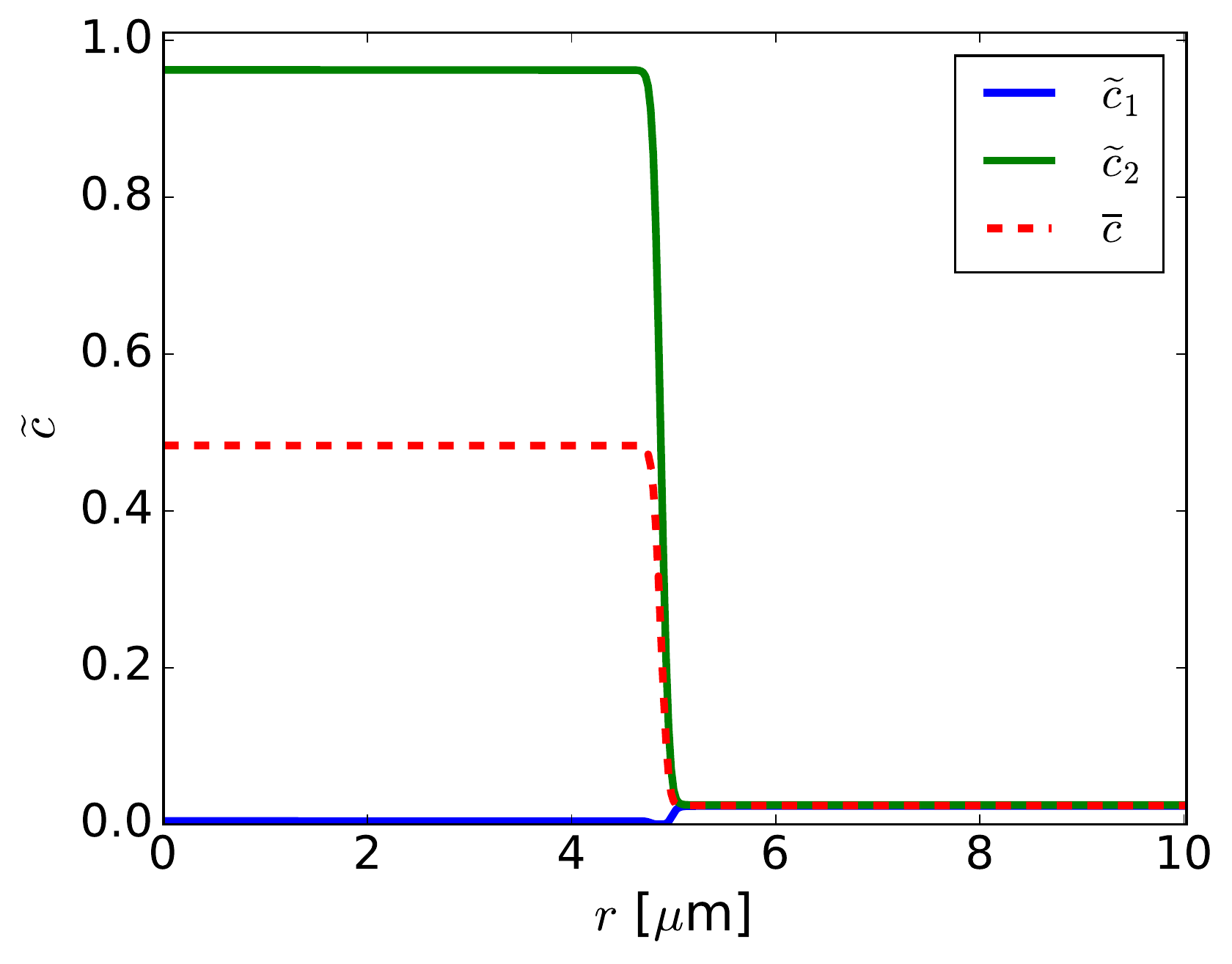}
    \ \
    (b)
    \includegraphics[width=0.4\textwidth]{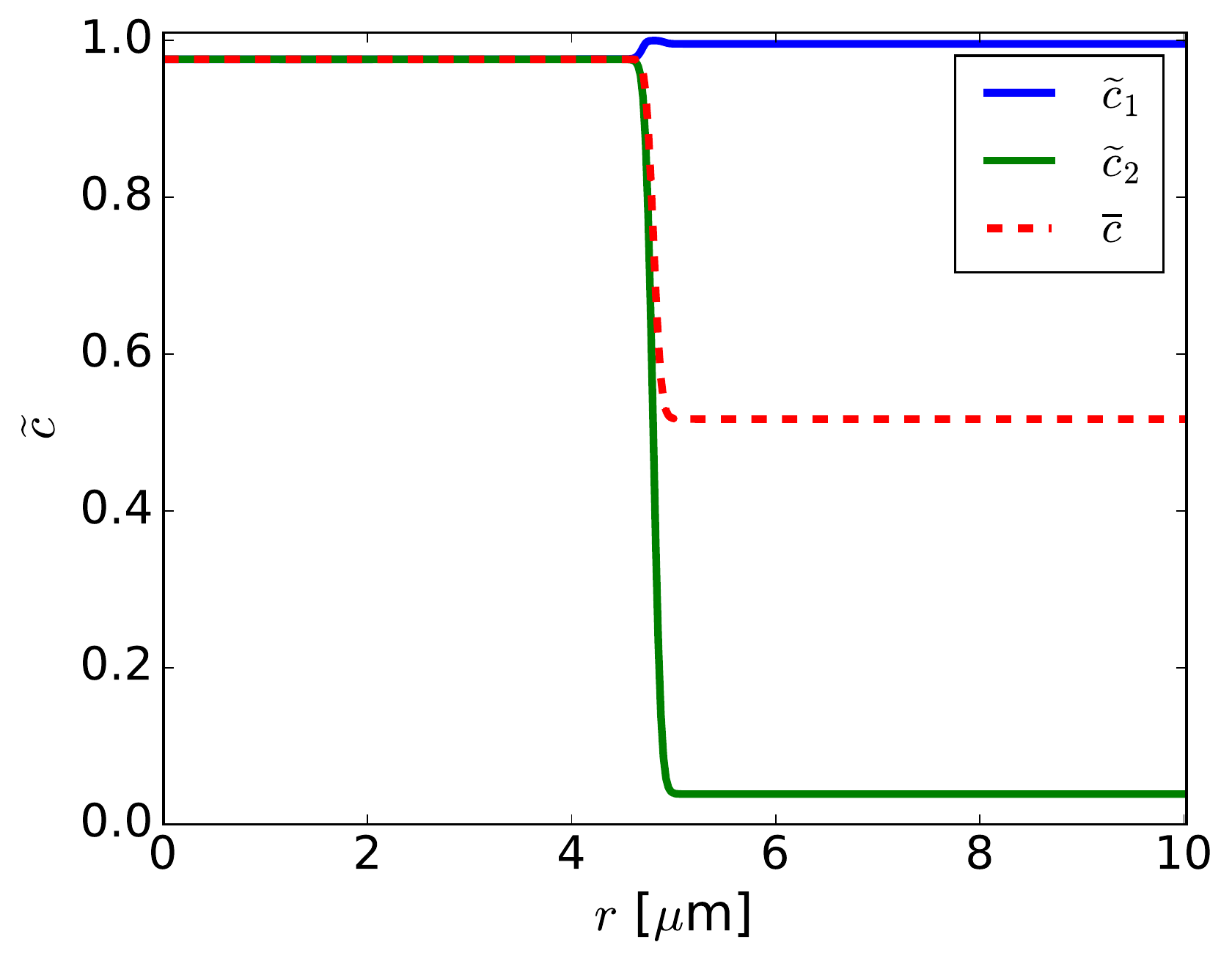}
    \caption{Steady state graphite concentration profiles at zero current. In (a) the concentrations at the edges closely match those of the calculated equilibrium concentrations from the common tangent plane construction for the lower half of the free energy function in Figure~\ref{fig:NRGD} (a). In (b) they match those calculated from the upper common tangent plane.}
    \label{fig:graphite_eqm}
\end{figure}

As described in Section~\ref{sec:model}, we can calculate equilibrium phase concentrations by making the homogeneous free energy density function convex using common tangent planes. Calculation of the planes was carried out numerically and provides values for the expected results of equilibrium structure calculations. In order to compare to results using the full model, we perform simulations to examine the simple cases of stage 1' - stage 2 equilibrium (Figure~\ref{fig:graphite_eqm}~(a)) and stage 1 - stage 2 equilibrium (Figure~\ref{fig:graphite_eqm}~(b)). These equilibrium structures are calculated by initializing the system with uniform high/low concentration in one layer and a step function for concentration in the other and simulating with no current until stage 2 - stage 1 (1') equilibrium is achieved. The transients of the process involve (1) forming the correct interface shape (with width $\sim\lambda_b$), (2) reaching the equilibrium concentrations in the phases, and (3) shifting the interface left/right according to the lever rule to preserve the total average concentration. Simulated final concentrations in each phase agree with those calculated directly from the common tangent construction, and show qualitative agreement with experimental images of lithiated graphite phase equilibrium~\cite{song1996microstructural}.

\subsubsection{Phase Diagram}
\label{sec:phaseDiagram}
One major advantage of models constructed from simpler thermodynamic models like the regular solution is that they make predictions about temperature dependence, unlike solid solution models in which the chemical potential as a function of filling fraction at a particular temperature is directly fit to an open circuit voltage at that temperature. Using the free energy presented here, we can construct phase diagrams following a similar procedure used to add the common tangent planes in Figure~\ref{fig:NRGD}.
We first find common tangent planes (if any) of the free energy surface at a fixed temperature. These indicate the presence of inter-layer phase separation, as demonstrated in Figure~\ref{fig:NRGD}. Then, in regions outside of the planes, common tangent lines are constructed along slices of the free energy at constant average filling fraction (e.g.\ see the stage 2 region in Figure~\ref{fig:NRGD}). The bounds of the stage 2 regions correspond to either (a) entering regions of intra-layer phase separation (entering a region with a common tangent plane), or (b) the largest/smallest filling fractions with a non-convex free energy slice possessing a common tangent.
\begin{figure}[h]
    \centering
    (a)
    \includegraphics[width=0.4\textwidth]{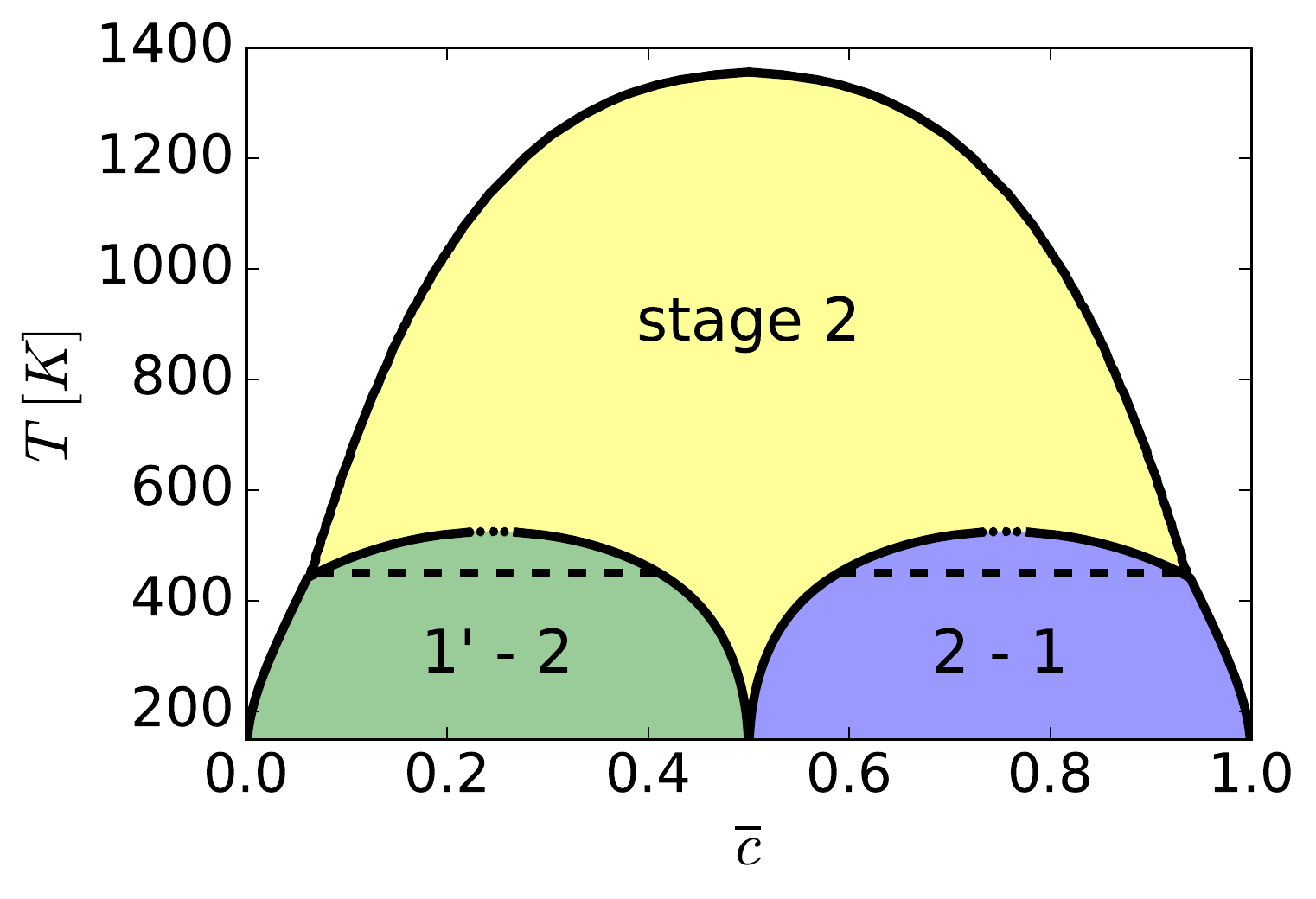}
    \ \
    (b)
    \includegraphics[width=0.4\textwidth]{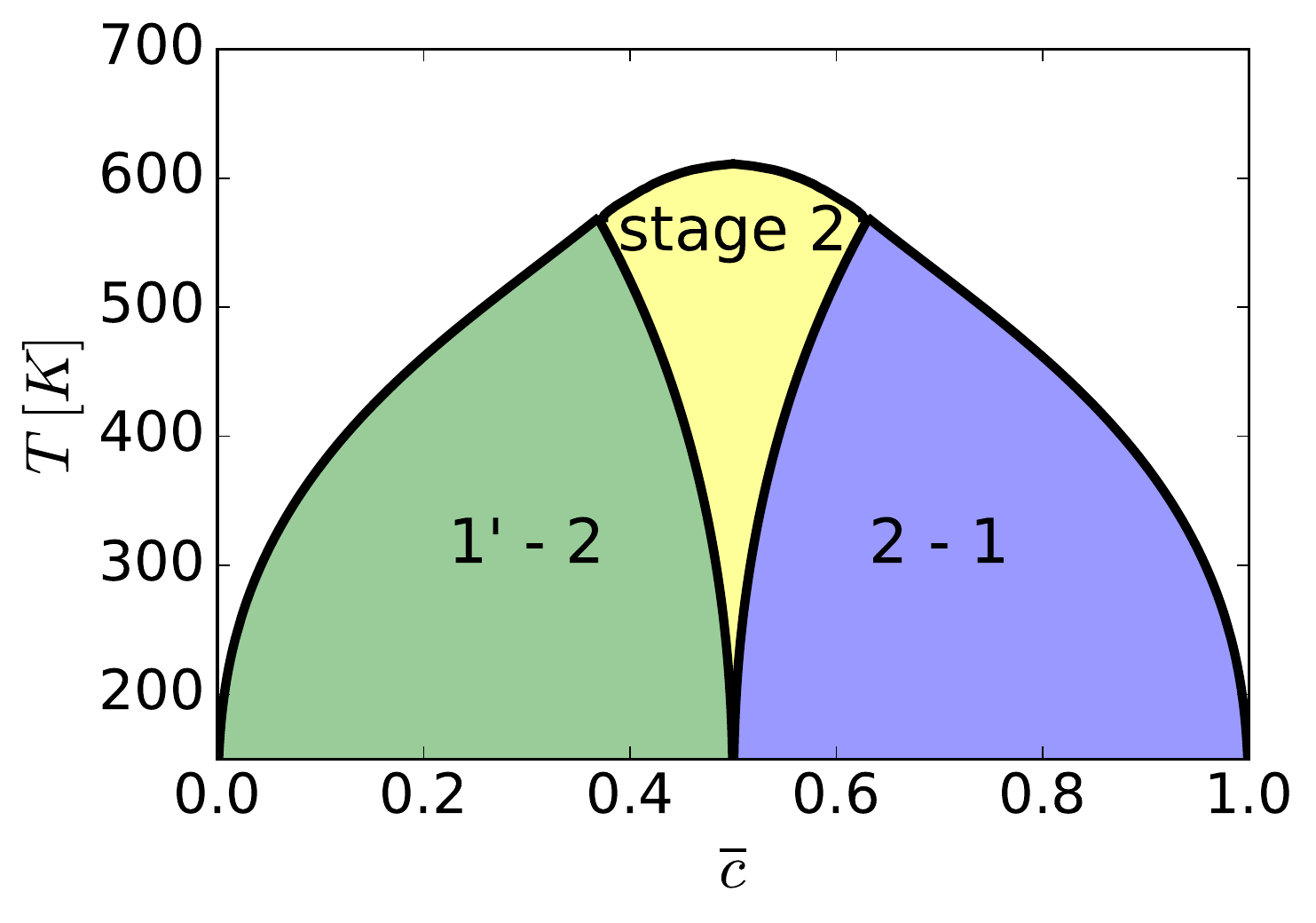}
    \caption{Phase diagram of graphite system with (a) thermodynamic parameters from Section~\ref{sec:implementation} and (b) the same but with $\Omega_c = 0$. The regions above the wide dashed lines in (a) represent stage~2-stage~2 equilibrium. The short dashed lines in (a) are approximations where numerical convergence is difficult.}
    \label{fig:graphite_phase}
\end{figure}

Doing this process over a range of temperatures, we obtain the diagrams in Figure~\ref{fig:graphite_phase}, in which we present the result both for the thermodynamic model parametrized as in Section~\ref{sec:implementation} and also for the same model neglecting the particle-vacancy mixing enthalpy (captured by the $\Omega_c$ term), which leads to similar predictions as made by Safran~\cite{safran1980}. Ferguson and Bazant~\cite{ferguson2014}, who originally introduced the term, did not carefully fit this parameter, as its value had little influence on their results. Here we can see that although the phase diagrams are similar at room temperature, this term strongly influences the predicted equilibrium behavior as temperature increases. Curiously, for the case in Figure~\ref{fig:graphite_phase} (a) near $T=500\ \mathrm{K}$, the common tangent planes intersect the free energy only at points off the $\wt{c}_1 = \wt{c}_2$ line, leaving regions of stage 2 both at intermediate filling fractions and also at low and high filling fractions between the intra-layer phase separation and the fully homogeneous region.

Comparing these diagrams to that of the near-room temperature LiC$_6$ system~\cite{dahn1991,fischer1983,kirczenow1990}, we see again that the model cannot capture the details of the low average filling (high stage number) phases. Also, neither model captures the details of the top line of the phase diagram including the high-temperature stable stage 2 region~\cite{fischer1988,woo1983,kirczenow1990}, although the case with $\Omega_c = 0$ more closely approximates the temperature values at which order disappears (top line of the diagram). Although neither model accurately captures the phase diagram, the result highlights the value of free energy based models as starting points to make predictive temperature-dependent models to capture electrochemical behavior in non-isothermal systems~\cite{reynier2004}. Natural modifications like inclusion of stresses or more careful representation of higher order cluster terms in the free energy would enable a better representation of the phase diagram, and including more layers with longer-range interactions leads to a rich set of phase diagram predictions~\cite{millman1982origin,millman1983study}. For consistency with Ferguson and Bazant~\cite{ferguson2014}, we will use the model as presented in Section~\ref{sec:implementation}, as the focus of this work is at room temperature where the differences between the two variants in Figure~\ref{fig:graphite_phase} are not significant. Importantly, this diagram is a reduction from the complete $\wt{c}_1,\ \wt{c}_2$ space for visual simplicity. For example, within the ``stage 2'' region, the system is actually phase separated with $\wt{c}_1 \ne \wt{c}_2$, and the model predicts the values of these equilibrium concentrations, but they are not shown here.

\subsection{Predicted Dynamics}
\label{sec:dynamics}
\subsubsection{Constant Voltage Intercalation}
\label{sec:CV}
\begin{figure}[h]
    \centering
    \includegraphics[width=0.4\textwidth]{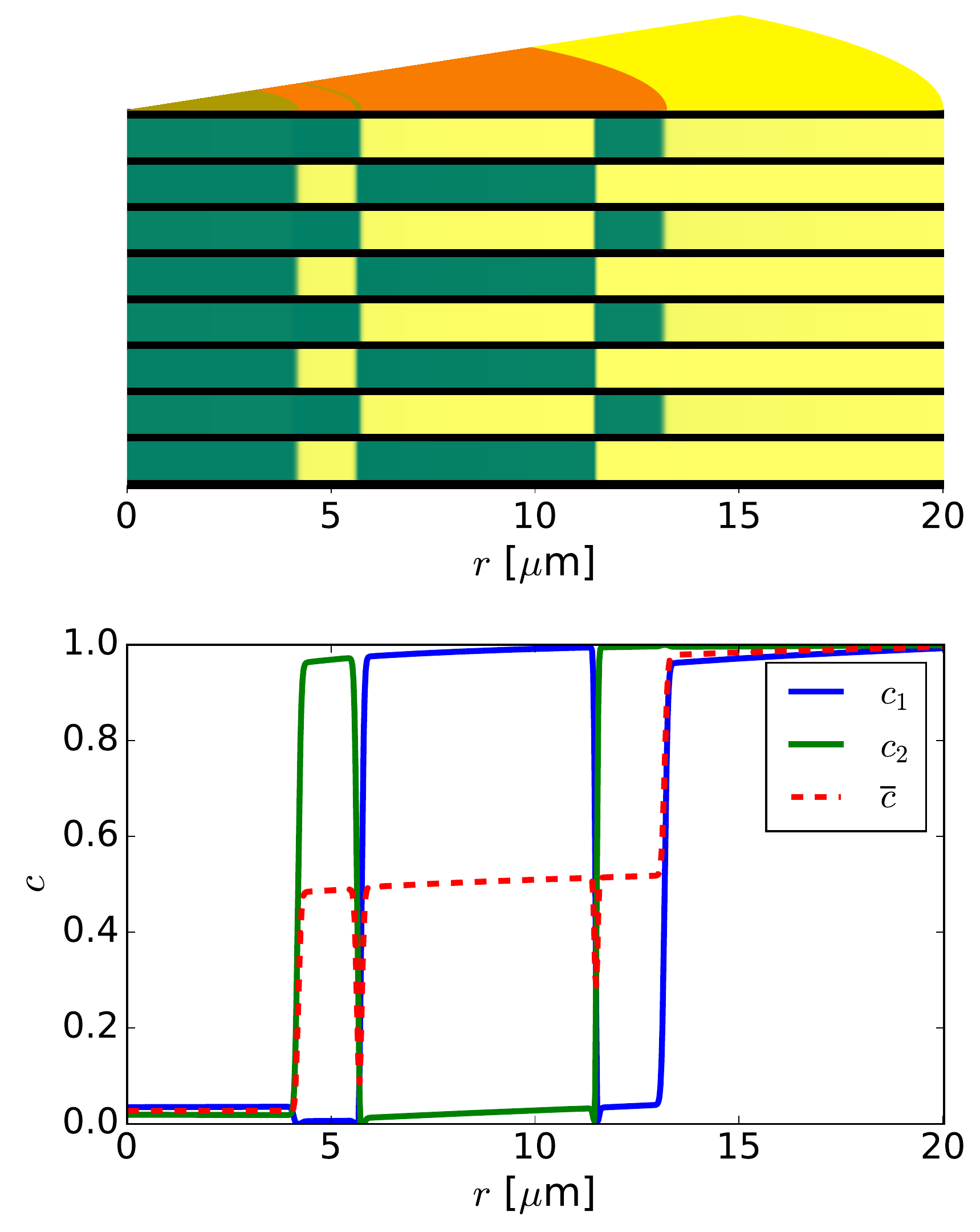}
    \caption{Simulated constant voltage discharge (filling) of a graphite particle and example conversion between simulated concentration profiles (bottom) and associated microstructure (top). Between black graphene planes in the microstructure, the yellow/green indicates high/low lithium content. The colors on the top of the microstructure are assigned based on whether the graphite is in stage 1 (yellow, $\overline{c} > 0.85$), 2 (red, $0.3 < \overline{c} < 0.85$), or 1' (green, $\overline{c} < 0.3$). Results are qualitatively insensitive to the arbitrarily chosen color cutoff values.}
    \label{fig:graphite_example}
\end{figure}

Here we apply a constant voltage to a particle with radius $R = 20\ \mu\mathrm{m}$ and set $k_0 = 0.1\ \mathrm{A/m}^2$. In this simulation, we choose $-0.38\ \mathrm{V}$ with respect to lithium metal, leading to a strong driving force for intercalation. This corresponds to the applied voltage simulated in ref.~\cite{guo2016}. Note that we are ignoring the competing reaction for lithium plating which is also thermodynamically favored under these conditions. To connect the concentration profiles to the microstructure and the associated visual particle colors~\cite{harris2010}, we find it informative to depict the simulated concentration profiles on a cross-sectional slice of the simulated cylindrical particle. For example, we plot a snapshot of this simulation in Figure~\ref{fig:graphite_example}, demonstrating the basic features of the model under these operating conditions.

The concentration profiles on the bottom correspond to the layers of green/yellow between the black graphene planes in the particle slice above. In addition, the particle surface (top) is colored as it would be seen experimentally, according to the internal stage number~\cite{harris2010}. This snapshot also demonstrates some key characteristics of the model. First, we see the anticipated behavior of phase separation within layers caused by the intra-layer regular solution model of Eq.~\ref{eq:layerFED}. Second, in the intermediate region ($4\ \mu\mathrm{m} < r < 13\ \mu\mathrm{m}$), the system forms a stage 2 structure with lithium organized in alternating full/empty layers, caused by the repulsive energetic parameters in Eq.~\ref{eq:totalFED}. Interestingly, in this stage 2 region, the simulation naturally predicts the internal ``checkerboard'' domains of stage 2 structures~\cite{krishnan2013,dimiev2013}. The overall concentration profile going from full near the particle edge to empty near the center is a result of transport limitations as the lithium is inserted from the edge of the particle. The gradual sloping of the concentration profiles within each stage region (including the half-full stage 2) demonstrates the diffusive profiles within the solid solution regimes of the particle between the phase interfaces. A movie of the simulated process is included in the supplement.

\subsubsection{Spinodal Decomposition}
\label{sec:spinodal}
\begin{figure}[h]
    \centering
    (a)
    \includegraphics[width=0.25\textwidth]{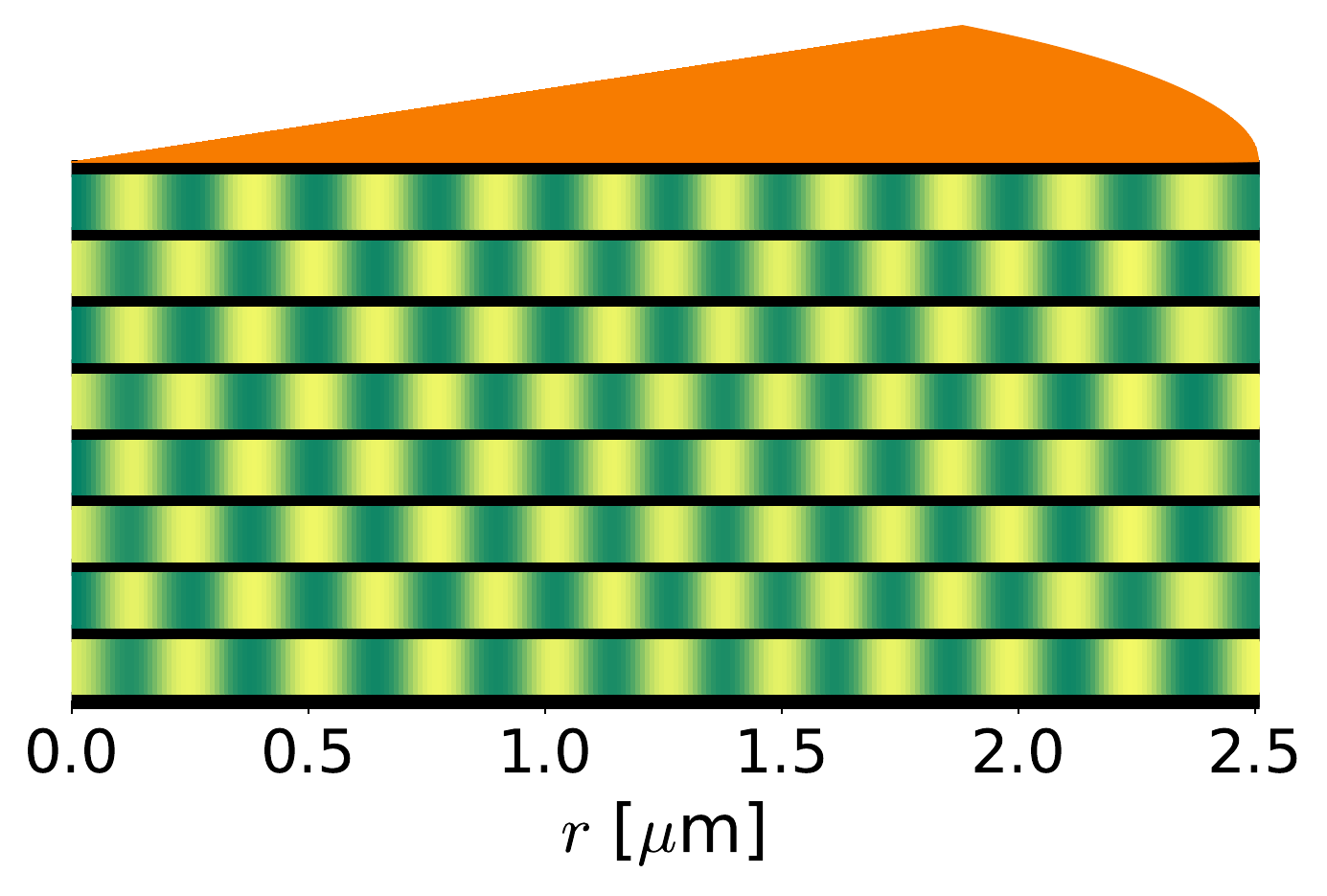}
    \ \
    (b)
    \includegraphics[width=0.25\textwidth]{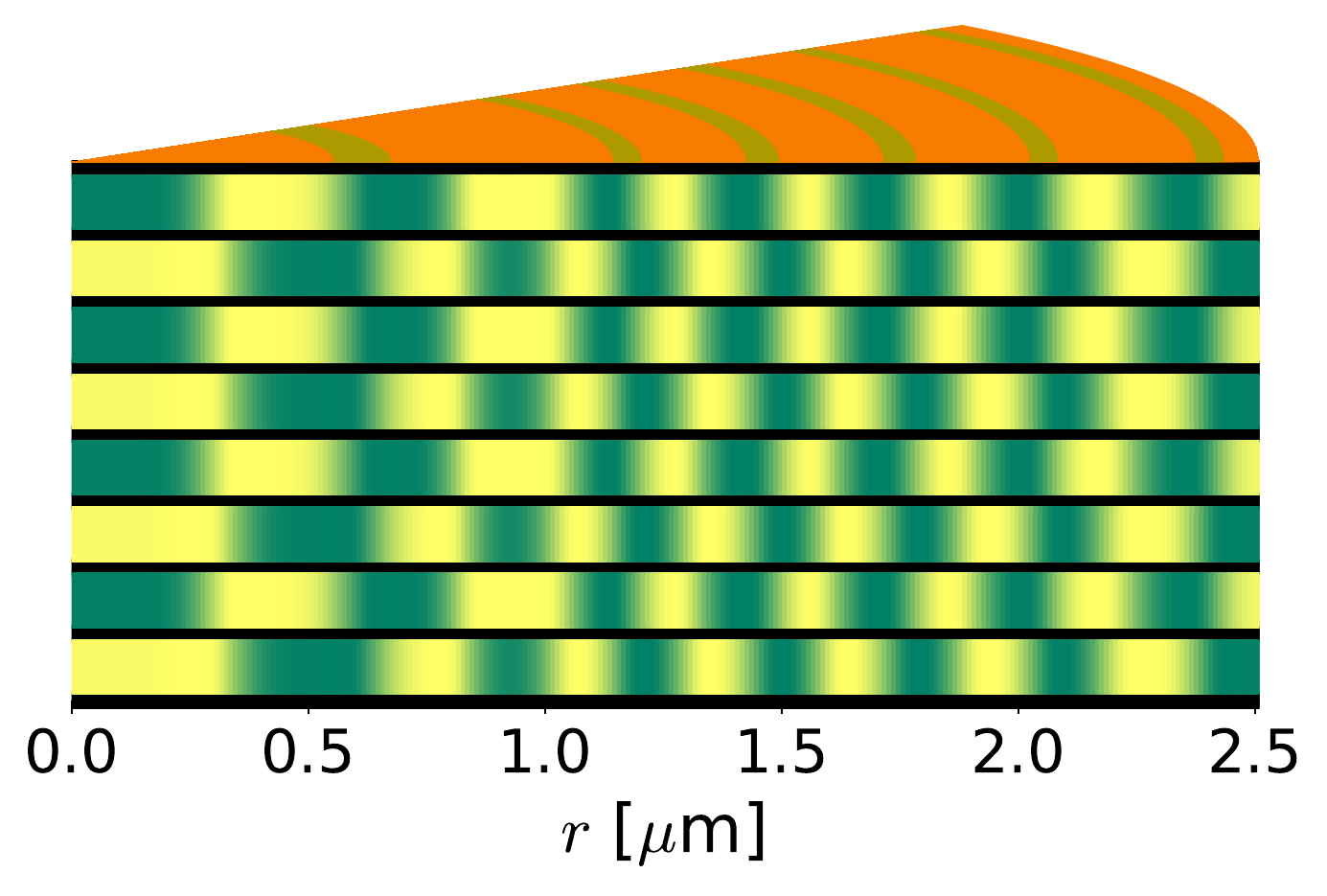}
    \ \
    (c)
    \includegraphics[width=0.25\textwidth]{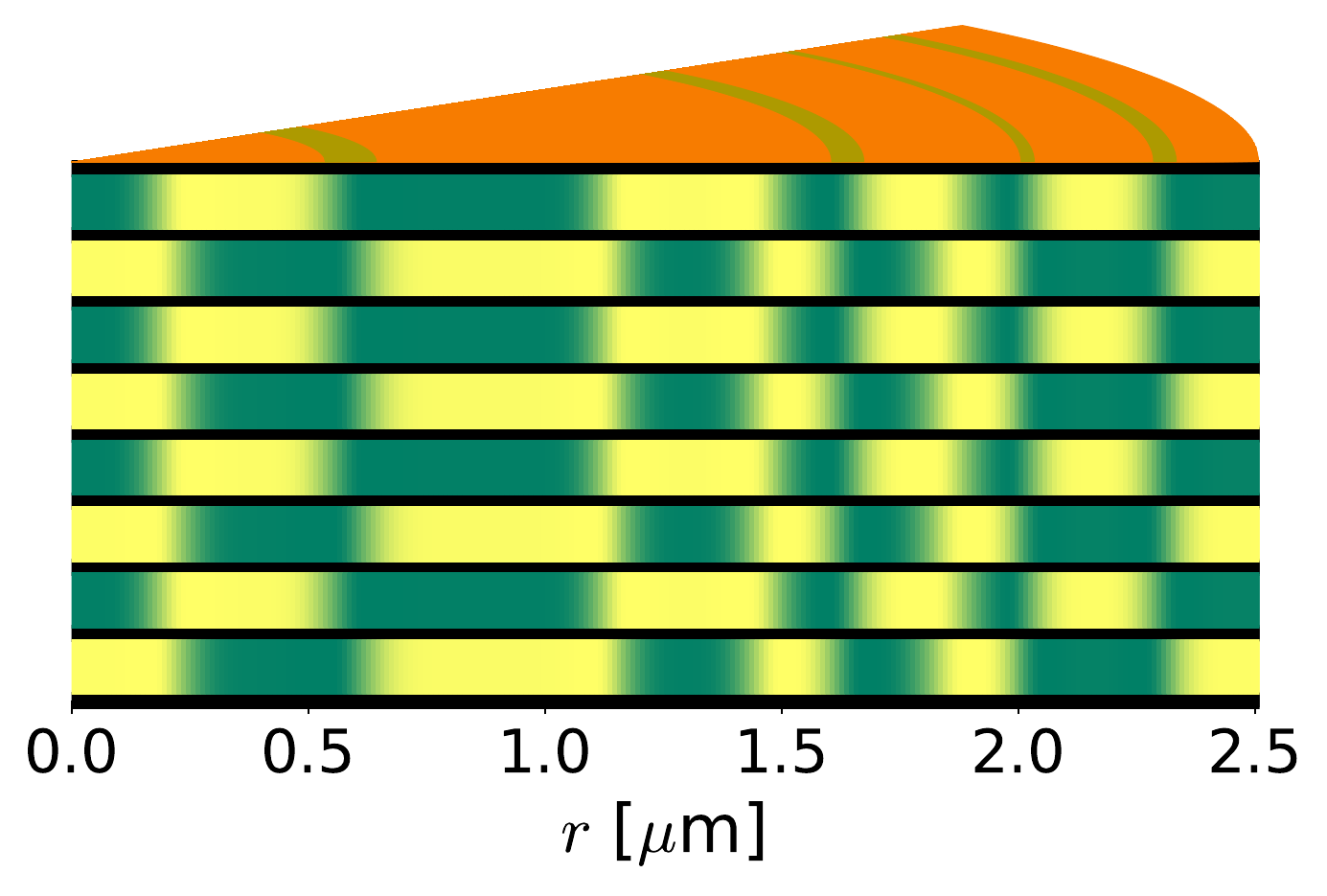}
    \\
    (d)
    \includegraphics[width=0.25\textwidth]{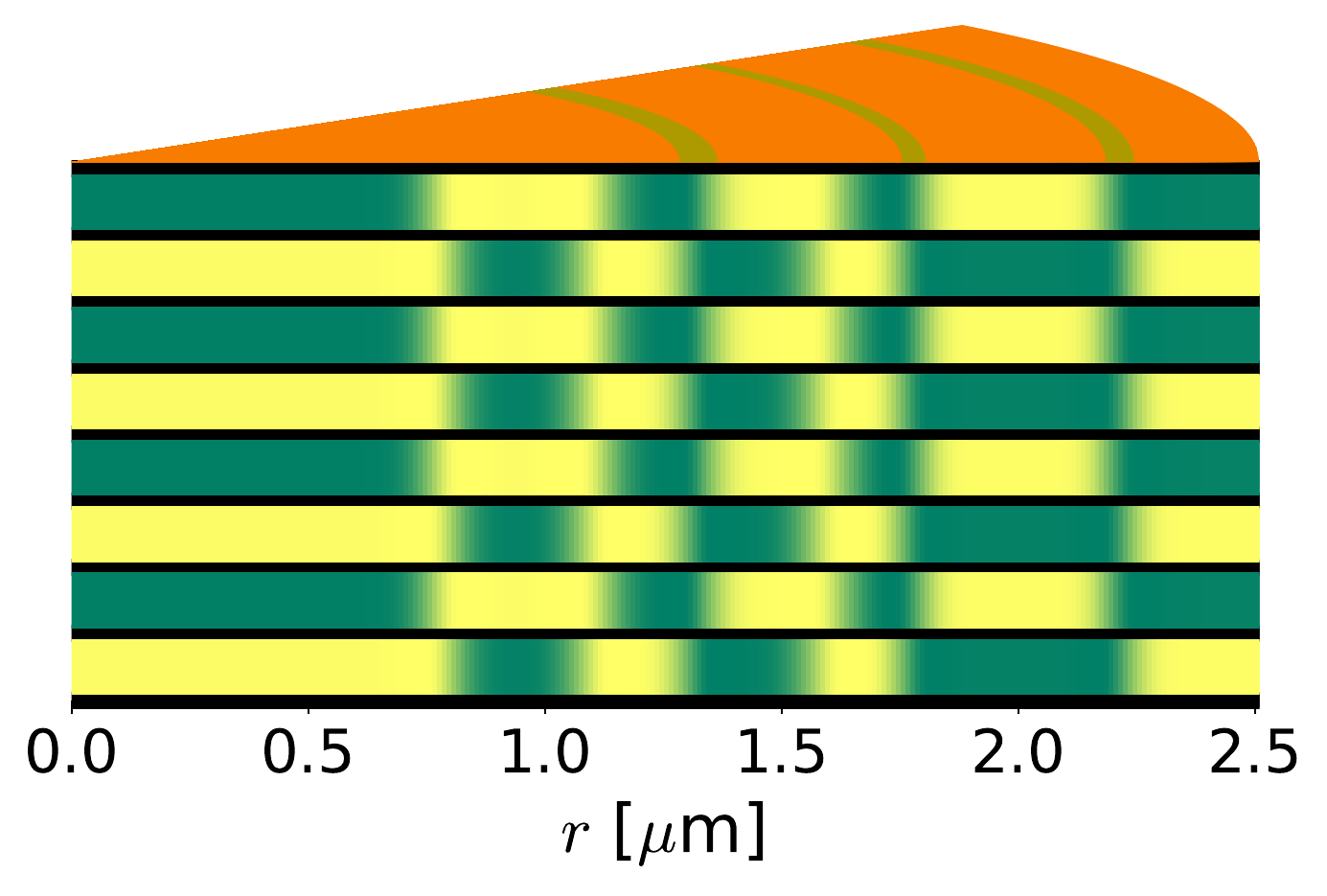}
    \ \
    (e)
    \includegraphics[width=0.25\textwidth]{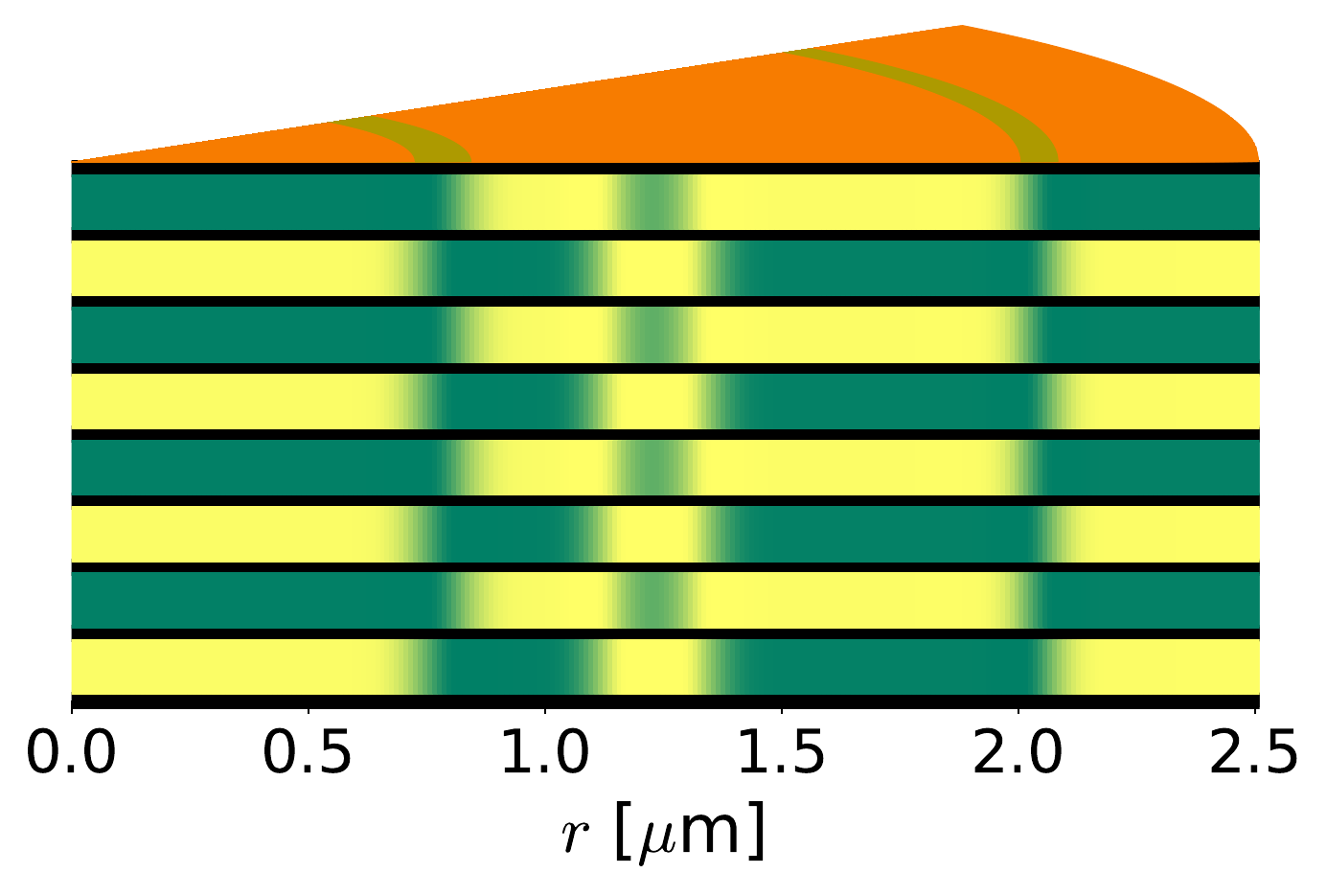}
    \ \
    (f)
    \includegraphics[width=0.25\textwidth]{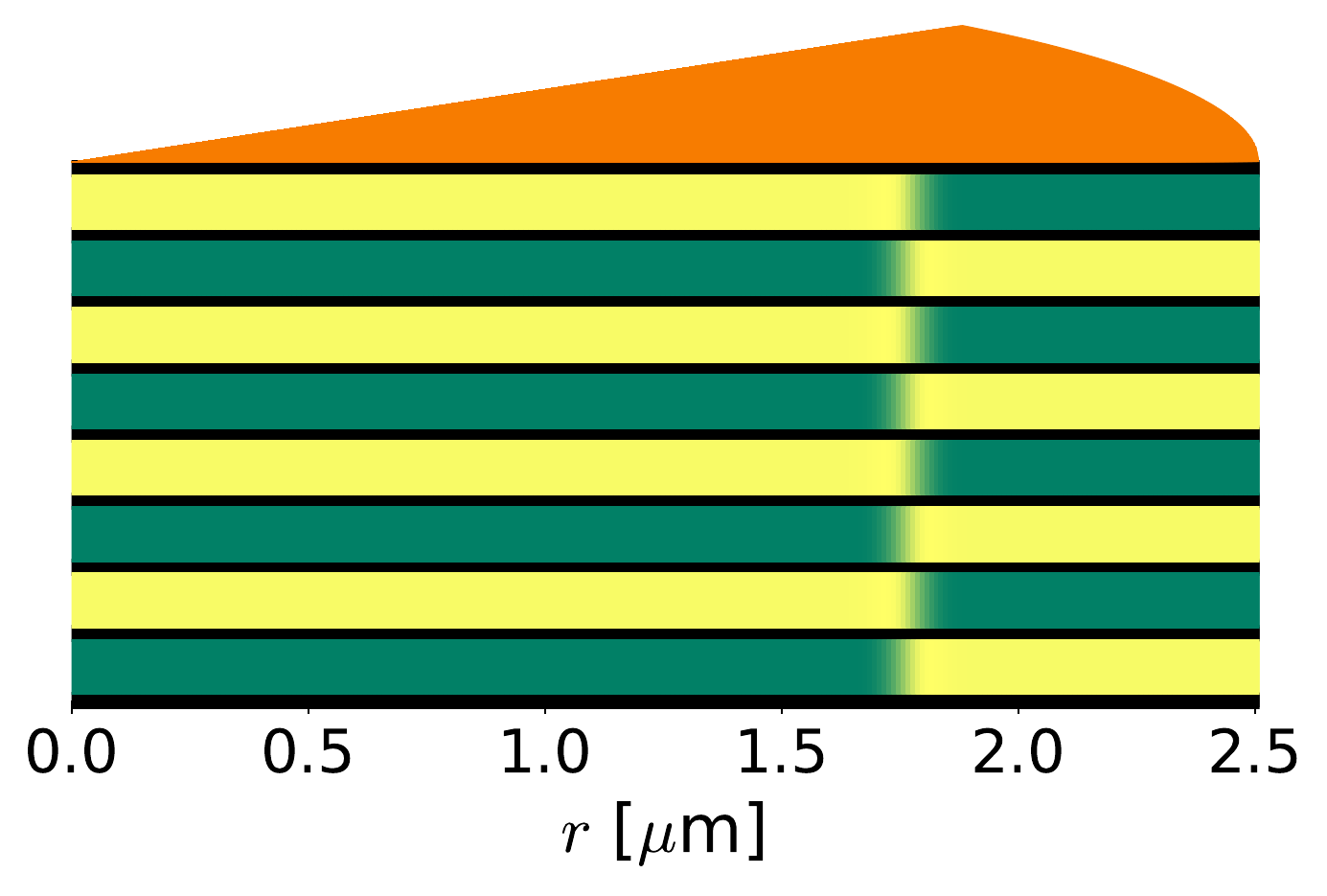}
    \caption{Graphite spinodal decomposition at (a) $t=0.01$~s, (b) $t=1$~s, (c) $t=15$~s, (d) $t=100$~s (e) $t=215$~s, and (f) $t=1000$~s (movie in supplement). Yellow/green regions between black graphene planes indicate high/low Li content, and the coloring on the top refers to the local stage as described in Figure~\ref{fig:graphite_example}.}
    \label{fig:graphite_decomp}
\end{figure}

Spinodal decomposition is the process of transitioning from a high energy, linearly unstable, homogeneous system to a phase separated system. It is characterized by a gradual process of domain coarsening in which initially small phase regions grow and coalesce in order to minimize energetically expensive interfaces between phases~\cite{balluffi1954}. By setting the rate constant to zero ($k_0=0$, which imposes zero current into each layer) and beginning with a randomly perturbed filling fraction of 0.5 in each layer, we can examine the spinodal decomposition process for this coupled, two-layer model. At early times, both layers undergo internal spinodal decomposition, much as predicted by typical Cahn-Hilliard models~\cite{balluffi1954}. However, because the layers are coupled via the overall free energy relation in Eq.~\ref{eq:totalFED} such that it is energetically unfavorable for both layers to be lithium-rich at the same position and time, the layers coordinate to form a checkerboard pattern as shown in Figure~\ref{fig:graphite_decomp}. Then, domain coarsening occurs during which the internal domains expand in size as the most central domain shrinks and disappears to minimize the circumferential interface between the domains. Throughout the process, interfaces move in concert to maintain near-equilibrium concentrations within each domain while preserving the average filling fraction within each layer. The final structure in Figure~\ref{fig:graphite_decomp}~(f) with a single interface in each layer is stable because the layers are unable to exchange lithium, forcing the average concentration within each layer to remain constant. Of note, the depicted rings of stage 1 (green) and stage 1' (yellow) in Figure~\ref{fig:graphite_decomp}~(b)~-~(f) are related to the large penalty for intermediate concentrations in both layers from the $\Omega_c$ term in Eq.~\ref{eq:totalFED}. Because of the strong penalty, the domains ``shift'' relative to each other on the adjacent planes such that they either overlap slightly or are separated in space to avoid both having an interface (and intermediate filling fractions) at the same location (Figure~\ref{fig:graphite_decomp}~(b)~-~(f)). When $\Omega_c$ is set to zero in an otherwise identical simulation, the rings do not appear.

\subsubsection{Constant Current Intercalation}
\label{sec:CC}
\begin{figure}[h]
    \centering
    (a)
    \includegraphics[width=0.4\textwidth]{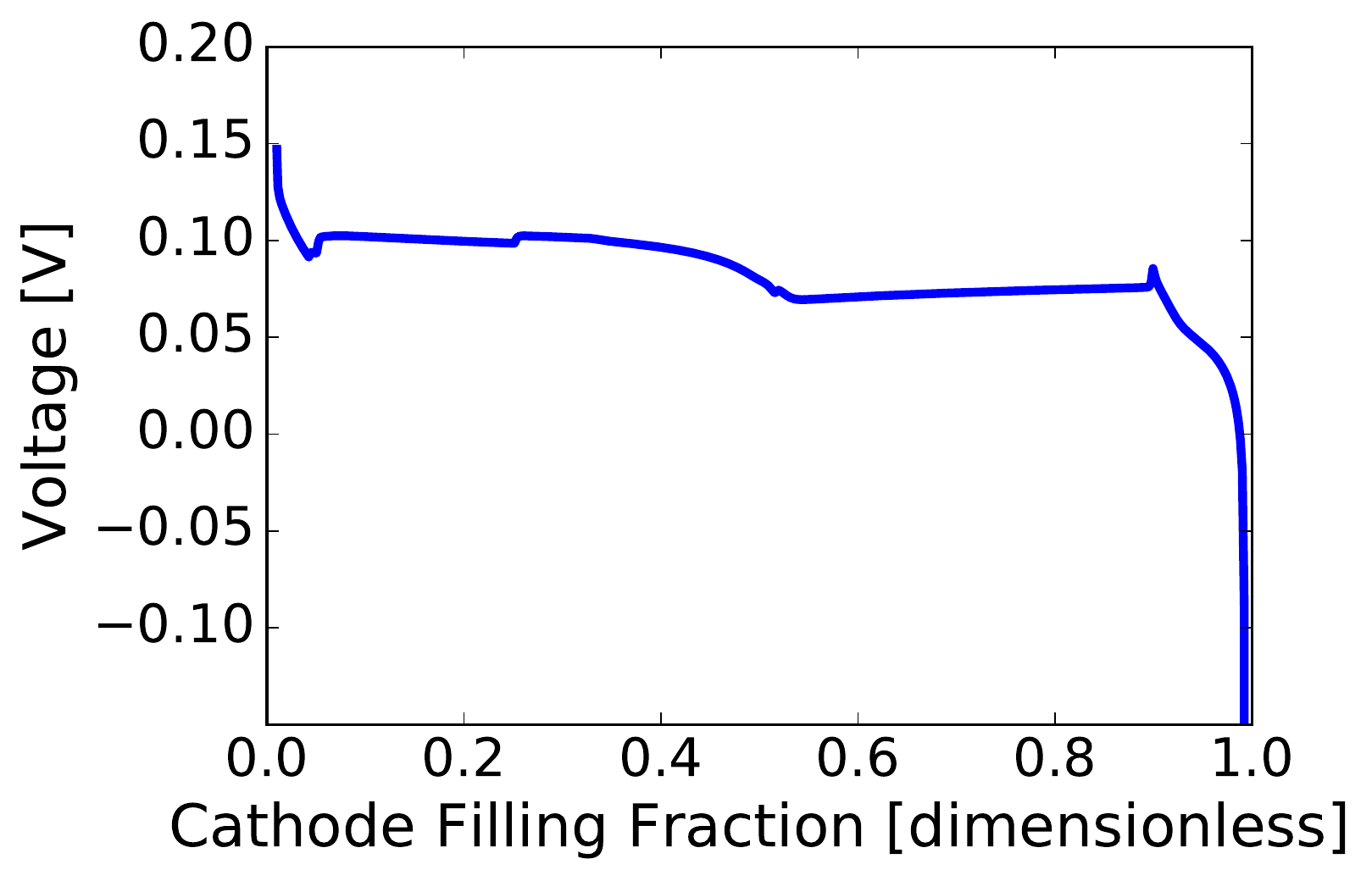}
    \ \
    (b)
%    \includegraphics[width=0.4\textwidth]{graphite_CC_20160902_125354_rxnp.pdf}
    % Note, 20161108_181419 is a repeat of the simulation actually used in the snapshots and
    % movies, 20160902_125354 but tracking the activity at the surface.
    \includegraphics[width=0.4\textwidth]{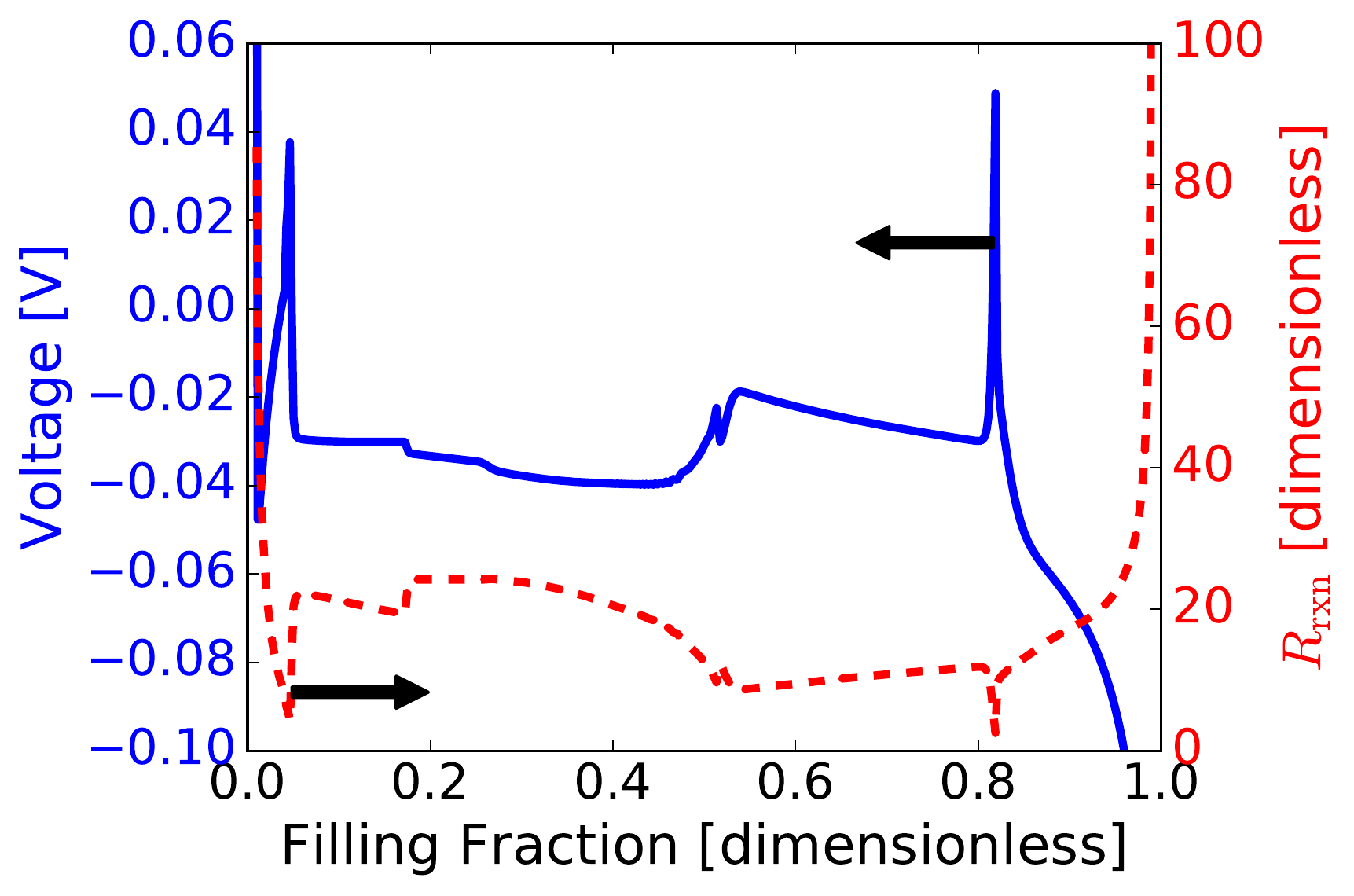}
    \caption{Constant current voltage profiles with (a) $\gamma_{\ddagger,i} = 1$ and (b) $\gamma_{\ddagger,i} = 1/\left(\wt{c}_i\left(1-\wt{c}_i\right)\right)$ as in Eq.~\ref{eq:gamma_ts} plotted with both voltage (solid) and reaction resistance (dashed).}
    \label{fig:graphite_CC_v_and_rxn}
\end{figure}

\begin{figure}[h]
    \centering
    (a)
    \includegraphics[width=0.4\textwidth]{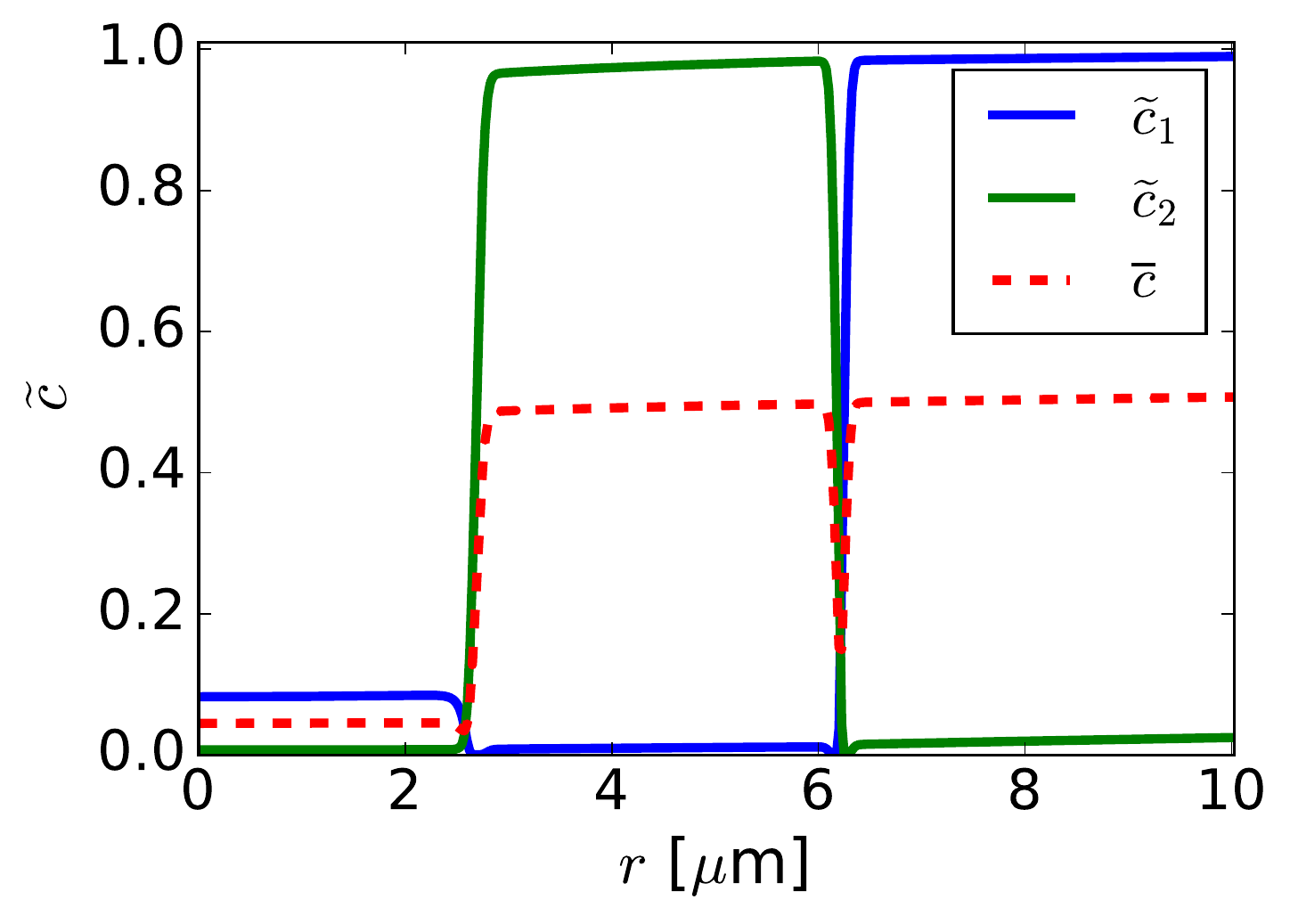}
    \ \
    (b)
    \includegraphics[width=0.4\textwidth]{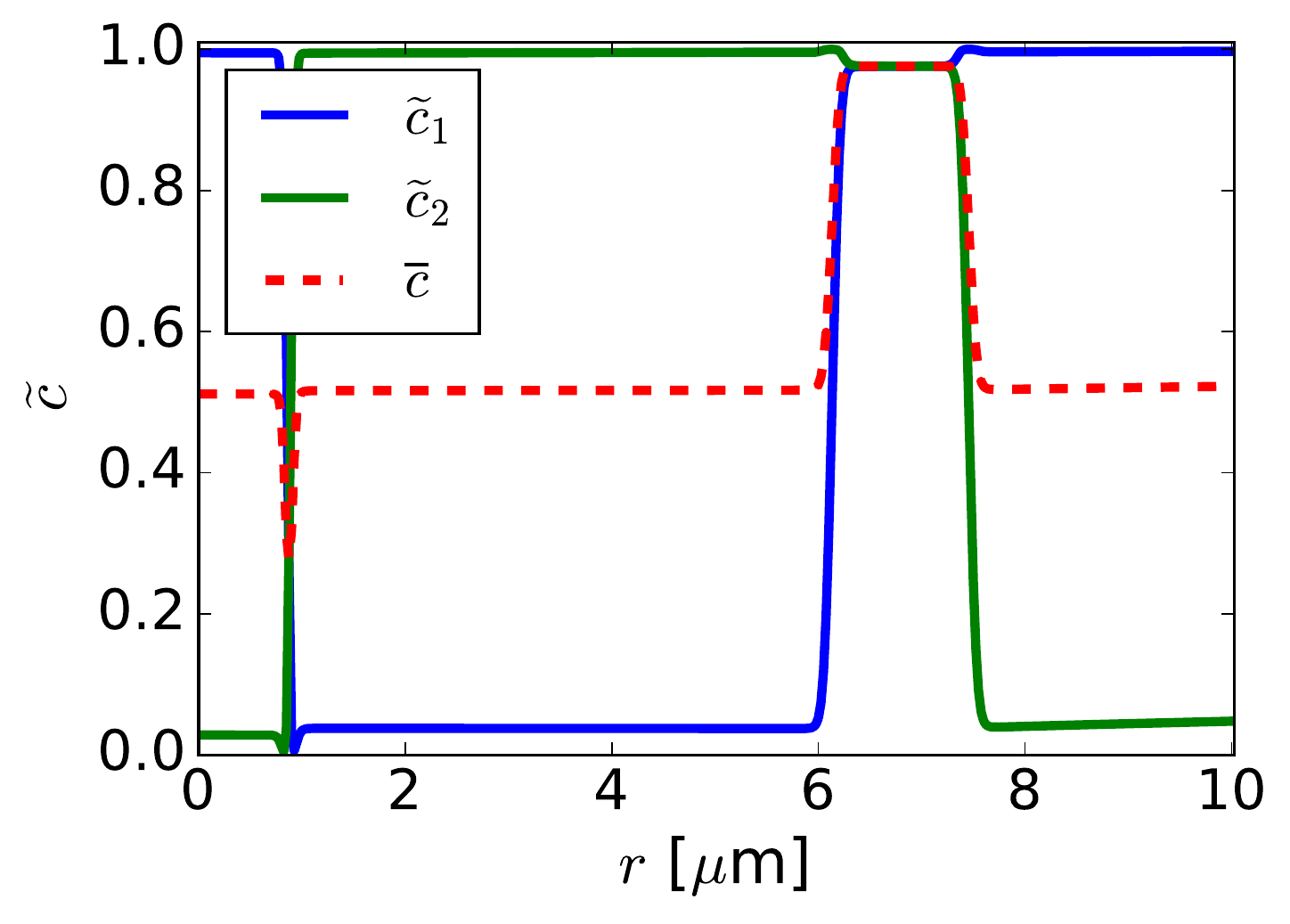}
    \caption{Example concentration profiles resulting from a constant current discharge (filling of the particle) and corresponding to the simulation in Figure~\ref{fig:graphite_CC_v_and_rxn}~(b). Concentration profiles are plotted at $t=1630$ s (a) and $t=2130$ s (b), at filling fractions near 0.45 and 0.6 respectively. See the supplement for the movies. In (a) both layers are receiving current. The internal layer 2 domain is shifting left as the superficial layer 1 domain expands toward the center of the particle. In (b), only layer 2 is sustaining the current, as the internal domain is growing toward the surface, causing a larger internal gradient to the surface, a more intermediate surface concentration, and lower reaction resistance.}
    \label{fig:graphite_parallelRxn}
\end{figure}

When (dis)charging a particle at a fixed current, we can observe multiple regimes. At low enough currents (defined by comparing system time scales), the observed voltage is set by the effective chemical potential of the solid, as discussed in Section~\ref{sec:OCV}. However, at larger currents, other losses in the system such as internal gradients and reaction losses contribute to the measured voltage. In particular, we find that when the imposed time scale from the specified current density approaches that of the reaction time scale, we enter a regime in which the behavior of the observed voltage is dominated by the reaction losses. As an example, we present the results of a $1\ \mathrm{C}$ discharge for a particle with radius $R = 10\ \mu\mathrm{m}$ and $k_0 = 10\ \mathrm{A/m}^2$. The value of $k_0$ was semi-arbitrarily chosen such that the reaction time scale was of similar magnitude to the imposed time scale of 1 hr. The reaction time scale for this system is $\tau_R \sim 10^3\ \mathrm{s}$. In Figure~\ref{fig:graphite_CC_v_and_rxn}, we consider two cases to compare the effect of the chosen model for the reaction transition state. In Figure~\ref{fig:graphite_CC_v_and_rxn}~(a), we set the value, $\gamma_{\ddagger,i} = 1$, whereas in Figure~\ref{fig:graphite_CC_v_and_rxn}~(b), we retain the model originally proposed in Eq.~\ref{eq:gamma_ts}. In the first, we obtain a voltage curve similar to that in Figure~\ref{fig:graphiteOCV} but shifted downward from reaction resistance and showing features of transients from transport losses between the surface and bulk of the particle. The mild increase in voltage over the second plateau is a result of a growing annulus of stage 1 graphite, causing the surface concentration to gradually approach a stable equilibrium value from a metastable value (see movies of concentration profiles in supplement). However, in Figure~\ref{fig:graphite_CC_v_and_rxn}~(b), we see the system has a voltage curve that is nearly uncorrelated with the effective chemical potential along the equilibrium path. As mentioned above, deviations from equilibrium voltage curves can come from both transport and reaction losses. Reaction losses arise from slow reaction kinetics leading to large overpotentials, $\eta_i$, which indicate the departure of the interfacial voltage from the equilibrium value~\cite{bazant2013}. Thus, we examine the reaction rate prefactor, called the exchange current density, which corresponds to a reaction conductance. We find the voltage is directly related to an effective reaction resistance, defined by taking two reaction resistances in parallel for the two layers. Noting that each reaction resistance scales as $\gamma_{\ddagger,i}/a_i^\alpha$ (from Eq.~\ref{eq:rxn}), we define the quantity
\begin{align}
%    R_\mathrm{rxn} \sim \frac{1}{\frac{1}{\gamma_{\ddagger,1}} +
%    \frac{1}{\gamma_{\ddagger,2}}}
    R_\mathrm{rxn}
%    = \frac{1}{\frac{1}{R_\mathrm{rxn,1}} + \frac{1}{R_\mathrm{rxn,2}}}
    = \frac{1}{\frac{a_1^\alpha}{\gamma_{\ddagger,1}} +
    \frac{a_2^\alpha}{\gamma_{\ddagger,2}}}
    ,
    \label{}
\end{align}
which differs functionally from the actual reaction resistance only because the reaction driving force, $\eta_i$, can be different for each layer. We compare this quantity with the system voltage in Figure~\ref{fig:graphite_CC_v_and_rxn}~(b) and observe that the two quantities are correlated: large reaction resistance, $R_\mathrm{rxn}$, corresponds to lower system voltage, indicating that this quantity (not changes in $\mu^\mathrm{eff}$ at the surface) is the primary contribution to departures of the observed voltage from the equilibrium value at the overall filling fraction. Surprisingly, the voltage has a sharp increase near half-filling rather than a decrease as in the equilibrium profile in Figure~\ref{fig:graphiteOCV}, something that could not be predicted by a simple diffusion-based model with imposed phase boundaries. Here, the voltage increases near half filling because the current transitions from being split by the two layers, both of which are enlarging internal domains, to being carried largely by layer 2. The associated steeper internal concentration gradient near the surface of layer 2 leads to values of $c_2$ closer to 0.5 and smaller overall reaction resistance, despite the fact that the current is carried by only half the surface area (see concentration profiles in Figure~\ref{fig:graphite_parallelRxn}). Throughout, the concentration at the surface is the result of coupled transport and reaction processes, and in particular the steep voltage drop off near complete filling is related to transport losses causing the surface filling fraction to approach unity which also leads to diverging reaction resistance. The large spikes in the voltage are related to surface concentration transitions from low to high values, leading to temporary low resistance at the intermediate filling fractions. It is worth mentioning that some of these features would be smoothed and not visible macroscopically in an ensemble of coupled particles as in a porous electrode.

Thus, we see that the choice of the reaction model can have significant impact on the predicted macroscopic properties. These observations underscore the importance of careful consideration of the reaction model when attempting to relate model predictions with experimental data, as this model makes strong connections between the microstructure and the single-particle discharge voltage profile. The single-particle discharge curve predicted in Figure~\ref{fig:graphite_CC_v_and_rxn} (b) bears little resemblance to experimental discharge curves of graphite electrodes~\cite{safari2011a}, which suggests that, at least for the secondary graphite particles used in electrodes~\cite{wissler2006}, Eq.~\ref{eq:gamma_ts} may not correctly describe the particle-electrolyte interface, and a reaction model more like that used in Figure~\ref{fig:graphite_CC_v_and_rxn} (a) or Eq.~\ref{eq:rxn_constecd} may be more appropriate, as we find in ref.~\cite{thomas-alyea2016}. We also note that in a porous electrode with many particles, some of the details of single-particle voltage curves can be masked by particle-particle interactions and system-scale transport losses~\cite{dreyer2010,ferguson2014}.

Because the term in the free energy involving $\Omega_c$ was originally chosen for a very different model and only more strongly penalizes states of intermediate filling in both layers (which is already unfavorable even with $\Omega_c = 0$), we chose this as a representative simulation to study the model without that term. In this case, the simulated concentration profiles look qualitatively similar (see video in supplement), but the simulation neglecting the $\Omega_c$ term ran considerably faster, so it may be reasonable to neglect this term in full spatially resolved models of graphite based on this two-layer framework.

\subsubsection{Inter-layer (Homogeneous) Reaction}
\label{sec:CCRxn}
\begin{figure}[h]
    \centering
    (a)
    \includegraphics[width=0.4\textwidth]{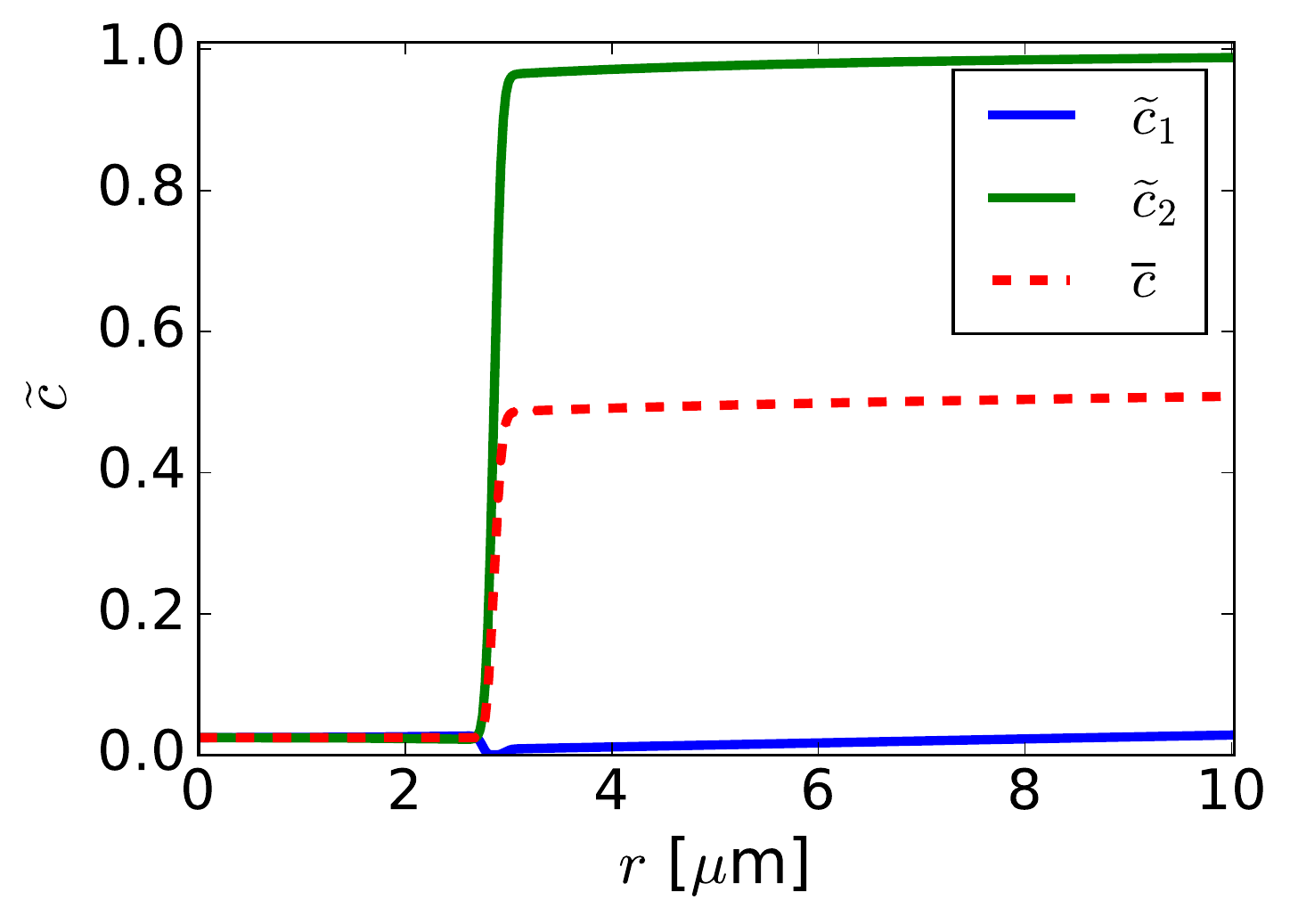}
    \ \
    (b)
    \includegraphics[width=0.4\textwidth]{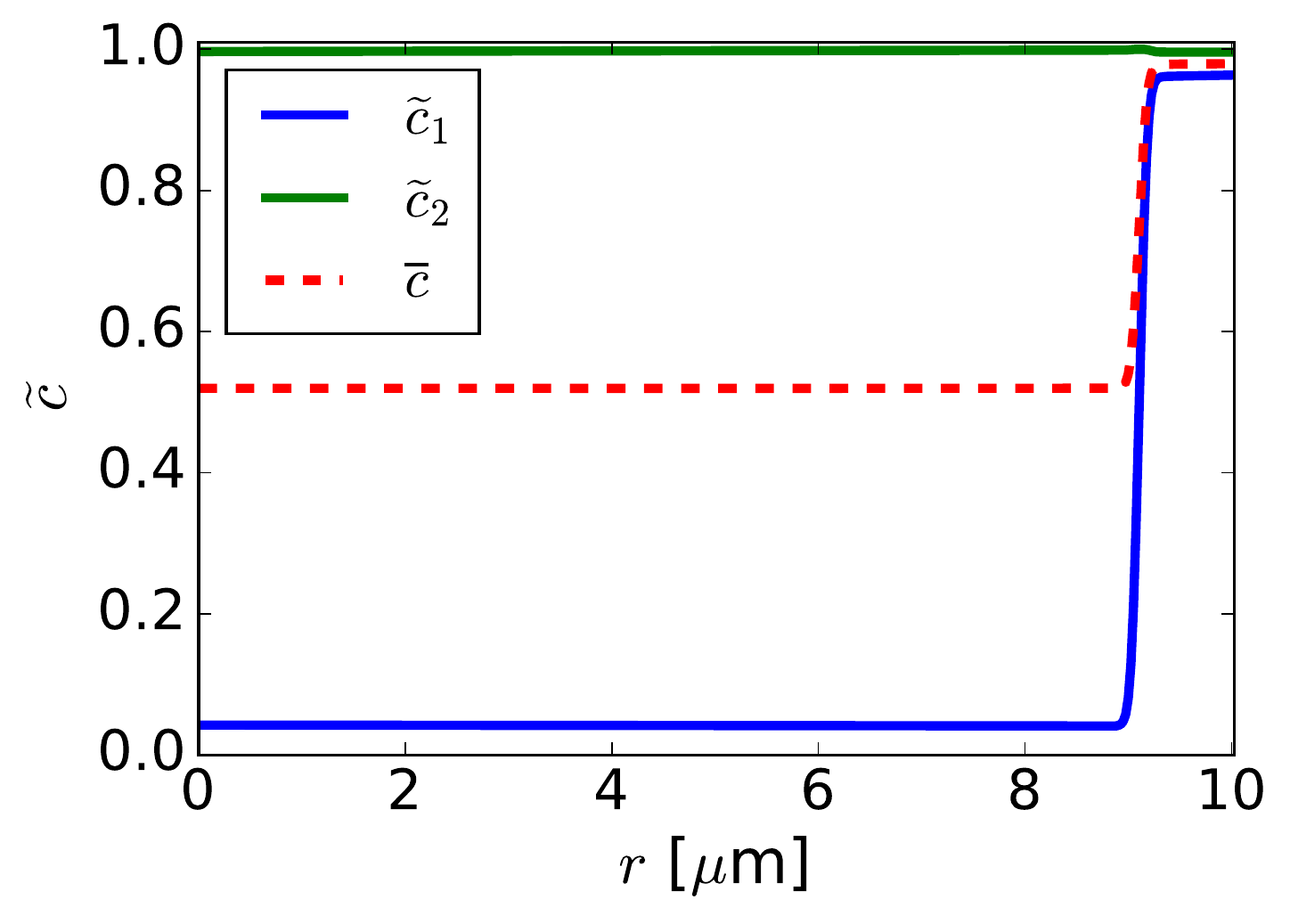}
    \caption{Repeat of the simulation in Figure~\ref{fig:graphite_parallelRxn} with homogeneous inter-layer reactions. Snapshots in (a) and (b) are taken at $t=1630$ s and $t=2130$ s as in the previous simulations. Movies are provided in the supplement.}
    \label{fig:graphite_interlayerRxn}
\end{figure}

As discussed above, various physical phenomena could enable exchange between the two concentration fields. In the case of the graphite model, a spatially regular presence of defects could allow internal exchange between layers of intercalated lithium~\cite{yao2012diffusion} to be modeled as a homogeneous reaction term. In order to properly formulate the reaction term in the context of non-equilibrium thermodynamics, simple mass action kinetics cannot apply, and the diffusional chemical potential must be the driving force for the reaction. Following Bazant~\cite{bazant2013} (Eq.~7), we note that the rate of inter-layer exchange from the first to second layer can be written as
\begin{align}
    R_{2} = -R_{1}
    = \frac{k_0^\mathrm{IL}}{\gamma_\ddagger^\mathrm{IL}}
    \left( a_1 - a_2 \right)
    \label{}
\end{align}
where $\gamma_\ddagger^\mathrm{IL}$ is the activity coefficient for the inter-layer reaction transition state, and we have made use of the fact that $\mu^\Theta$ is identical for each layer to absorb reference chemical potential factors into the inter-layer rate constant, $k_0^\mathrm{IL}$. For asymmetric lattices, the above should be modified to include the reference chemical potentials for each layer, biasing the reaction relative to equal activity in both lattices. Again, we are left a choice in the activity coefficient of the transition state. Here, we consider a reaction in which a species hops from a site on lattice 1 to a corresponding site on lattice 2. A natural choice for the transition state, then, is to assume that it excludes a single site in both layers, such that it is given simply by
\begin{align}
    \gamma_\ddagger^\mathrm{IL} = \frac{1}{\left( 1-\wt{c}_1 \right)\left( 1-\wt{c}_2 \right)}.
    \label{}
\end{align}
Other choices could include terms weighting enthalpic contributions from initial and final states (similar to the model proposed in Eq.~\ref{eq:TS_d}); however, the above seems to be a simple reasonable model of the reaction, such that
\begin{align}
    R_{2} = k_0^\mathrm{IL}
    \left( 1-\wt{c}_1 \right)\left( 1-\wt{c}_2 \right)\left( a_1 - a_2 \right).
    \label{}
\end{align}
Thus, we have a slightly modified version of the conservation equation in each layer, such that instead of having zero homogeneous reaction term in Eq.~\ref{eq:massCons}, we have expressions for the source term, $R_i$. In Figure~\ref{fig:graphite_interlayerRxn}, we compare the concentration profiles predicted during constant current intercalation with $k_0^\mathrm{IL}/c_\mathrm{ref} = 1\times10^{-4}\ \mathrm{s}^{-1}$ for the same set of conditions as presented in Figure~\ref{fig:graphite_parallelRxn} in which $k_0^\mathrm{IL}$ was set to zero. This corresponds to a Damk\"{o}hler number of
\begin{align}
    \mathrm{Da} = \frac{k_0^\mathrm{IL}R^2}{c_\mathrm{ref}D_0} \approx 10^{-2},
    \label{}
\end{align}
indicating relatively slow inter-layer reactions compared to the bulk transport. Nevertheless, even with this small exchange rate, we see qualitative differences in the model predictions.

With the inter-layer reaction, because the layers can exchange species, the system is able to fill a single layer completely before the other layer fills, thus minimizing the number of interfaces to follow a lower energy path during the filling. Physically, this occurs because as the lithium enters at the surface of either layer, it raises the diffusional chemical potential (and activity) in that layer. However, as soon as layer 2 phase separates, it is able to support increasing amounts of lithium without a significant change in its diffusional chemical potential by moving the phase boundary. In layer 1, as the concentration increases, it can go beyond the binodal point into the metastable region and reach a higher diffusional chemical potential than that in layer 2 (see Figure~\ref{fig:NRGD} (c)), thus causing species from layer 1 to transfer to the lower activity layer 2, which simply advances the position of the interface. Once layer 2 is completely full, layer 1 fills independently, moving a phase boundary from the edge of the particle toward the center, much like the profiles predicted in similar systems with only one concentration variable~\cite{zeng2014}.

\subsection{Comparison to Solid Solution Particles}
\label{sec:fickCompar}
Although graphite phase separates upon intercalation with lithium, it is commonly modeled in battery electrodes using various simplifications to avoid the challenges of properly capturing phase separation dynamics. For example, it is common to approximate the transport of lithium in the material as a solid solution~\cite{yu1999determination,verbrugge2003,srinivasan2004design,safari2011a,bernardi2011,baker2012}, which has the advantage of being computationally very straightforward. By employing a concentration-dependent diffusivity, these models can even predict steep concentration gradients similar to phase boundaries~\cite{baker2012}, but these gradients relax to uniform concentrations under zero current, even within spinodal regions. It is also possible to describe the intercalation as multiple processes involving solid-solution diffusion and phase boundary motion which enables use of electroanalytical techniques with simple expressions to determine values of transport coefficients~\cite{bard2001,aurbach1998,levi1997simultaneous}, particularly in the solid solution regimes~\cite{funabiki1998impedance}, but doing so can lead to results in which the chemical diffusivity remains positive even in spinodal regions~\cite{levi1997mechanism2,levi1997diffusion} where uphill diffusion should occur. This can be interpreted as an ``effective'' transport parameter.

Another approach involves explicitly solving for motion of a phase boundary via a Stefan condition assuming phase equilibrium at the interface and solid-solution diffusion elsewhere~\cite{funabiki1999stage,baker2012,hess2013}. However, when solving numerically, the creation, elimination, and tracking of arbitrary numbers of phase boundaries becomes cumbersome. Moving boundary models also neglect physical contributions of interfacial energy and stresses (not modeled here but naturally a part of this framework~\cite{bazant2013}), both of which influence the location of phase interfaces~\cite{cogswell2012}. They also would not capture the uphill diffusion process involved in spinodal decomposition (Section~\ref{sec:spinodal}), metastable concentrations reached when minimal gradients are present (e.g.\@ the overshoot in Figure~\ref{fig:graphiteOCV}), or the effects of surface (de)wetting properties~\cite{zeng2014}. Using a free energy approach like that developed here, open circuit voltages are emergent properties of the free energy, and phase interfaces develop naturally and do not need to be created artificially and tracked numerically. Nevertheless, for near-equilibrium systems without significant surface (de)wetting, the moving boundary and free energy approaches can lead to similar results.

Levi et al.\ compared the solid solution and moving boundary approaches in ref.~\cite{levi2005comparison} and found that for many cases, the two lead to similar fit transport parameters. This similarity in predicted electrochemical outputs could help explain the success of solid solution models of graphite when applied to porous battery electrodes and their ability to fit macroscopic electrochemical data~\cite{verbrugge2003,safari2011a,bernardi2011,baker2012,thomas-alyea2016}. In each case, transport parameters fit using the above models could be interpreted as describing ``effective'' transport within the active material particles, perhaps a combination of transport within primary particles and along grain boundaries of highly polycrystalline graphite secondary particle agglomerates~\cite{wissler2006}.

Although a comprehensive comparison of solid solution and phase separating models is beyond the scope of this work, we develop and present a simple example which highlights the differences in the predictions of a simple solid solution model of graphite and the phase separating model developed here -- a current pulse-relaxation process in which the average concentration remains within the miscibility gap. We will not present the details of solid solution models, as that has been extensively studied elsewhere~\cite{newman2004,doyle1993,fuller1994}. Because the structures of the models are different, some care must be taken to obtain comparisons with few enough differences to reasonably interpret the outputs. First, we consider the reaction model. In both models, the driving force is the thermodynamic driving force, $\eta$, which depends on the diffusional chemical potential at the surface of the solid. In the solid solution models, this depends only on the concentration at the surface, whereas the free-energy based models permit dependence on additional factors such as gradients of concentration. To match exchange current densities in the simplest way, we replace Eq.~\ref{eq:rxn} with a form using a constant exchange current density with arbitrarily set $k_0 = 10\ \mathrm{A}/\mathrm{m}^2$,
\begin{align}
    \frac{J_i}{e} = f_{i}k_{0}\left[ \exp\left(-\alpha e\eta_i/k_\mathrm{B}T\right) -
    \exp\left(\left(1-\alpha\right)e\eta_i/k_\mathrm{B}T\right) \right]
    \label{eq:rxn_constecd}
\end{align}
with $f_i = 0.5$ for each layer in the phase separating model and unity in the solid solution model. This could be interpreted as a particular functional form assumed for $\gamma_{\ddagger,i}$. The solid solution diffusional chemical potential is defined entirely by the open circuit voltage, and we use a function which matches the blue dashed $V_\mathrm{PS}$ curve in Figure~\ref{fig:simp1param} (b) to match the stable-equilibrium open circuit voltage of the phase separating model.

For the transport processes, we want to compare situations with similar diffusive behavior, at least outside of the miscibility gap. For the solid solution model, we use a constant and uniform solid solution chemical diffusivity, $D^\mathrm{ss}_\mathrm{chem}$ such that the solid solution flux is given simply by $\mathbf{F}^\mathrm{ss} = -D^\mathrm{ss}_\mathrm{chem}\bnab c$. As presented in Appendix~\ref{sec:SSPSdiffn}, rather than matching the flux prefactors, $D_0$ and $D^\mathrm{ss}_\mathrm{chem}$, we approximately match the concentration dependence (but not the magnitude) of the diffusive behavior in the solid solution regimes of the particles (far from the interfaces in the phase separating particle). To do this we retain the flux expression in Eq.~\ref{eq:flux_with_prefactor} with $D_0=1.25\times10^{-12}\ \mathrm{m}^2/\mathrm{s}$ and use $D^\mathrm{ss}_\mathrm{chem} = 3\times10^{-14}\ \mathrm{m}^2/\mathrm{s}$. The value of $D^\mathrm{ss}_\mathrm{chem}$ is insignificant to the interpretation of the results, as described below. We simulate both particles as cylinders of radius $10\ \mu\mathrm{m}$ with a uniform mesh of 400 volumes.

\begin{figure}[h]
    \centering
    (a)
    \includegraphics[width=0.28\textwidth]{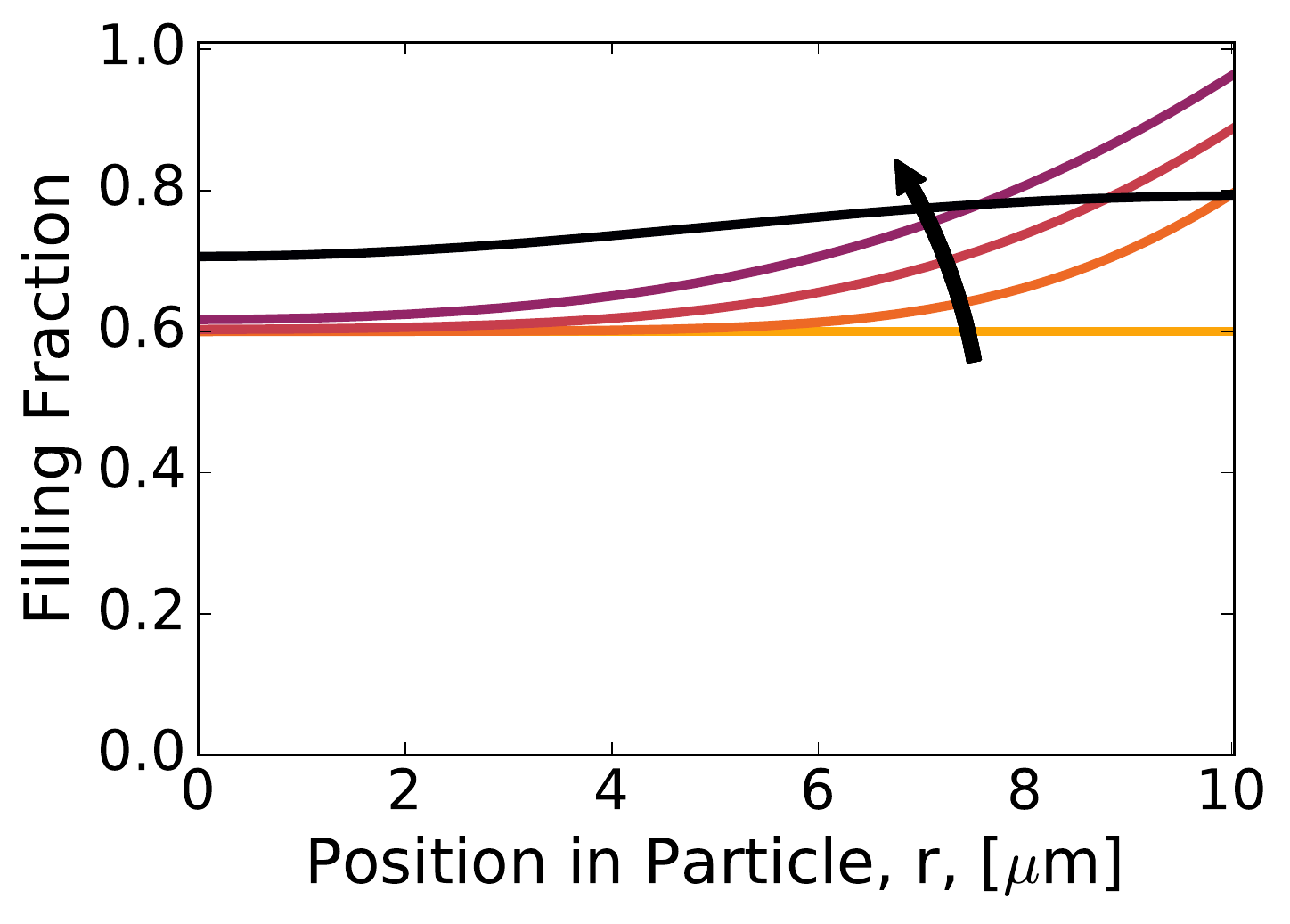}
    \ \
    (b)
    \includegraphics[width=0.28\textwidth]{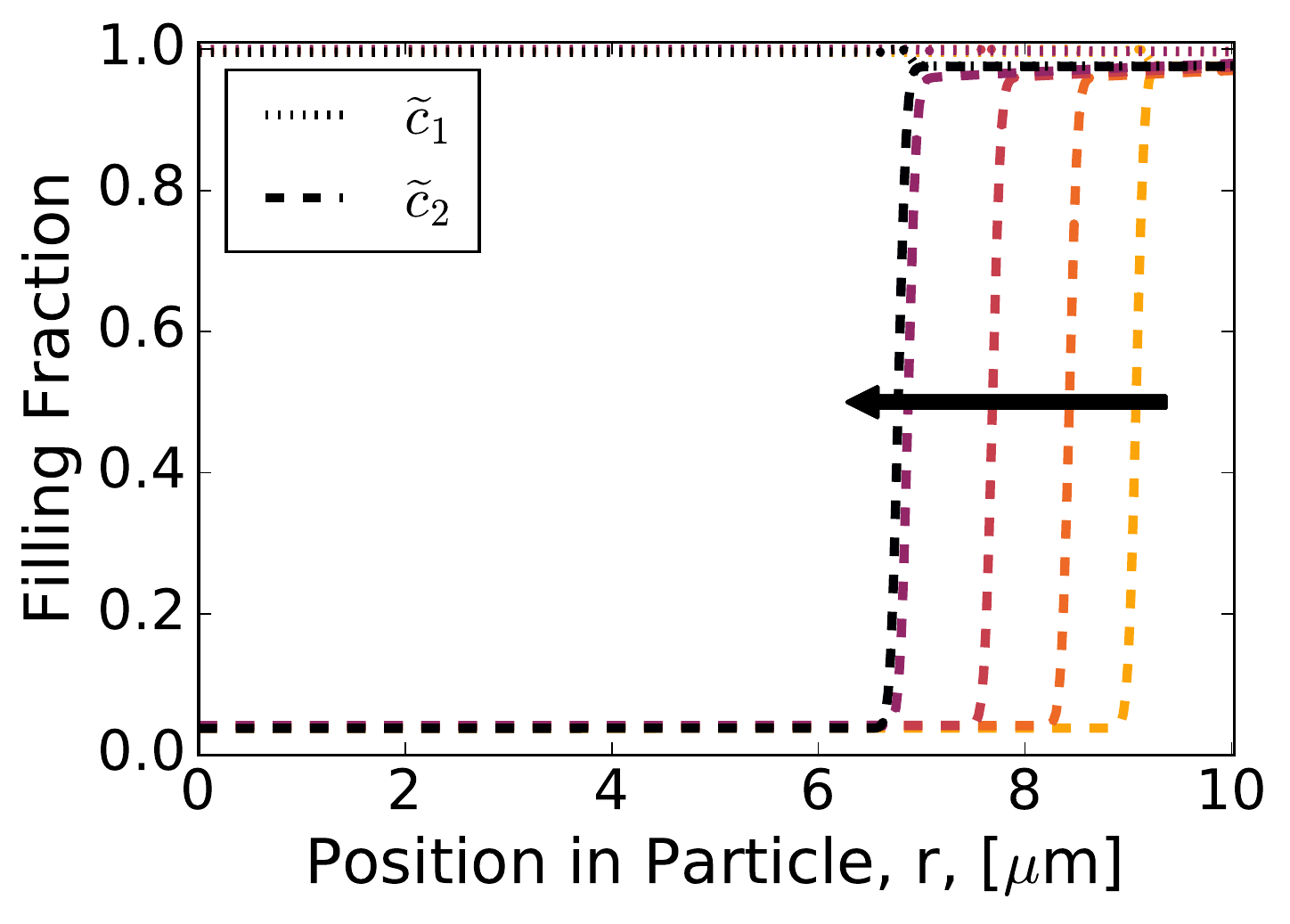}
    \ \
    (c)
    \includegraphics[width=0.28\textwidth]{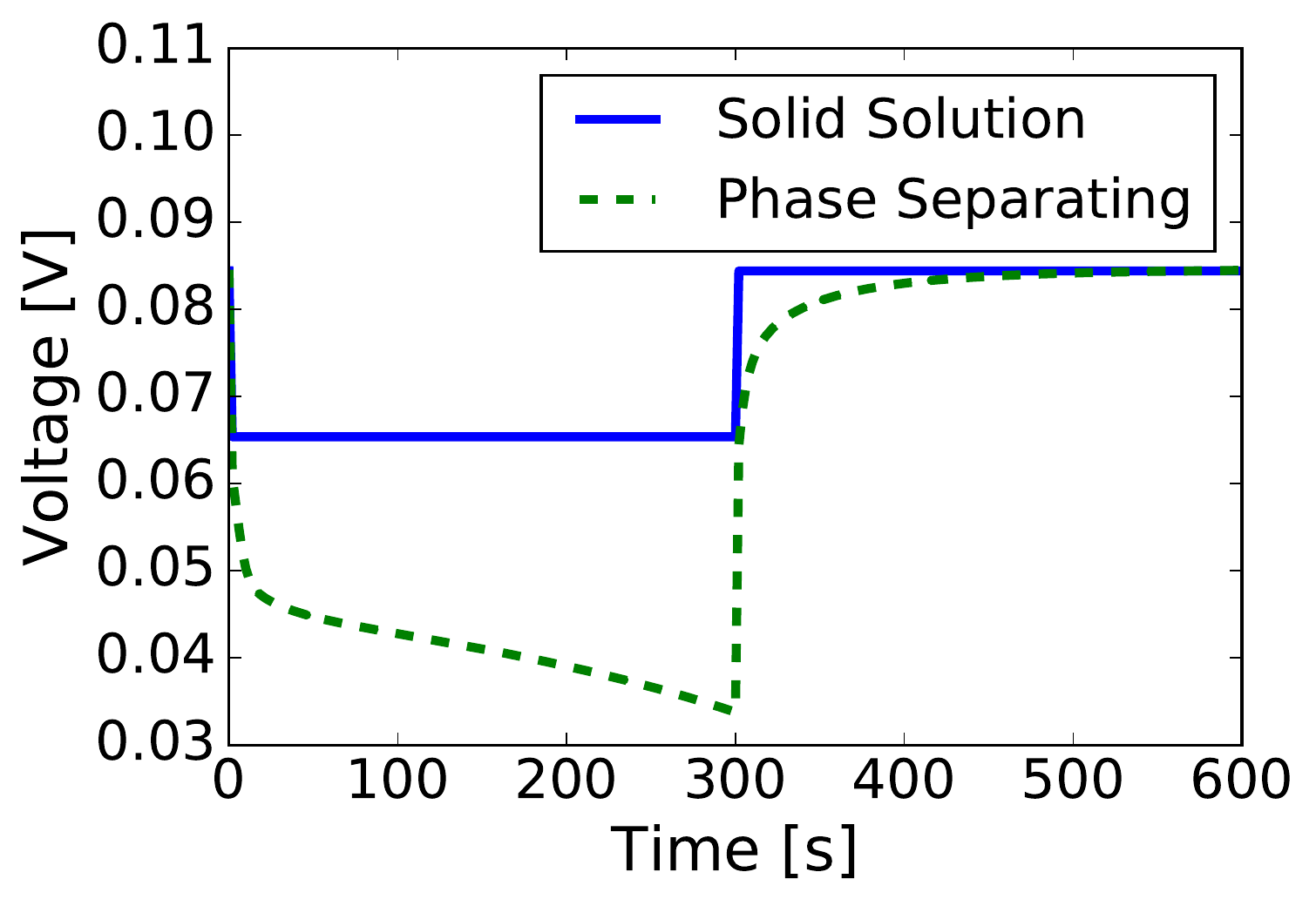}
%    \ \
%    (d)
%    Insert (similar comparison) output voltage(t) plot
    \caption{Comparison of simulated single particle concentration profiles in response to a 300~s current pulse at 2~C followed by a zero-current relaxation. Concentration profiles for the solid solution and phase separating models are presented in (a) and (b) respectively at times $t=0$~s, $t=100$~s, $t=200$~s, $t=300$~s, and $t=600$~s. The arrows and increasing line darkness indicate increasing time. The voltage profiles in (c) differ because of transport losses which only affect the voltage for the phase separating model.}
    \label{fig:graphite_compare}
\end{figure}
In Figure~\ref{fig:graphite_compare}, we compare simulations of the solid solution and phase separating particles exposed to a 2 C current pulse for 300 s beginning after a low-current filling to reach an overall filling fraction of 0.6. We immediately see differences between the initial states, the lightest orange lines in Figure~\ref{fig:graphite_compare} (a) and (b). The solid solution model begins this pulse from a uniform 0.6 filling fraction, whereas the phase separating model begins with a stable phase interface and a high surface concentration. The concentration profiles for the solid solution and phase separating particles (Figure~\ref{fig:graphite_compare} (a) and (b) respectively) are qualitatively different. The solid solution model leads to gradually sloping concentration profiles, which relax toward a uniform concentration again after the current pulse ends. The phase separating model fills by propagating a phase interface in toward the center of the particle and retains a stable interface after relaxation.

The simulated voltage profiles in Figure~\ref{fig:graphite_compare} (c) show that the phase separating particle has transport losses during the current pulse because the surface diffusional chemical potential can rise. In contrast, the solid solution particle voltage is fixed only by the reaction losses because the equilibrium voltage is only a function of concentration and is within the voltage plateau over the range of filling fractions studied. As a result, the predicted voltage profile for the solid solution model with this current pulse is relatively insensitive to the solid solution diffusivity (as long as the surface concentration remains within the miscibility gap). That is, for this current pulse, any value of $D^\mathrm{ss}_\mathrm{chem}$ larger than the simulated value would lead to more uniform concentration profiles and give an identical voltage profile. Of note, when using the phase separating flux prefactor defined by Eq.~\ref{eq:Dfuncconst}, as we use in ref.~\cite{thomas-alyea2016}, in an otherwise similar simulation the voltage profiles are more similar because that form leads to larger flux prefactors in the stable phases (and smaller gradients), but the phase separating voltage profile still retains some decreasing slope. In multi-particle porous electrode simulations with particle-particle interactions~\cite{chung2014particle,stephenson2007modeling}, both models could predict similar transport losses because of non-uniform particle filling and other losses~\cite{newman2004,garcia2005}, but in the case of single particles, the models make qualitatively distinct predictions, both of the concentration profiles in the particles and their associated voltage profiles.

Here, we should reemphasize that solid solution models of graphite have had good success capturing macroscopic porous electrode behavior. The present model is designed to accurately capture the single crystal dynamics, which may not strongly affect the macroscopic current and voltage measurements of full porous electrodes with complex secondary electrode particles. Nevertheless, consistently coupling a phase separating single-crystal model to a hierarchy of kinetic processes including grain boundary diffusion should more consistently predict overall electrode lithium distribution. This would lead to more accurate surface concentration predictions, which are important to describe accurately when developing models to predict lithium plating risk~\cite{harris2010} as well as a more consistent coupling to stresses~\cite{dileo2014,schiffer2016strain}, which we have neglected here but are often studied using solid solution models~\cite{christensen2006,cheng2010,grantab2011,woodford2010}.

\section{Reduced Model for Graphite Battery Electrodes}
\label{sec:modSimp}
By enforcing two-layer periodicity, our free-energy model is able to capture the thermodynamics of lithium intercalation at high filling fractions, but it cannot describe the plethora of stable or metastable phases in graphite at low filling fractions (not only for lithium), which exhibit longer-range periodicity across three or more layers. The complex phase behavior is reflected in the irregular rise of the ``voltage staircase''  at low filling fractions, which cannot be easily described by the two-variable model (Figure~\ref{fig:graphiteOCV}). Moreover, even with these physical limitations, the model is much more computationally expensive than phase-field intercalation models with a single concentration variable, such as that studied by Zeng and Bazant~\cite{zeng2014}. More realistic models with three or more concentration variables to capture low-density phase behavior or including other effects such as elastic coherency strain~\cite{cogswell2012} would be even more costly. 
Similar concerns would apply to the modeling of other multilayer multiphase materials mentioned in the Introduction. There is a trade-off between explicitly describing the microscopic phase behavior with multiple concentration variables (for periodically repeating layers or interpenetrating lattices) and accurately fitting the chemical potential or voltage with a computationally efficient model.

\begin{figure}[h]
    \centering
    (a)
    \includegraphics[width=0.4\textwidth]{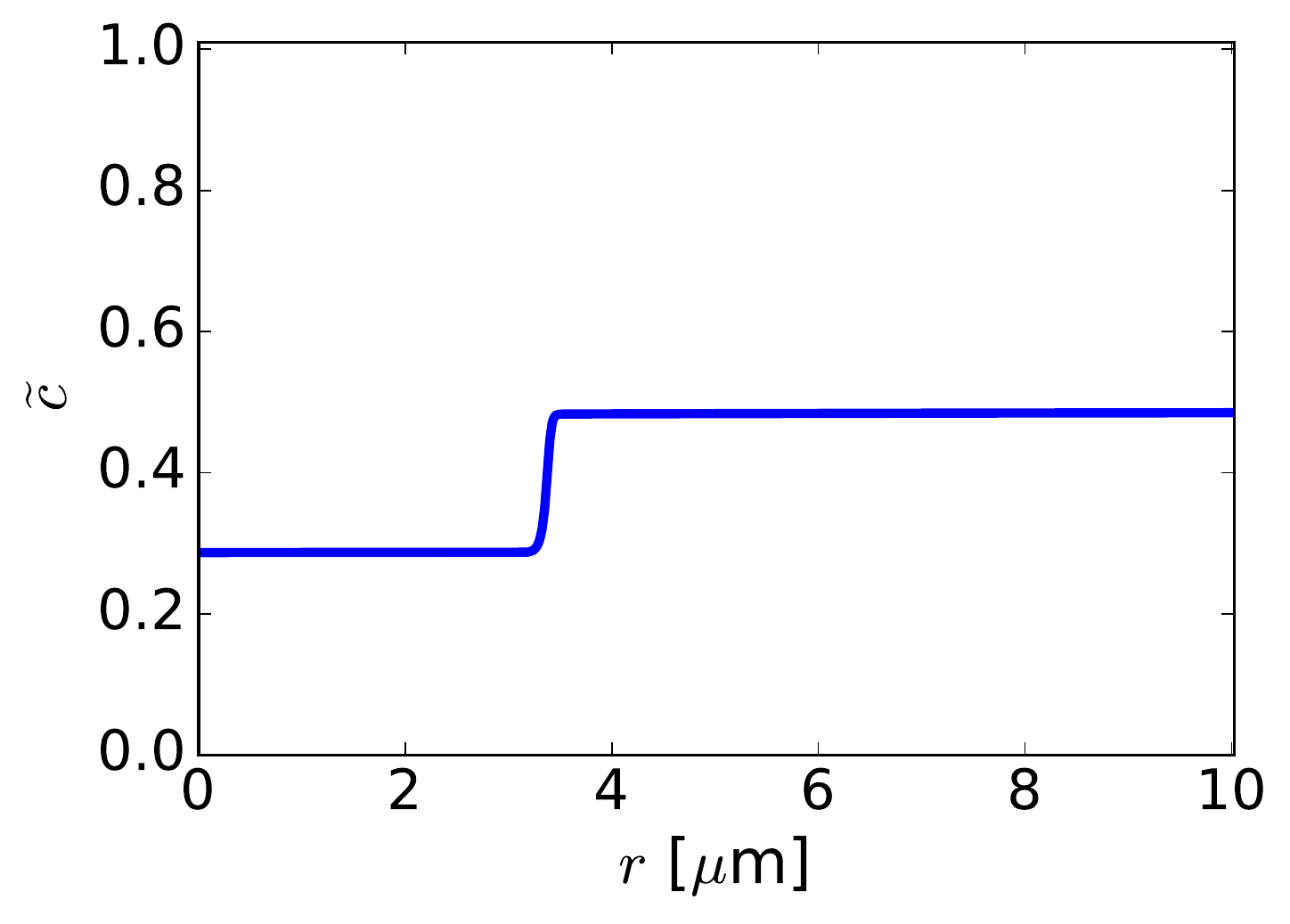}
    \ \
    (b)
    \includegraphics[width=0.4\textwidth]{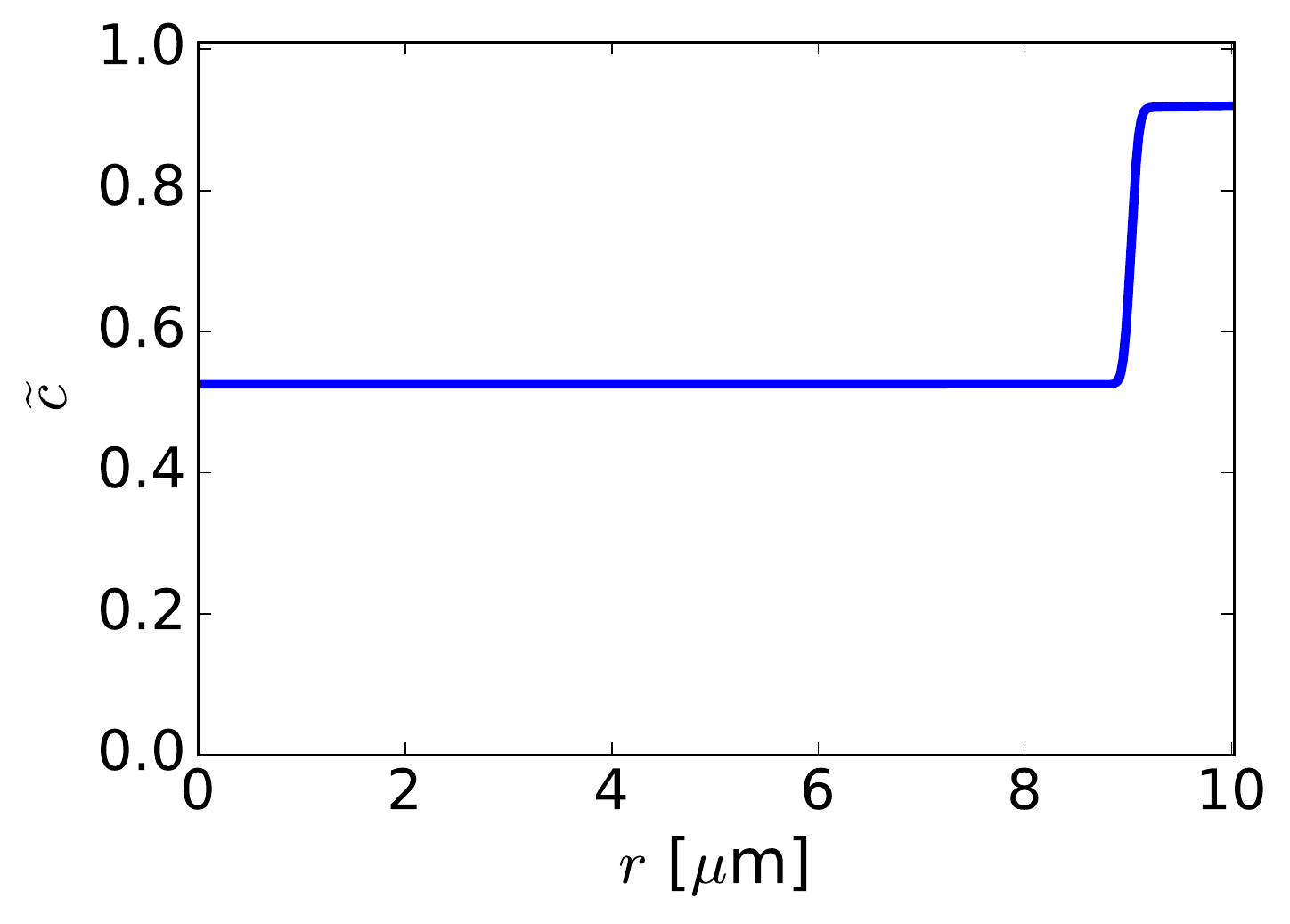}
    \caption{Repeat of the simulation in Figure~\ref{fig:graphite_parallelRxn} with the simplified single-parameter model representing only the average local filling fraction rather than that in each of two repeating layers. (a) and (b) are taken at $t=1630$ s and $t=2130$ s as in the previous simulations.}
    \label{fig:graphite_1paramCC}
\end{figure}

To illustrate the construction of such a reduced model, we develop a single-variable free-energy model for lithium intercalation in graphite, which fits the open circuit voltage at low filling fractions as an effective solid solution, while still capturing the two primary voltage plateaus at high filling fractions, albeit without allowing for symmetry breaking (e.g.\ checkerboard patterns) or any prediction of temperature dependence of the staging behavior. Because of the large energetic barrier to fill along the fully homogeneous yellow paths in Figure~\ref{fig:NRGD}, particles typically follow along the lower energy blue path. To simplify the model, we began with a function which follows the red free energy path to capture the energetics of the unstable phases as a function of only an average concentration. We further made the assumption that below a filling fraction of $1/3$, the high stage numbers can be effectively treated as a solid solution, which allowed us to modify the homogeneous free energy function (or equivalently, the homogeneous diffusional chemical potential) to match the voltage in this low-filling-fraction region. Doing so ignores any higher order staging and stage transitions, but leads to a simple model which is better suited to scale up within multi-particle simulations, e.g.\ for porous, phase separating battery simulation as we explore in a companion work~\cite{thomas-alyea2016}. We also adjusted the high-filling behavior to have a more gradual drop off, as observed in various experiments~\cite{verbrugge2003,thomas2003heats,dahn1990}. Finally, we adjusted the decreasing regions of diffusional chemical potential (increasing equilibrium voltage) to reduce the metastability region which causes open circuit voltage hysteresis~\cite{levi1997simultaneous,dreyer2010}. The diffusional chemical potential function we used which satisfies the criteria above is presented in Appendix~\ref{sec:app_mu1param} and in Figure~\ref{fig:simp1param} noting that $V_{\mathrm{op},H} = E^\Theta - \mu_\mathrm{op}/e$. The free energy permits construction of two common tangents between filling fractions near $0.3$ and $0.5$ and another between filling fractions near $0.5$ and $1$ which leads to the two clear voltage plateaus. We emphasize that, although this approach predicts phase separation based on an average filling fraction, it cannot capture checkerboard microstructures or make predictions about temperature dependence of the phase diagram (Section~\ref{sec:phaseDiagram}), which are possible with the two-variable model above.
\begin{figure}[h]
    \centering
    (a)
    \includegraphics[width=0.4\textwidth]{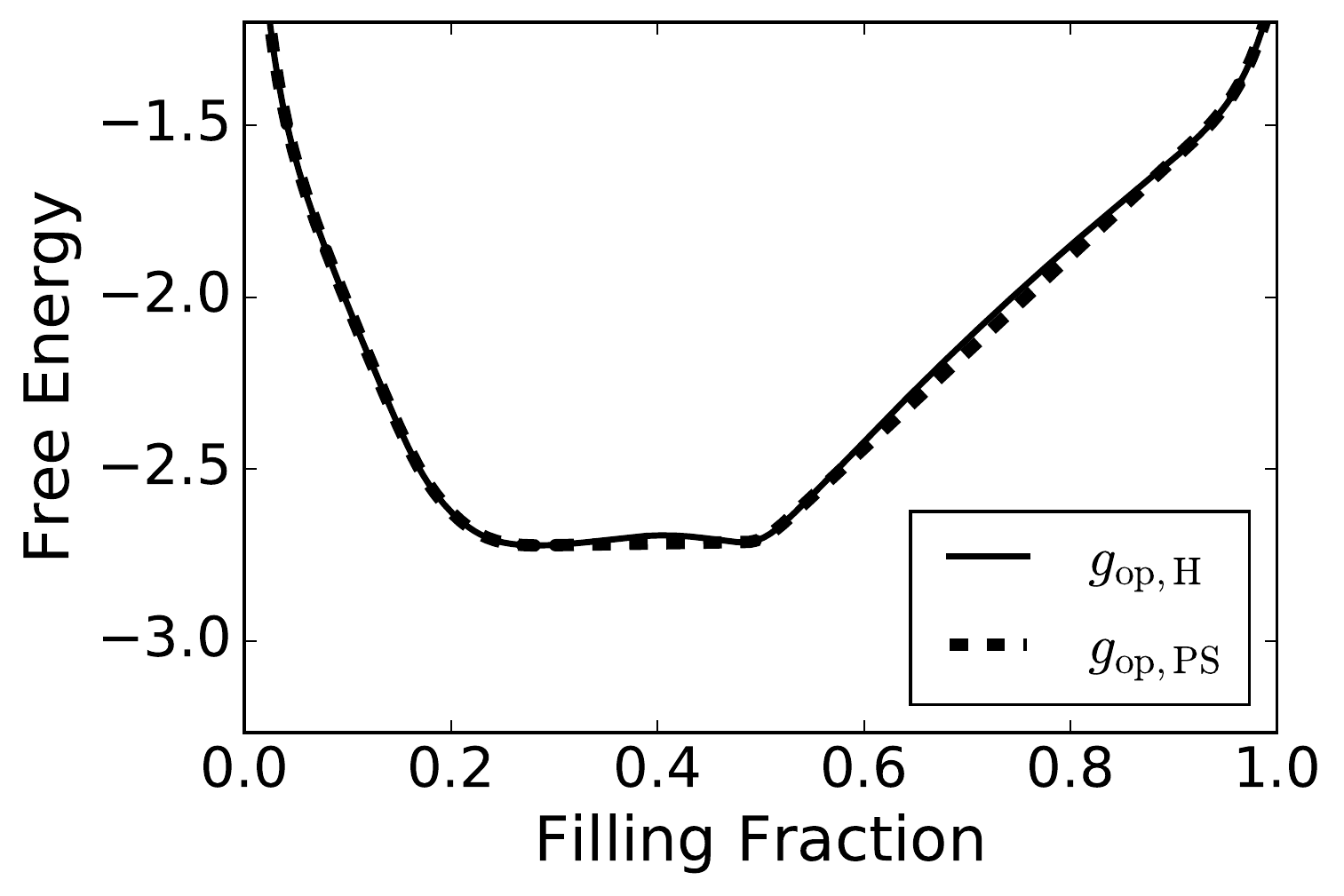}
    \ \
    (b)
    \includegraphics[width=0.4\textwidth]{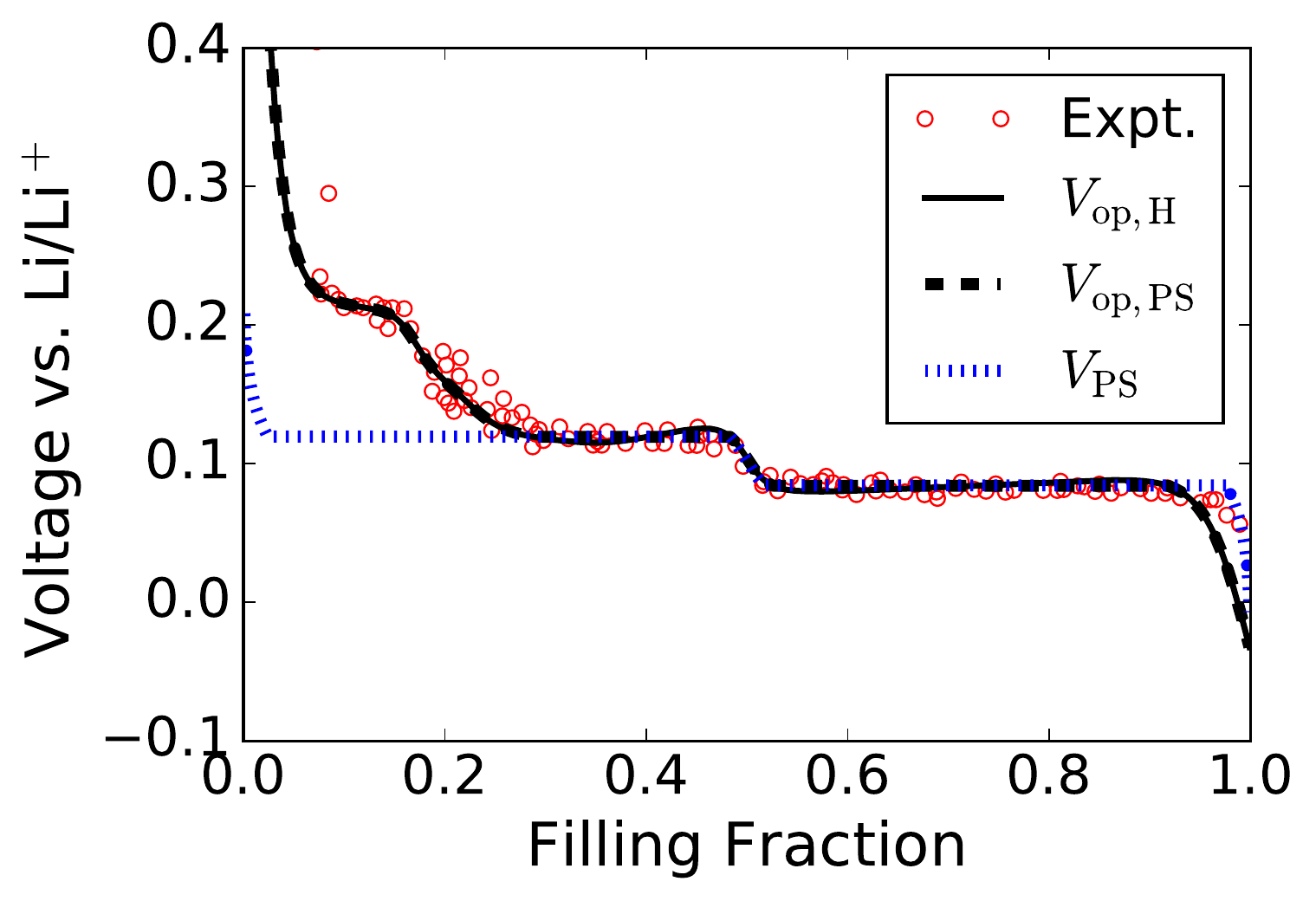}
    \caption{Plot of the reduced, one-parameter model of graphite which captures the experimental open circuit voltage data (from ref.~\cite{ohzuku1993}) over the full range of filling fractions. The free energy in (a) is scaled to $c_\mathrm{ref}k_\mathrm{B}T$ and presents both the homogeneous free energy per particle ($g_\mathrm{op,H}$) and that of a phase separating system constructed with common tangents ($g_\mathrm{op,PS}$). The resultant voltage of the two functions are presented in (b) along with the voltage resulting from the equilibrium blue paths in Figure~\ref{fig:NRGD} ($V_\mathrm{PS}$).}
    \label{fig:simp1param}
\end{figure}
Simulations using this free energy lead to predicted concentration profiles with stable average filling fractions near $0.5$ and $1$ in the high-filling states as in the two-parameter model, but the low-filling state has an average filling fraction near $0.3$, which is clearly demonstrated in a constant current discharge like those presented in Figure~\ref{fig:graphite_1paramCC} (compare to Figures~\ref{fig:graphite_parallelRxn} and~\ref{fig:graphite_interlayerRxn}). The predicted average concentration profiles look similar to those with inter-layer reaction but miss the formation of internal rings of stage 1 ($\overline{c}\approx 1$) predicted by the original 2-parameter model.

We apply this model within a porous electrode simulation in ref.~\cite{thomas-alyea2016} and demonstrate good agreement with both macroscopic voltage transients and visual indicators of lithium distribution within a graphite electrode responding to current pulses. We also find that using Eq.~\ref{eq:flux_with_prefactor} and the reaction model described in Eqs.~\ref{eq:rxn} and~\ref{eq:gamma_ts} does not provide reasonable fits to the experimental data, but we are able to fit data using the flux expression defined in Eqs.~\ref{eq:flux_base},~\ref{eq:flux_prefactor_gammas}, and~\ref{eq:Dfuncconst}. The reaction model we use to fit the data is a Butler-Volmer expression dominated by film resistance. The general framework developed here is inspired by simple thermodynamic arguments developed for single crystals, whereas practical graphite electrode particles are secondary agglomerates composed of many smaller crystals joined by grain boundaries~\cite{wissler2006}. This suggests that these modifications to the general model are capturing effective properties of the secondary particles and provide a better starting point to describe graphite electrodes.

\section{Summary and Conclusions}
\label{sec:concl}
We have developed and studied a model for the kinetics of intercalation for layered, phase-separating compounds that tend to form staged structures. Using the particular example of lithium in graphite, the most common lithium-ion battery anode, we specialized the model to describe the most visually distinct low-stage number structures of lithiated graphite as well as the staircase character of its equilibrium voltage. Having applied the model to a single particle in ref.~\cite{guo2016}, we explored the generalization of the model as well as some of the unique predictions it makes which differ from other models commonly employed to describe lithiated graphite. We found the specifics of the choice for reaction model to be critically important in affecting practical macroscopic simulation outputs such as predicted voltage under constant current intercalation (discharge). In particular, the reaction model used for the single-crystal simulations and experiment studied in ref.~\cite{guo2016} does not lead to qualitative agreement with typical graphite electrode discharge curves, which supports the idea that for secondary graphite electrode particles as we study in ref.~\cite{thomas-alyea2016}, the reaction model described in Section~\ref{sec:modSimp} is more appropriate. Simulations directly comparing the model to more commonly employed solid solution models highlight some of the key differences between the models, including concentration profiles of particles under current pulses and the associated voltage profiles. In addition, we developed a simplified model designed to accurately match observed graphite open circuit voltage profiles while retaining the overall phase separating characteristics of the 2-layer model at a substantially reduced computational cost. This facilitated the work in ref.~\cite{thomas-alyea2016}, in which we show that the model can reproduce electrochemical data as well as visual indications of concentration carried out \emph{in situ} during current pulses. Natural extensions of the work include more carefully capturing the temperature dependence of the phase diagram by refining the free energy model. We have also neglected here the effects of elastic contributions to the energetics and dynamics of the materials, and incorporating them is a natural next step to make the model more physically descriptive of graphite~\cite{dahn1982elastic,qi2010threefold,qi2010insitu,bower2011finite,sethuraman2012realtime,qi2014lithium} in a way that consistently couples the stresses to the phase separating concentration profiles.

While much of the focus here was on lithiated graphite, the general form of the model could be useful for other materials exhibiting staging or similar behavior, such as some layered double hydroxides~\cite{pisson2003}, or metal borocarbides~\cite{joshi2015}. MXenes~\cite{wang2015pseudocapacitance,naguib2014} have also shown behavior analogous to staging, indicating the general framework developed here may help capture their interactions with intercalants. Because many intercalants in graphite exhibit similar behavior~\cite{dresselhaus1981}, the model could likely be adapted to describe other graphitic electrodes such as those intercalated with aluminum~\cite{lin2015} or sodium~\cite{kim2015sodium}. It may also be extensible to other sodium-based layered systems with non-trivial phase behavior~\cite{wu2015sodium}. Beyond the layered framework developed here, the model's general structure of spatially overlapping lattices may be useful for similar phenomena such as that of different occupational sites available for Li in the electrode material anatase TiO$_2$~\cite{zachau-christiansen1988,lafont2010,shin2011,shen2014,deklerk2016anatase_draft} or different interstitial sites available for hydrogen in metals~\cite{dileo2013}.

The choice to develop this model with a thermodynamically consistent connection to electrochemical reaction kinetics makes it particularly suited to describe systems with internal phase separation as well as staging and the associated implications on battery and supercapacitor performance, but the generality of the model enables adaptation to describe intercalation kinetics of other related behaviors. Because the model is based on physical understanding of system microstructure, it is well suited to parameterization by \emph{ab initio} calculations and can be used to infer microscopic physical mechanisms for transport and intercalation from experimental data. This allows it to provide meso-scale understanding of system behavior while retaining a clear connection to both microscopic modeling and macroscopic experiments with the generality to be specialized and applied to a wide range of systems.

\section{Acknowledgements}
R. B. Smith and M. Z. Bazant acknowledge support from the Samsung Advanced Institute of Technology. E. Khoo acknowledges support from the National Science Scholarship (PhD) funded by Agency for Science, Technology and Research, Singapore (A*STAR). The authors gratefully thank K. E. Thomas-Alyea for helpful discussions throughout, guidance toward the one-parameter model reduction, and very thoughtful criticisms of the manuscript. We also thank E. Rejovitzky for insightful criticisms of the manuscript.

\appendix

\section{Transport coefficient for diffusion on a lattice}
\label{sec:appTransCoeff}
The activity coefficients are related to excess chemical potentials, $\mu_i^\mathrm{ex} = \mu_i-k_\mathrm{B}T\ln{\wt{c}_i}$, so using Eq.~\ref{eq:chemPot},
\begin{align}
    \gamma_i &= \exp\left( \frac{\mu_i^\mathrm{ex} - \mu^\Theta}{k_\mathrm{B}T}\right)
%    = \exp\left[ \frac{\left(\mu_i - k_\mathrm{B}T\ln c_i\right) - \mu_i^\Theta}
%    {k_\mathrm{B}T}\right]
    \\
    &= \frac{1}{1-\wt{c}_i}\exp\left\{ \frac{1}{k_\mathrm{B}T}\left[
    \Omega_a(1-2\wt{c}_i) -
    \frac{\kappa}{c_{\textrm{ref}}}\nabla^2\wt{c}_i
    + \Omega_b\wt{c}_j + \Omega_c(1-2\wt{c}_i)\wt{c}_j(1-\wt{c}_j)
    \right] \right\}.
    \label{}
\end{align}
Then, we postulate a diffusion transition state activity coefficient in which the diffusing species excludes two sites and maintains the bulk enthalpic contributions to its diffusional chemical potential~\cite{bazant2013},
\begin{align}
    \gamma_{\ddagger,i}^d = \frac{1}{{\left(1-\wt{c}_i\right)}^2}
    \exp\left\{ \frac{1}{k_\mathrm{B}T}\left[
    \Omega_a(1-2\wt{c}_i) -
    \frac{\kappa}{c_{\textrm{ref}}}\nabla^2\wt{c}_i
    + \Omega_b\wt{c}_j + \Omega_c(1-2\wt{c}_i)\wt{c}_j(1-\wt{c}_j)
    \right] \right\}
    ,
    \label{eq:TS_d}
\end{align}
which, combined with Eq.~\ref{eq:flux_prefactor_gammas} leads to Eq.~\ref{eq:flux_with_prefactor}.

\section{Simulation details for examining axial symmetry assumption}
\label{sec:appAxsym}
\begin{figure}[!h]
    \centering
    (a)
    \includegraphics[width=0.4\textwidth]{2d_comsol_c1_snapshot_1.png}
    \ \
    (b)
    \includegraphics[width=0.4\textwidth]{2d_comsol_c1_snapshot_2.png}
    \\
    (c)
    \includegraphics[width=0.4\textwidth]{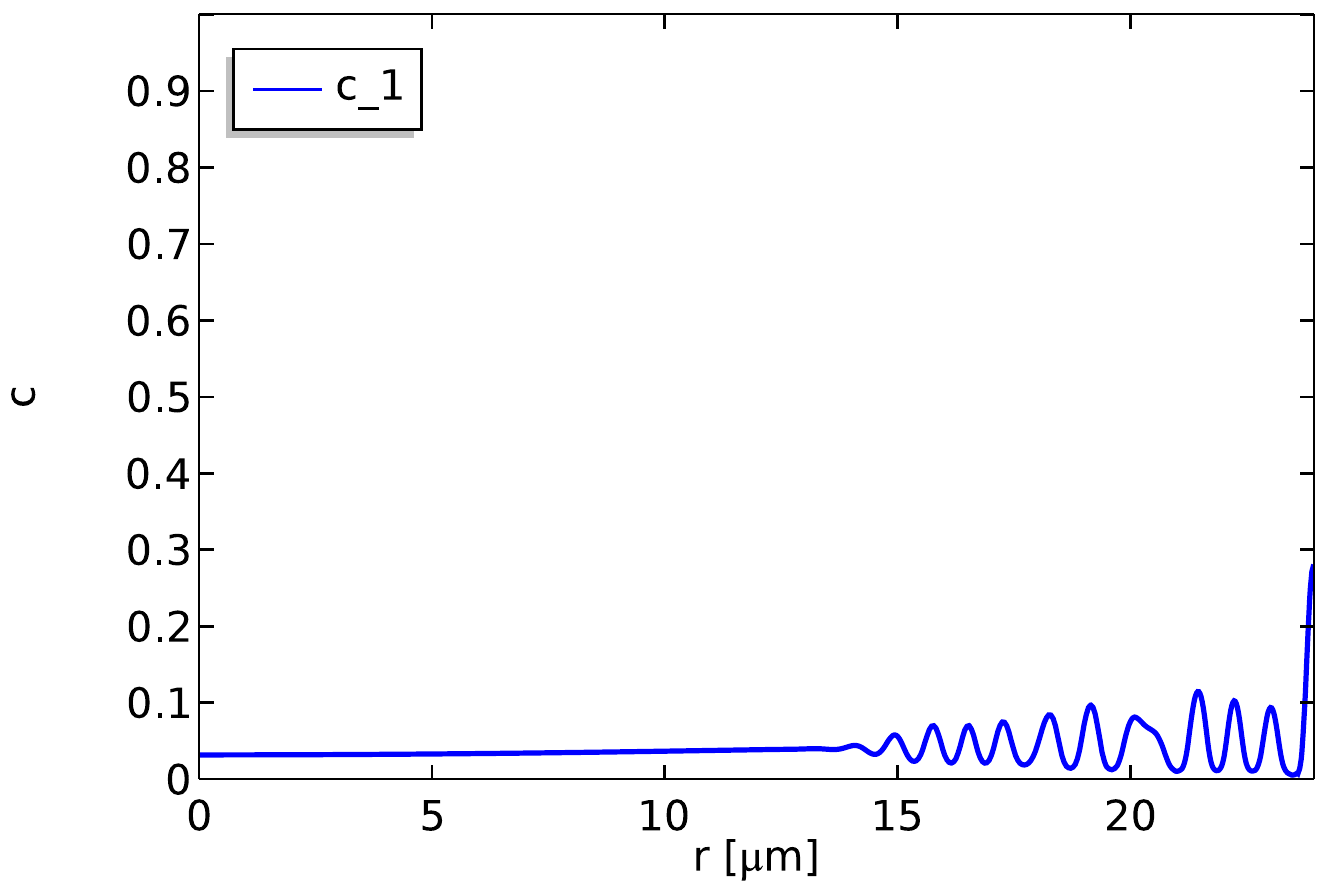}
    \ \
    (d)
    \includegraphics[width=0.4\textwidth]{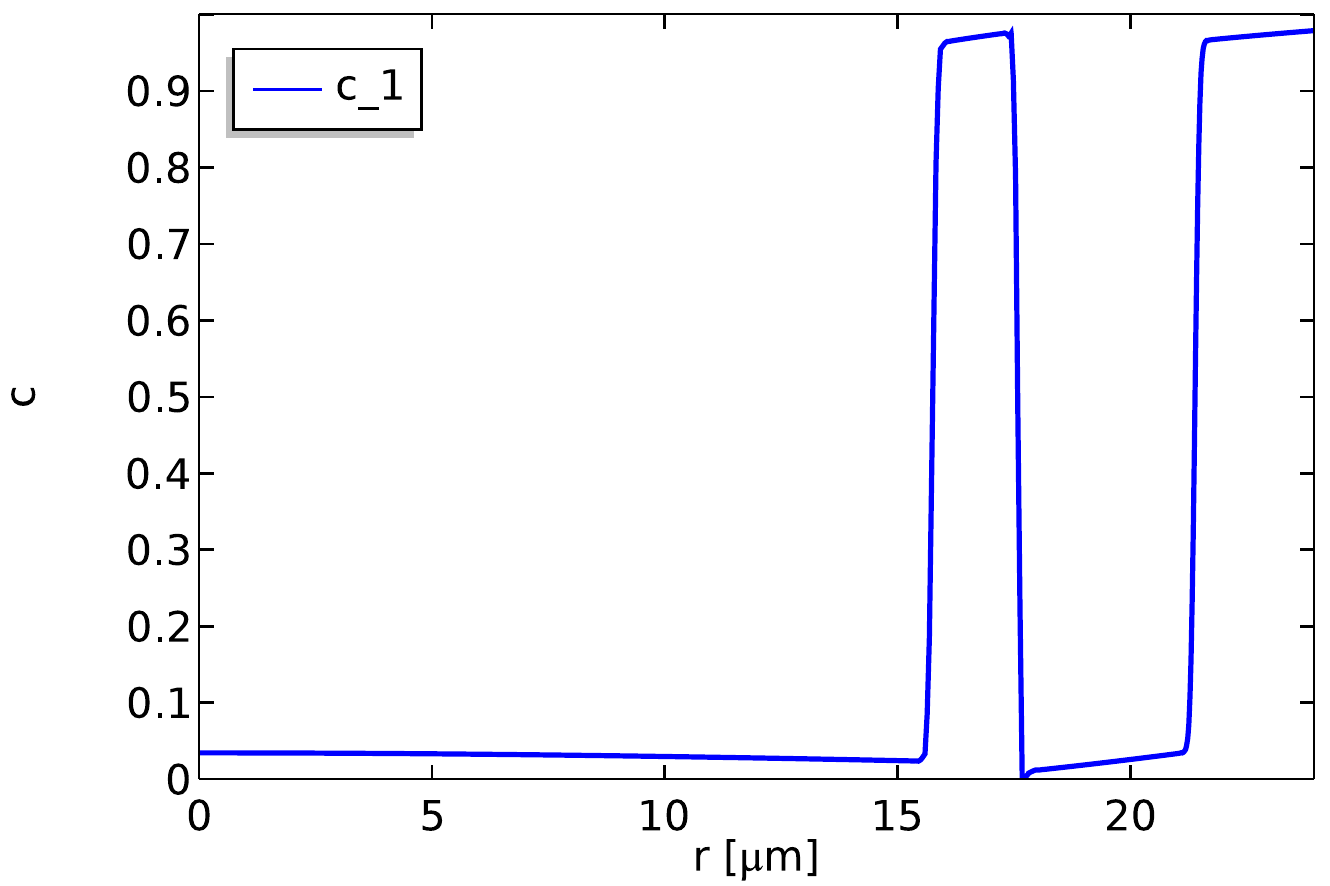}
    \\
    (e)
    \includegraphics[width=0.4\textwidth]{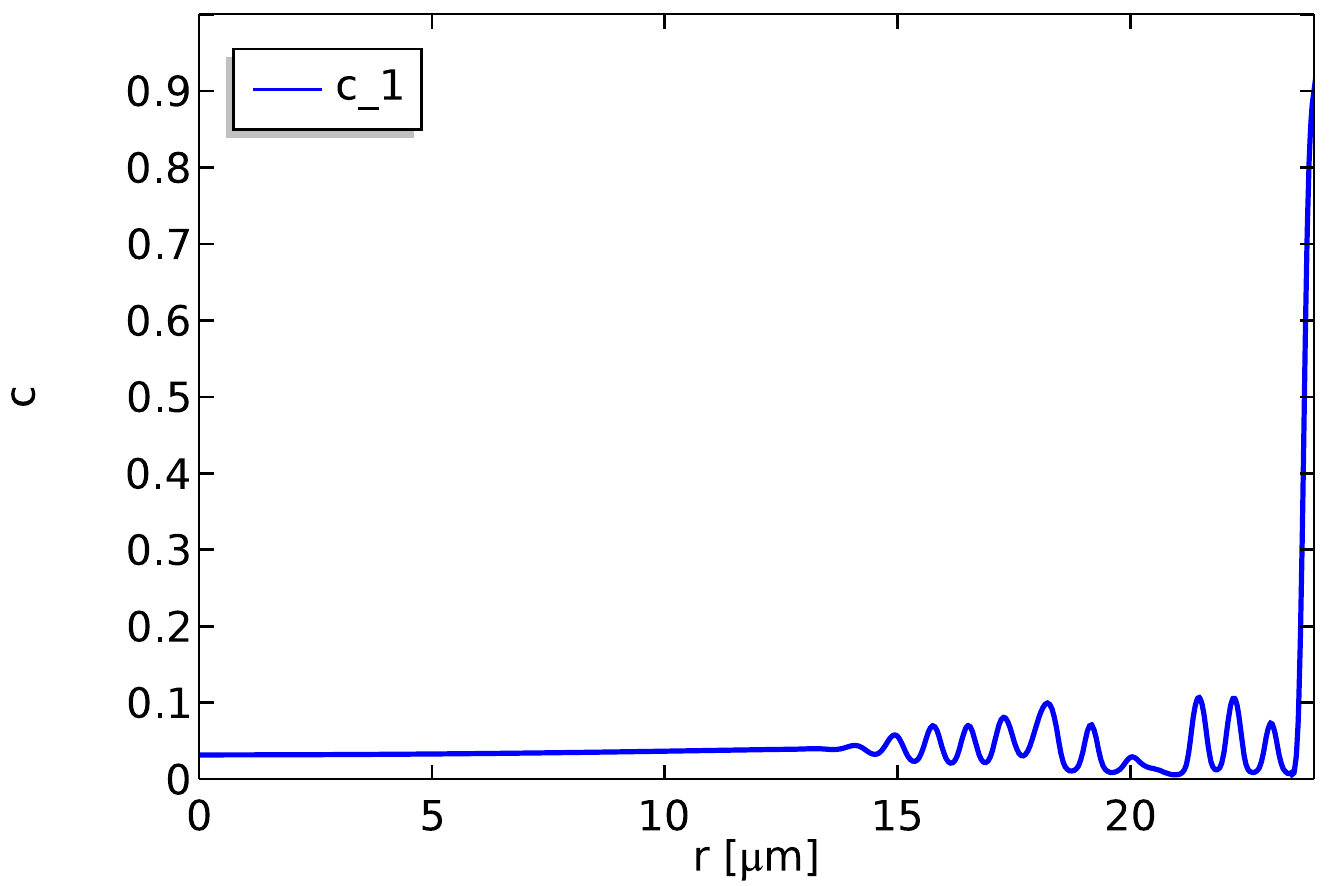}
    \ \
    (f)
    \includegraphics[width=0.4\textwidth]{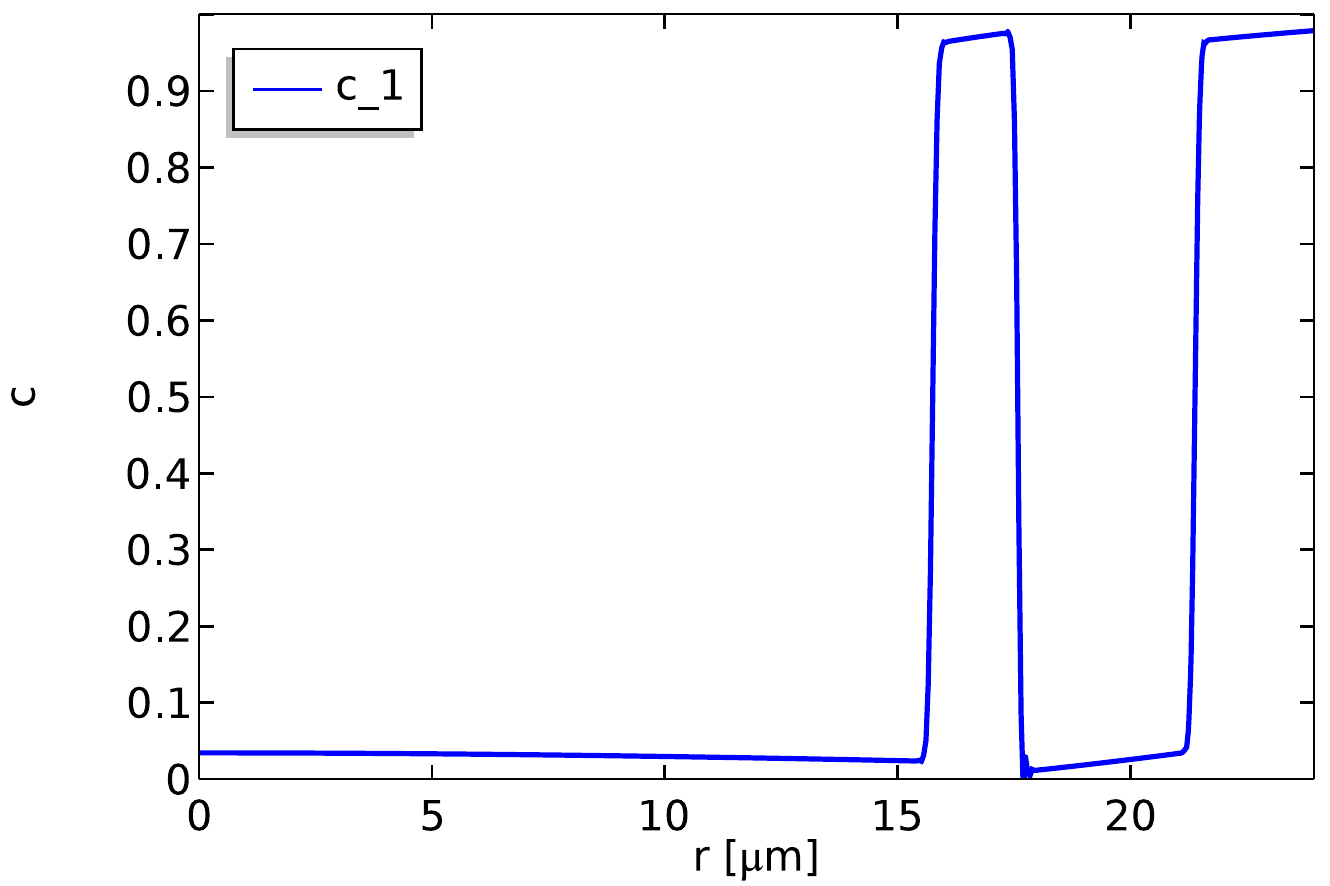}
    \\
    (g)
    \includegraphics[width=0.4\textwidth]{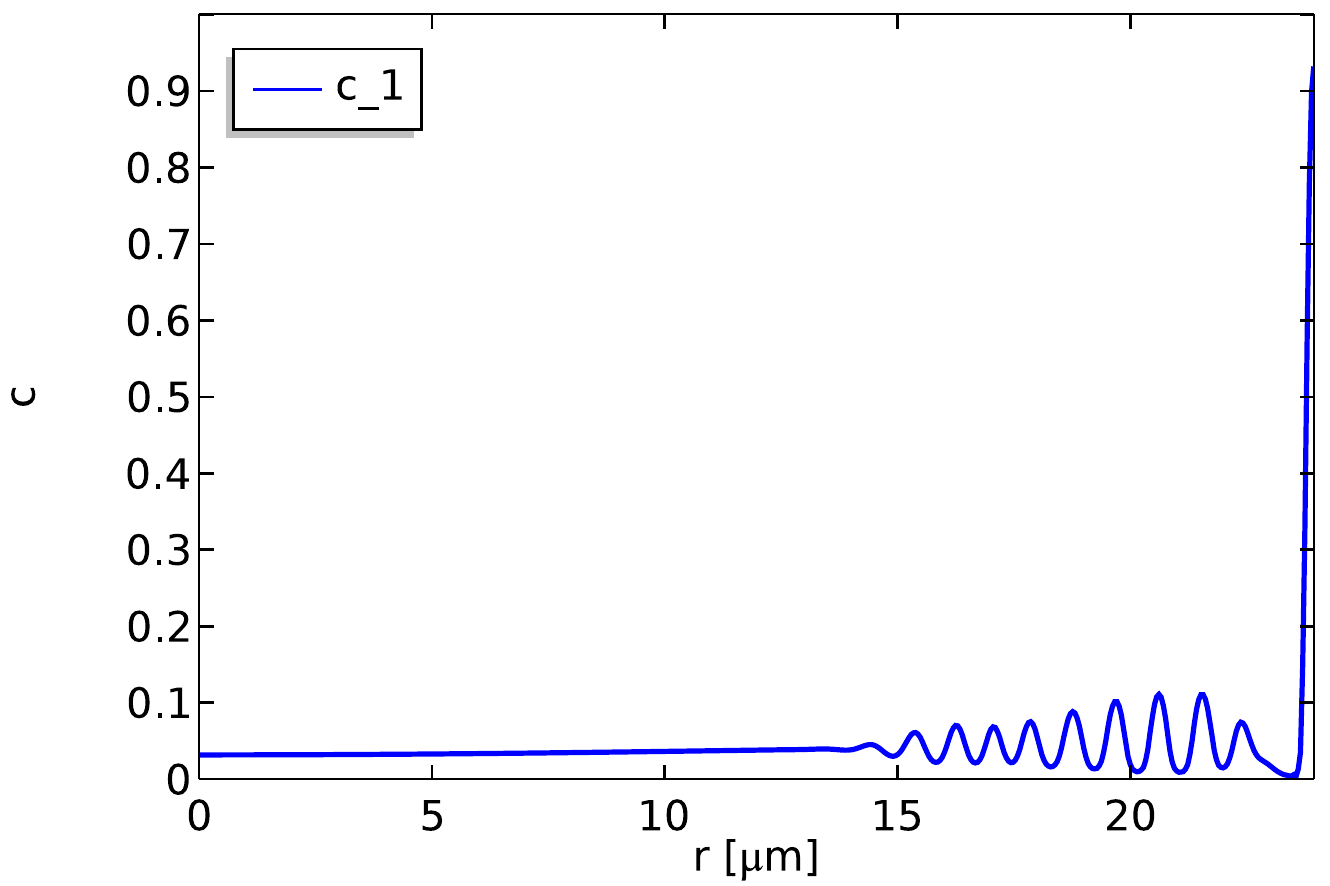}
    \ \
    (h)
    \includegraphics[width=0.4\textwidth]{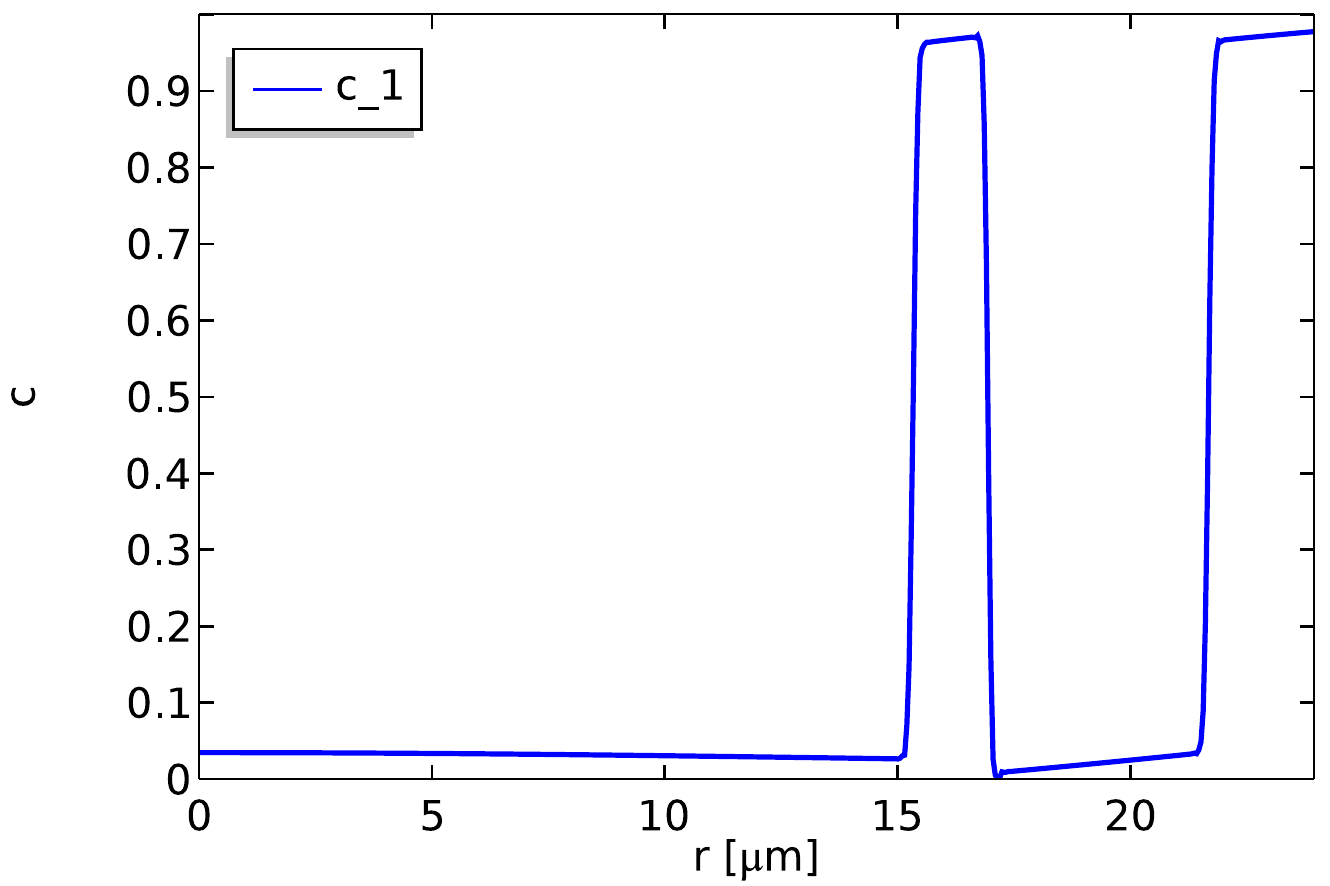}
    \caption{2D and 1D simulations from ref.~\cite{guo2016} performed using COMSOL Multiphysics (finite element discretization). (a) and (b) are a repeat of Figure~\ref{fig:2d_sim} for easy comparison. Snapshots in (a), (c), (e) and (g) are taken at $t = 669\ \textrm{s}$ and those in (b), (d), (f) and (h) are taken at $t = 1553\ \textrm{s}$. Snapshots from the 2D simulation in (c) and (d) are taken at $\theta = 0\degree$ and those in (e) and (f) are taken at $\theta = 0.352\degree$. Snapshots in (g) and (h) are taken from the 1D simulation.}
    \label{fig:2d_comparison}
\end{figure}

In the interest of keeping the 2D simulation time to a few hours, we perform the simulation up to $t = 1553\ \textrm{s}$ instead of $t = 6210\ \textrm{s}$ as in ref.~\cite{guo2016}. We impose no-flux boundary conditions on the straight edges of the sector and lithium intercalates along the arc of the sector. In Figure~\ref{fig:2d_comparison}, we plot $\wt{c}_1$ at $t = 669\ \textrm{s}$ where axial symmetry is broken, and at $t = 1553\ \textrm{s}$ where axial symmetry is not broken. We provide plots of $\wt{c}_1$ against $r$ at $\theta = 0\degree$ and $\theta = 0.352\degree$ to examine how axial symmetry is broken or not broken. In addition, we compare these $\wt{c}_1$ plots with those obtained from a 1D simulation in COMSOL Multiphysics using the same grid spacing of $1.43 \times 10^{-7}\ \textrm{m}$ to examine how well the 1D model approximates the 2D model. This grid spacing is larger than the approximate value of the interface width in these simulations ($7.1\times10^{-8}\ \mathrm{m}$), which is a compromise on accuracy to speed up the simulations and enable a larger central angle for the 2D sector. We retain the same grid spacing in both cases in Figure~\ref{fig:2d_comparison} to focus on the comparison rather than accurate model predictions. Full movies are provided in the supplement.

In Figure~\ref{fig:2d_comparison}, comparing plots (a), (c) and (e) for $t = 669\ \textrm{s}$, we observe that axial symmetry in the $\wt{c}_1$ profile is broken; for instance, at the arc of the sector, $\wt{c}_1$ increases with increasing $\theta$. On the other hand, comparing plots (b), (d) and (f) for $t = 1553\ \textrm{s}$, $\wt{c}_1$ does not vary as a function of $\theta$ and axial symmetry is not broken. More generally, even though we initialize $\wt{c}_1$ and $\wt{c}_2$ to be $10^{-2}$ uniformly, we observe that axial symmetry is broken immediately, turns ``less broken'' as $t$ increases, first becomes unbroken at $t = 776\ \textrm{s}$ (not shown in Figure~\ref{fig:2d_comparison}, but can be seen in the full videos in the supplement), and then remains unbroken till $t = 1553\ \textrm{s}$. Although we did not perform 2D simulations up to $t = 6210\ \textrm{s}$ as was done in ref.~\cite{guo2016}, we expect that axial symmetry remains unbroken from $t = 1553\ \textrm{s}$ to $t = 6210\ \textrm{s}$. Comparing plots (c) and (e) with (g), the broken axial symmetry at $t = 669\ \textrm{s}$ causes some deviations between the $\wt{c}_1$ profiles for $\theta = 0\degree$, $\theta = 0.352\degree$ and the true 1D simulation. However, comparing plots (d) and (f) with (h), these $\wt{c}_1$ profiles agree reasonably well with each other when axial symmetry is not broken at $t = 1553\ \textrm{s}$; this also holds true when axial symmetry first becomes unbroken at $t = 776\ \textrm{s}$. Crucially, the four internal domains for $\wt{c}_1$ in the 2D simulation are also reproduced by the 1D simulation. Therefore, the 1D simulation approximates the 2D simulation reasonably well except at early times where there are minor deviations that do not significantly affect the accuracy of the 1D simulation at later times.

\section{Matching diffusive behavior in solid solution and phase separating simulations}
\label{sec:SSPSdiffn}
To approximately match the transport behavior of the two models, it is helpful to rearrange the flux expression in Eqs.~\ref{eq:flux_base} and~\ref{eq:flux_prefactor_gammas} in terms of a chemical diffusivity, which is valid in regions where the gradient term is negligible (far from phase interfaces),
\begin{align}
    \mathbf{F}_i \approx -\frac{D_0}{k_\mathrm{B}T}\frac{c_i\gamma_i}{\gamma_{\ddagger,i}^d}\frac{\partial\mu_i}{\partial \wt{c}_i}\bnab\wt{c}_i = -D_{\mathrm{chem},i}\bnab c_i
    \label{}
\end{align}
where
\begin{align}
    D_{\mathrm{chem},i} &= \frac{D_0}{k_\mathrm{B}T}\frac{\wt{c}_i\gamma_i}{\gamma_{\ddagger,i}^d}\frac{\partial\mu_i}{\partial \wt{c}_i}
    \\
    &= \frac{D_0}{k_\mathrm{B}T}\frac{\wt{c}_i\gamma_i}{\gamma_{\ddagger,i}^d}
    \left[ \frac{k_\mathrm{B}T}{\wt{c}_i\left( 1-\wt{c}_i \right)} - 2\Omega_a - 2\Omega_c\wt{c}_j\left( 1-\wt{c}_j \right) \right].
    \label{}
\end{align}
Here, we are constrained on our choice for the ratio $\gamma_i/\gamma_{\ddagger,i}^d > 0$ by the second law of thermodynamics~\cite{groot1962}, so we cannot choose it such that we have perfectly constant $D_{\mathrm{chem},i}$. We note that the choice originally proposed in Appendix~\ref{sec:appTransCoeff} and simplified in Eq.~\ref{eq:flux_with_prefactor} gives us~\cite{guo2016}
\begin{align}
    D_{\mathrm{chem},i} &= \frac{D_0}{k_\mathrm{B}T}\left( k_\mathrm{B}T - 2\Omega_a\wt{c}_i\left( 1-\wt{c}_i \right)
    - 2\Omega_c\wt{c}_i\left( 1-\wt{c}_i \right)\wt{c}_j\left( 1-\wt{c}_j \right)\right),
    \\
    &\approx D_0 \quad\quad\text{far from phase interfaces}.
    \label{}
\end{align}
Another natural choice may be to assume constant prefactor in front of the $\bnab\mu_i$, or
\begin{align}
    \frac{\wt{c}_i\gamma_i}{\gamma_{\ddagger,i}^d} = 1,
    \label{eq:Dfuncconst}
\end{align}
leading to
\begin{align}
    D_{\mathrm{chem},i} = \frac{D_0}{k_\mathrm{B}T}
    \left[ \frac{k_\mathrm{B}T}{\wt{c}_i\left( 1-\wt{c}_i \right)} - 2\Omega_a - 2\Omega_c\wt{c}_j\left( 1-\wt{c}_j \right) \right]
    \label{}
\end{align}
which has been used in a similar model~\cite{hawrylak1984} and has diverging chemical diffusivity near full and empty states. We also find this model specified by Eq.~\ref{eq:Dfuncconst} to match porous electrode data in our companion paper in which the particle models are describing effective properties of secondary (polycrystalline) graphite particles~\cite{thomas-alyea2016}. However, to retain the most similar behavior of chemical diffusivity between the solid solution and phase separating models, we focus here on the model proposed in Eq.~\ref{eq:flux_with_prefactor}.

\section{Diffusional chemical potential for single-variable model reduction}
\label{sec:app_mu1param}
In the reduced thermodynamic model for lithium intercalation in graphite, the function describing the diffusional chemical potential as a function of a single total filling fraction variable is given by
\begin{align}
    \frac{\mu_\mathrm{op}\left( \wt{c} \right)}{k_\mathrm{B}T_\mathrm{ref}} =
    0.18 + \wt{\mu}_A\left( \wt{c} \right) + \wt{\mu}_B\left( \wt{c} \right)
    + \wt{\mu}_C\left( \wt{c} \right) + \wt{\mu}_D\left( \wt{c} \right)
    + \wt{\mu}_E\left( \wt{c} \right)
    - \frac{\kappa}{c_\mathrm{ref}k_\mathrm{B}T_\mathrm{ref}}\nabla^2\wt{c}
    \label{}
\end{align}
where
\begin{align}
    \wt{\mu}_A\left( \wt{c} \right) &= \bigg( -40\exp\left( -\frac{\wt{c}}{0.015} \right)
    + 0.75\left( \tanh\left( \frac{\wt{c} - 0.17}{0.02} \right) - 1 \right)
    \\ \nonumber
    &\quad\quad
    + \left( \tanh\left( \frac{\wt{c} - 0.22}{0.04} \right) - 1 \right) \bigg)
    * S_D\left( \wt{c}, 0.35, 0.05 \right)
    \\
    \wt{\mu}_B\left( \wt{c} \right) &= -\frac{0.05}{\wt{c}^{0.85}}
    \\
    \wt{\mu}_C\left( \wt{c} \right) &= 10 * S_U\left( \wt{c}, 1, 0.045 \right)
    \\
    \wt{\mu}_D\left( \wt{c} \right) &= 6.12\left( 0.4 - \wt{c}^{0.98} \right)
    * S_D\left( \wt{c}, 0.49, 0.045 \right) * S_U\left( \wt{c}, 0.35, 0.05 \right)
    \\
    \wt{\mu}_E\left( \wt{c} \right) &= \left( 1.36\left( 0.74-\wt{c} \right) + 1.26 \right)
    * S_U\left( \wt{c}, 0.5, 0.02 \right)
    \label{}
\end{align}
and step up and step down functions respectively defined by
\begin{align}
    S_U\left(x, x_c, \delta\right) &= 0.5\left( \tanh\left( \frac{x-x_c}{\delta} \right) + 1 \right)
    \\
    S_D\left(x, x_c, \delta\right) &= 0.5\left( -\tanh\left( \frac{x-x_c}{\delta} \right) + 1 \right).
    \label{}
\end{align}
The homogeneous free energy can be computed from an integral of the homogeneous contribution to the diffusional chemical potential.

\bibliography{graphite}

\begin{thebibliography}{124}%
\makeatletter
\providecommand \@ifxundefined [1]{%
 \@ifx{#1\undefined}
}%
\providecommand \@ifnum [1]{%
 \ifnum #1\expandafter \@firstoftwo
 \else \expandafter \@secondoftwo
 \fi
}%
\providecommand \@ifx [1]{%
 \ifx #1\expandafter \@firstoftwo
 \else \expandafter \@secondoftwo
 \fi
}%
\providecommand \natexlab [1]{#1}%
\providecommand \enquote  [1]{``#1''}%
\providecommand \bibnamefont  [1]{#1}%
\providecommand \bibfnamefont [1]{#1}%
\providecommand \citenamefont [1]{#1}%
\providecommand \href@noop [0]{\@secondoftwo}%
\providecommand \href [0]{\begingroup \@sanitize@url \@href}%
\providecommand \@href[1]{\@@startlink{#1}\@@href}%
\providecommand \@@href[1]{\endgroup#1\@@endlink}%
\providecommand \@sanitize@url [0]{\catcode `\\12\catcode `\$12\catcode
  `\&12\catcode `\#12\catcode `\^12\catcode `\_12\catcode `\%12\relax}%
\providecommand \@@startlink[1]{}%
\providecommand \@@endlink[0]{}%
\providecommand \url  [0]{\begingroup\@sanitize@url \@url }%
\providecommand \@url [1]{\endgroup\@href {#1}{\urlprefix }}%
\providecommand \urlprefix  [0]{URL }%
\providecommand \Eprint [0]{\href }%
\providecommand \doibase [0]{http://dx.doi.org/}%
\providecommand \selectlanguage [0]{\@gobble}%
\providecommand \bibinfo  [0]{\@secondoftwo}%
\providecommand \bibfield  [0]{\@secondoftwo}%
\providecommand \translation [1]{[#1]}%
\providecommand \BibitemOpen [0]{}%
\providecommand \bibitemStop [0]{}%
\providecommand \bibitemNoStop [0]{.\EOS\space}%
\providecommand \EOS [0]{\spacefactor3000\relax}%
\providecommand \BibitemShut  [1]{\csname bibitem#1\endcsname}%
\let\auto@bib@innerbib\@empty
%</preamble>
\bibitem [{\citenamefont {Wells}\ \emph {et~al.}(1996)\citenamefont {Wells},
  \citenamefont {Birgeneau}, \citenamefont {Chou}, \citenamefont {Endoh},
  \citenamefont {Johnston}, \citenamefont {Kastner}, \citenamefont {Lee},
  \citenamefont {Shirane}, \citenamefont {Tranquada},\ and\ \citenamefont
  {Yamada}}]{wells1996}%
  \BibitemOpen
  \bibfield  {author} {\bibinfo {author} {\bibfnamefont {B.~O.}\ \bibnamefont
  {Wells}}, \bibinfo {author} {\bibfnamefont {R.~J.}\ \bibnamefont
  {Birgeneau}}, \bibinfo {author} {\bibfnamefont {F.~C.}\ \bibnamefont {Chou}},
  \bibinfo {author} {\bibfnamefont {Y.}~\bibnamefont {Endoh}}, \bibinfo
  {author} {\bibfnamefont {D.~C.}\ \bibnamefont {Johnston}}, \bibinfo {author}
  {\bibfnamefont {M.~A.}\ \bibnamefont {Kastner}}, \bibinfo {author}
  {\bibfnamefont {Y.~S.}\ \bibnamefont {Lee}}, \bibinfo {author} {\bibfnamefont
  {G.}~\bibnamefont {Shirane}}, \bibinfo {author} {\bibfnamefont {J.~M.}\
  \bibnamefont {Tranquada}}, \ and\ \bibinfo {author} {\bibfnamefont
  {K.}~\bibnamefont {Yamada}},\ }\href {\doibase 10.1007/s002570050158}
  {\bibfield  {journal} {\bibinfo  {journal} {Zeitschrift f{\"u}r Physik B
  Condensed Matter}\ }\textbf {\bibinfo {volume} {100}},\ \bibinfo {pages}
  {535} (\bibinfo {year} {1996})}\BibitemShut {NoStop}%
\bibitem [{\citenamefont {Mohottala}\ \emph {et~al.}(2006)\citenamefont
  {Mohottala}, \citenamefont {Wells}, \citenamefont {Budnick}, \citenamefont
  {Hines}, \citenamefont {Niedermayer}, \citenamefont {Udby}, \citenamefont
  {Bernhard}, \citenamefont {Moodenbaugh},\ and\ \citenamefont
  {Chou}}]{mohottala2006}%
  \BibitemOpen
  \bibfield  {author} {\bibinfo {author} {\bibfnamefont {H.~E.}\ \bibnamefont
  {Mohottala}}, \bibinfo {author} {\bibfnamefont {B.~O.}\ \bibnamefont
  {Wells}}, \bibinfo {author} {\bibfnamefont {J.~I.}\ \bibnamefont {Budnick}},
  \bibinfo {author} {\bibfnamefont {W.~A.}\ \bibnamefont {Hines}}, \bibinfo
  {author} {\bibfnamefont {C.}~\bibnamefont {Niedermayer}}, \bibinfo {author}
  {\bibfnamefont {L.}~\bibnamefont {Udby}}, \bibinfo {author} {\bibfnamefont
  {C.}~\bibnamefont {Bernhard}}, \bibinfo {author} {\bibfnamefont {A.~R.}\
  \bibnamefont {Moodenbaugh}}, \ and\ \bibinfo {author} {\bibfnamefont {F.-C.}\
  \bibnamefont {Chou}},\ }\href {\doibase 10.1038/nmat1633} {\bibfield
  {journal} {\bibinfo  {journal} {Nature Materials}\ }\textbf {\bibinfo
  {volume} {5}},\ \bibinfo {pages} {377} (\bibinfo {year} {2006})}\BibitemShut
  {NoStop}%
\bibitem [{\citenamefont {Ijdo}\ and\ \citenamefont
  {Pinnavaia}(1998)}]{ijdo1998}%
  \BibitemOpen
  \bibfield  {author} {\bibinfo {author} {\bibfnamefont {W.~L.}\ \bibnamefont
  {Ijdo}}\ and\ \bibinfo {author} {\bibfnamefont {T.~J.}\ \bibnamefont
  {Pinnavaia}},\ }\href@noop {} {\bibfield  {journal} {\bibinfo  {journal}
  {Journal of Solid State Chemistry}\ }\textbf {\bibinfo {volume} {139}},\
  \bibinfo {pages} {281} (\bibinfo {year} {1998})}\BibitemShut {NoStop}%
\bibitem [{\citenamefont {Iyi}\ \emph {et~al.}(2002)\citenamefont {Iyi},
  \citenamefont {Kurashima},\ and\ \citenamefont {Fujita}}]{iyi2002}%
  \BibitemOpen
  \bibfield  {author} {\bibinfo {author} {\bibfnamefont {N.}~\bibnamefont
  {Iyi}}, \bibinfo {author} {\bibfnamefont {K.}~\bibnamefont {Kurashima}}, \
  and\ \bibinfo {author} {\bibfnamefont {T.}~\bibnamefont {Fujita}},\ }\href
  {\doibase 10.1021/cm0105211} {\bibfield  {journal} {\bibinfo  {journal}
  {Chemistry of Materials}\ }\textbf {\bibinfo {volume} {14}},\ \bibinfo
  {pages} {583} (\bibinfo {year} {2002})}\BibitemShut {NoStop}%
\bibitem [{\citenamefont {Pisson}\ \emph {et~al.}(2003)\citenamefont {Pisson},
  \citenamefont {Taviot-Gueho}, \citenamefont {Isra{\"e}li}, \citenamefont
  {Leroux}, \citenamefont {Munsch}, \citenamefont {Itie}, \citenamefont
  {Briois}, \citenamefont {Morel-Desrosiers},\ and\ \citenamefont
  {Besse}}]{pisson2003}%
  \BibitemOpen
  \bibfield  {author} {\bibinfo {author} {\bibfnamefont {J.}~\bibnamefont
  {Pisson}}, \bibinfo {author} {\bibfnamefont {C.}~\bibnamefont
  {Taviot-Gueho}}, \bibinfo {author} {\bibfnamefont {Y.}~\bibnamefont
  {Isra{\"e}li}}, \bibinfo {author} {\bibfnamefont {F.}~\bibnamefont {Leroux}},
  \bibinfo {author} {\bibfnamefont {P.}~\bibnamefont {Munsch}}, \bibinfo
  {author} {\bibfnamefont {J.-P.}\ \bibnamefont {Itie}}, \bibinfo {author}
  {\bibfnamefont {V.}~\bibnamefont {Briois}}, \bibinfo {author} {\bibfnamefont
  {N.}~\bibnamefont {Morel-Desrosiers}}, \ and\ \bibinfo {author}
  {\bibfnamefont {J.-P.}\ \bibnamefont {Besse}},\ }\href@noop {} {\bibfield
  {journal} {\bibinfo  {journal} {The Journal of Physical Chemistry B}\
  }\textbf {\bibinfo {volume} {107}},\ \bibinfo {pages} {9243} (\bibinfo {year}
  {2003})}\BibitemShut {NoStop}%
\bibitem [{\citenamefont {Naguib}\ \emph {et~al.}(2012)\citenamefont {Naguib},
  \citenamefont {Come}, \citenamefont {Dyatkin}, \citenamefont {Presser},
  \citenamefont {Taberna}, \citenamefont {Simon}, \citenamefont {Barsoum},\
  and\ \citenamefont {Gogotsi}}]{naguib2012}%
  \BibitemOpen
  \bibfield  {author} {\bibinfo {author} {\bibfnamefont {M.}~\bibnamefont
  {Naguib}}, \bibinfo {author} {\bibfnamefont {J.}~\bibnamefont {Come}},
  \bibinfo {author} {\bibfnamefont {B.}~\bibnamefont {Dyatkin}}, \bibinfo
  {author} {\bibfnamefont {V.}~\bibnamefont {Presser}}, \bibinfo {author}
  {\bibfnamefont {P.-L.}\ \bibnamefont {Taberna}}, \bibinfo {author}
  {\bibfnamefont {P.}~\bibnamefont {Simon}}, \bibinfo {author} {\bibfnamefont
  {M.~W.}\ \bibnamefont {Barsoum}}, \ and\ \bibinfo {author} {\bibfnamefont
  {Y.}~\bibnamefont {Gogotsi}},\ }\href {\doibase 10.1016/j.elecom.2012.01.002}
  {\bibfield  {journal} {\bibinfo  {journal} {Electrochemistry Communications}\
  }\textbf {\bibinfo {volume} {16}},\ \bibinfo {pages} {61} (\bibinfo {year}
  {2012})}\BibitemShut {NoStop}%
\bibitem [{\citenamefont {Wang}\ \emph {et~al.}(2015)\citenamefont {Wang},
  \citenamefont {Kajiyama}, \citenamefont {Iinuma}, \citenamefont {Hosono},
  \citenamefont {Oro}, \citenamefont {Moriguchi}, \citenamefont {Okubo},\ and\
  \citenamefont {Yamada}}]{wang2015pseudocapacitance}%
  \BibitemOpen
  \bibfield  {author} {\bibinfo {author} {\bibfnamefont {X.}~\bibnamefont
  {Wang}}, \bibinfo {author} {\bibfnamefont {S.}~\bibnamefont {Kajiyama}},
  \bibinfo {author} {\bibfnamefont {H.}~\bibnamefont {Iinuma}}, \bibinfo
  {author} {\bibfnamefont {E.}~\bibnamefont {Hosono}}, \bibinfo {author}
  {\bibfnamefont {S.}~\bibnamefont {Oro}}, \bibinfo {author} {\bibfnamefont
  {I.}~\bibnamefont {Moriguchi}}, \bibinfo {author} {\bibfnamefont
  {M.}~\bibnamefont {Okubo}}, \ and\ \bibinfo {author} {\bibfnamefont
  {A.}~\bibnamefont {Yamada}},\ }\href {\doibase 10.1038/ncomms7544} {\bibfield
   {journal} {\bibinfo  {journal} {Nature Communications}\ }\textbf {\bibinfo
  {volume} {6}},\ \bibinfo {pages} {6544} (\bibinfo {year} {2015})}\BibitemShut
  {NoStop}%
\bibitem [{\citenamefont {Joshi}\ \emph {et~al.}(2015)\citenamefont {Joshi},
  \citenamefont {Ozdemir}, \citenamefont {Barone},\ and\ \citenamefont
  {Peralta}}]{joshi2015}%
  \BibitemOpen
  \bibfield  {author} {\bibinfo {author} {\bibfnamefont {R.~P.}\ \bibnamefont
  {Joshi}}, \bibinfo {author} {\bibfnamefont {B.}~\bibnamefont {Ozdemir}},
  \bibinfo {author} {\bibfnamefont {V.}~\bibnamefont {Barone}}, \ and\ \bibinfo
  {author} {\bibfnamefont {J.~E.}\ \bibnamefont {Peralta}},\ }\href {\doibase
  10.1021/acs.jpclett.5b01110} {\bibfield  {journal} {\bibinfo  {journal} {The
  Journal of Physical Chemistry Letters}\ }\textbf {\bibinfo {volume} {6}},\
  \bibinfo {pages} {2728} (\bibinfo {year} {2015})}\BibitemShut {NoStop}%
\bibitem [{\citenamefont {{Van der Ven}}\ \emph {et~al.}(1998)\citenamefont
  {{Van der Ven}}, \citenamefont {Aydinol},\ and\ \citenamefont
  {Ceder}}]{vanderven1998a}%
  \BibitemOpen
  \bibfield  {author} {\bibinfo {author} {\bibfnamefont {A.}~\bibnamefont {{Van
  der Ven}}}, \bibinfo {author} {\bibfnamefont {M.~K.}\ \bibnamefont
  {Aydinol}}, \ and\ \bibinfo {author} {\bibfnamefont {G.}~\bibnamefont
  {Ceder}},\ }\href@noop {} {\bibfield  {journal} {\bibinfo  {journal} {Journal
  of The Electrochemical Society}\ }\textbf {\bibinfo {volume} {145}},\
  \bibinfo {pages} {2149} (\bibinfo {year} {1998})}\BibitemShut {NoStop}%
\bibitem [{\citenamefont {Chen}\ \emph {et~al.}(2002)\citenamefont {Chen},
  \citenamefont {Lu},\ and\ \citenamefont {Dahn}}]{chen2002}%
  \BibitemOpen
  \bibfield  {author} {\bibinfo {author} {\bibfnamefont {Z.}~\bibnamefont
  {Chen}}, \bibinfo {author} {\bibfnamefont {Z.}~\bibnamefont {Lu}}, \ and\
  \bibinfo {author} {\bibfnamefont {J.~R.}\ \bibnamefont {Dahn}},\ }\href
  {\doibase 10.1149/1.1519850} {\bibfield  {journal} {\bibinfo  {journal}
  {Journal of The Electrochemical Society}\ }\textbf {\bibinfo {volume}
  {149}},\ \bibinfo {pages} {A1604} (\bibinfo {year} {2002})}\BibitemShut
  {NoStop}%
\bibitem [{\citenamefont {Dresselhaus}\ and\ \citenamefont
  {Dresselhaus}(1981)}]{dresselhaus1981}%
  \BibitemOpen
  \bibfield  {author} {\bibinfo {author} {\bibfnamefont {M.}~\bibnamefont
  {Dresselhaus}}\ and\ \bibinfo {author} {\bibfnamefont {G.}~\bibnamefont
  {Dresselhaus}},\ }\href {\doibase 10.1080/00018738100101367} {\bibfield
  {journal} {\bibinfo  {journal} {Advances in Physics}\ }\textbf {\bibinfo
  {volume} {30}},\ \bibinfo {pages} {139} (\bibinfo {year} {1981})}\BibitemShut
  {NoStop}%
\bibitem [{\citenamefont {Di~Leo}\ and\ \citenamefont
  {Anand}(2013)}]{dileo2013}%
  \BibitemOpen
  \bibfield  {author} {\bibinfo {author} {\bibfnamefont {C.~V.}\ \bibnamefont
  {Di~Leo}}\ and\ \bibinfo {author} {\bibfnamefont {L.}~\bibnamefont {Anand}},\
  }\href {\doibase 10.1016/j.ijplas.2012.11.005} {\bibfield  {journal}
  {\bibinfo  {journal} {International Journal of Plasticity}\ }\textbf
  {\bibinfo {volume} {43}},\ \bibinfo {pages} {42} (\bibinfo {year}
  {2013})}\BibitemShut {NoStop}%
\bibitem [{\citenamefont {{de Klerk}}\ \emph {et~al.}(2016)\citenamefont {{de
  Klerk}}, \citenamefont {Vasileiadis}, \citenamefont {Smith}, \citenamefont
  {Bazant},\ and\ \citenamefont {Wagemaker}}]{deklerk2016anatase_draft}%
  \BibitemOpen
  \bibfield  {author} {\bibinfo {author} {\bibfnamefont {N.}~\bibnamefont {{de
  Klerk}}}, \bibinfo {author} {\bibfnamefont {A.}~\bibnamefont {Vasileiadis}},
  \bibinfo {author} {\bibfnamefont {R.~B.}\ \bibnamefont {Smith}}, \bibinfo
  {author} {\bibfnamefont {M.~Z.}\ \bibnamefont {Bazant}}, \ and\ \bibinfo
  {author} {\bibfnamefont {M.}~\bibnamefont {Wagemaker}},\ }\href@noop {}
  {\enquote {\bibinfo {title} {Explaining key properties of lithiation in
  {{TiO2}}-anatase using a phase-field model without fitted parameters (in
  preparation)},}\ } (\bibinfo {year} {2016})\BibitemShut {NoStop}%
\bibitem [{\citenamefont {Lafont}\ \emph {et~al.}(2010)\citenamefont {Lafont},
  \citenamefont {Carta}, \citenamefont {Mountjoy}, \citenamefont {Chadwick},\
  and\ \citenamefont {Kelder}}]{lafont2010}%
  \BibitemOpen
  \bibfield  {author} {\bibinfo {author} {\bibfnamefont {U.}~\bibnamefont
  {Lafont}}, \bibinfo {author} {\bibfnamefont {D.}~\bibnamefont {Carta}},
  \bibinfo {author} {\bibfnamefont {G.}~\bibnamefont {Mountjoy}}, \bibinfo
  {author} {\bibfnamefont {A.~V.}\ \bibnamefont {Chadwick}}, \ and\ \bibinfo
  {author} {\bibfnamefont {E.~M.}\ \bibnamefont {Kelder}},\ }\href {\doibase
  10.1021/jp908786t} {\bibfield  {journal} {\bibinfo  {journal} {The Journal of
  Physical Chemistry C}\ }\textbf {\bibinfo {volume} {114}},\ \bibinfo {pages}
  {1372} (\bibinfo {year} {2010})}\BibitemShut {NoStop}%
\bibitem [{\citenamefont {Shin}\ \emph {et~al.}(2011)\citenamefont {Shin},
  \citenamefont {Samuelis},\ and\ \citenamefont {Maier}}]{shin2011}%
  \BibitemOpen
  \bibfield  {author} {\bibinfo {author} {\bibfnamefont {J.-Y.}\ \bibnamefont
  {Shin}}, \bibinfo {author} {\bibfnamefont {D.}~\bibnamefont {Samuelis}}, \
  and\ \bibinfo {author} {\bibfnamefont {J.}~\bibnamefont {Maier}},\ }\href
  {\doibase 10.1002/adfm.201002527} {\bibfield  {journal} {\bibinfo  {journal}
  {Advanced Functional Materials}\ }\textbf {\bibinfo {volume} {21}},\ \bibinfo
  {pages} {3464} (\bibinfo {year} {2011})}\BibitemShut {NoStop}%
\bibitem [{\citenamefont {Shen}\ \emph {et~al.}(2014)\citenamefont {Shen},
  \citenamefont {Chen}, \citenamefont {Klaver}, \citenamefont {Mulder},\ and\
  \citenamefont {Wagemaker}}]{shen2014}%
  \BibitemOpen
  \bibfield  {author} {\bibinfo {author} {\bibfnamefont {K.}~\bibnamefont
  {Shen}}, \bibinfo {author} {\bibfnamefont {H.}~\bibnamefont {Chen}}, \bibinfo
  {author} {\bibfnamefont {F.}~\bibnamefont {Klaver}}, \bibinfo {author}
  {\bibfnamefont {F.~M.}\ \bibnamefont {Mulder}}, \ and\ \bibinfo {author}
  {\bibfnamefont {M.}~\bibnamefont {Wagemaker}},\ }\href {\doibase
  10.1021/cm4037346} {\bibfield  {journal} {\bibinfo  {journal} {Chemistry of
  Materials}\ }\textbf {\bibinfo {volume} {26}},\ \bibinfo {pages} {1608}
  (\bibinfo {year} {2014})}\BibitemShut {NoStop}%
\bibitem [{\citenamefont {Zachau-Christiansen}\ \emph
  {et~al.}(1988)\citenamefont {Zachau-Christiansen}, \citenamefont {West},
  \citenamefont {Jacobsen},\ and\ \citenamefont
  {Atlung}}]{zachau-christiansen1988}%
  \BibitemOpen
  \bibfield  {author} {\bibinfo {author} {\bibfnamefont {B.}~\bibnamefont
  {Zachau-Christiansen}}, \bibinfo {author} {\bibfnamefont {K.}~\bibnamefont
  {West}}, \bibinfo {author} {\bibfnamefont {T.}~\bibnamefont {Jacobsen}}, \
  and\ \bibinfo {author} {\bibfnamefont {S.}~\bibnamefont {Atlung}},\ }\href
  {\doibase http://dx.doi.org/10.1016/0167-2738(88)90352-9} {\bibfield
  {journal} {\bibinfo  {journal} {Solid State Ionics}\ }\textbf {\bibinfo
  {volume} {28-30}},\ \bibinfo {pages} {1176} (\bibinfo {year}
  {1988})}\BibitemShut {NoStop}%
\bibitem [{\citenamefont {Ohzuku}\ \emph {et~al.}(1993)\citenamefont {Ohzuku},
  \citenamefont {Iwakoshi},\ and\ \citenamefont {Sawai}}]{ohzuku1993}%
  \BibitemOpen
  \bibfield  {author} {\bibinfo {author} {\bibfnamefont {T.}~\bibnamefont
  {Ohzuku}}, \bibinfo {author} {\bibfnamefont {Y.}~\bibnamefont {Iwakoshi}}, \
  and\ \bibinfo {author} {\bibfnamefont {K.}~\bibnamefont {Sawai}},\
  }\href@noop {} {\bibfield  {journal} {\bibinfo  {journal} {Journal of The
  Electrochemical Society}\ }\textbf {\bibinfo {volume} {140}},\ \bibinfo
  {pages} {2490} (\bibinfo {year} {1993})}\BibitemShut {NoStop}%
\bibitem [{\citenamefont {Scrosati}\ and\ \citenamefont
  {Garche}(2010)}]{scrosati2010}%
  \BibitemOpen
  \bibfield  {author} {\bibinfo {author} {\bibfnamefont {B.}~\bibnamefont
  {Scrosati}}\ and\ \bibinfo {author} {\bibfnamefont {J.}~\bibnamefont
  {Garche}},\ }\href {\doibase 10.1016/j.jpowsour.2009.11.048} {\bibfield
  {journal} {\bibinfo  {journal} {Journal of Power Sources}\ }\textbf {\bibinfo
  {volume} {195}},\ \bibinfo {pages} {2419} (\bibinfo {year}
  {2010})}\BibitemShut {NoStop}%
\bibitem [{\citenamefont {Tarascon}\ and\ \citenamefont
  {Armand}(2001)}]{tarascon2001}%
  \BibitemOpen
  \bibfield  {author} {\bibinfo {author} {\bibfnamefont {J.-M.}\ \bibnamefont
  {Tarascon}}\ and\ \bibinfo {author} {\bibfnamefont {M.}~\bibnamefont
  {Armand}},\ }\href@noop {} {\bibfield  {journal} {\bibinfo  {journal}
  {Nature}\ }\textbf {\bibinfo {volume} {414}},\ \bibinfo {pages} {359}
  (\bibinfo {year} {2001})}\BibitemShut {NoStop}%
\bibitem [{\citenamefont {Dahn}(1991)}]{dahn1991}%
  \BibitemOpen
  \bibfield  {author} {\bibinfo {author} {\bibfnamefont {J.~R.}\ \bibnamefont
  {Dahn}},\ }\href {\doibase 10.1103/PhysRevB.44.9170} {\bibfield  {journal}
  {\bibinfo  {journal} {Physical Review B}\ }\textbf {\bibinfo {volume} {44}},\
  \bibinfo {pages} {9170} (\bibinfo {year} {1991})}\BibitemShut {NoStop}%
\bibitem [{\citenamefont {Bazant}(2013)}]{bazant2013}%
  \BibitemOpen
  \bibfield  {author} {\bibinfo {author} {\bibfnamefont {M.~Z.}\ \bibnamefont
  {Bazant}},\ }\href {\doibase 10.1021/ar300145c} {\bibfield  {journal}
  {\bibinfo  {journal} {Accounts of Chemical Research}\ }\textbf {\bibinfo
  {volume} {46}},\ \bibinfo {pages} {1144} (\bibinfo {year}
  {2013})}\BibitemShut {NoStop}%
\bibitem [{\citenamefont {Maire}\ \emph {et~al.}(2008)\citenamefont {Maire},
  \citenamefont {Evans}, \citenamefont {Kaiser}, \citenamefont {Scheifele},\
  and\ \citenamefont {Nov{\'a}k}}]{maire2008}%
  \BibitemOpen
  \bibfield  {author} {\bibinfo {author} {\bibfnamefont {P.}~\bibnamefont
  {Maire}}, \bibinfo {author} {\bibfnamefont {A.}~\bibnamefont {Evans}},
  \bibinfo {author} {\bibfnamefont {H.}~\bibnamefont {Kaiser}}, \bibinfo
  {author} {\bibfnamefont {W.}~\bibnamefont {Scheifele}}, \ and\ \bibinfo
  {author} {\bibfnamefont {P.}~\bibnamefont {Nov{\'a}k}},\ }\href {\doibase
  10.1149/1.2979696} {\bibfield  {journal} {\bibinfo  {journal} {Journal of The
  Electrochemical Society}\ }\textbf {\bibinfo {volume} {155}},\ \bibinfo
  {pages} {A862} (\bibinfo {year} {2008})}\BibitemShut {NoStop}%
\bibitem [{\citenamefont {Harris}\ \emph {et~al.}(2010)\citenamefont {Harris},
  \citenamefont {Timmons}, \citenamefont {Baker},\ and\ \citenamefont
  {Monroe}}]{harris2010}%
  \BibitemOpen
  \bibfield  {author} {\bibinfo {author} {\bibfnamefont {S.~J.}\ \bibnamefont
  {Harris}}, \bibinfo {author} {\bibfnamefont {A.}~\bibnamefont {Timmons}},
  \bibinfo {author} {\bibfnamefont {D.~R.}\ \bibnamefont {Baker}}, \ and\
  \bibinfo {author} {\bibfnamefont {C.}~\bibnamefont {Monroe}},\ }\href@noop {}
  {\bibfield  {journal} {\bibinfo  {journal} {Chemical Physics Letters}\
  }\textbf {\bibinfo {volume} {485}},\ \bibinfo {pages} {265} (\bibinfo {year}
  {2010})}\BibitemShut {NoStop}%
\bibitem [{\citenamefont {Daumas}\ and\ \citenamefont
  {Herold}(1969)}]{daumas1969}%
  \BibitemOpen
  \bibfield  {author} {\bibinfo {author} {\bibfnamefont {N.}~\bibnamefont
  {Daumas}}\ and\ \bibinfo {author} {\bibfnamefont {A.}~\bibnamefont
  {Herold}},\ }\href@noop {} {\bibfield  {journal} {\bibinfo  {journal}
  {Comptes Rendus de l'Acad{\'e}mie des Sciences de Paris}\ }\bibinfo {series}
  {C},\ \textbf {\bibinfo {volume} {268}},\ \bibinfo {pages} {373} (\bibinfo
  {year} {1969})}\BibitemShut {NoStop}%
\bibitem [{\citenamefont {Safran}(1980{\natexlab{a}})}]{safran1980}%
  \BibitemOpen
  \bibfield  {author} {\bibinfo {author} {\bibfnamefont {S.~A.}\ \bibnamefont
  {Safran}},\ }\href@noop {} {\bibfield  {journal} {\bibinfo  {journal}
  {Synthetic Metals}\ }\textbf {\bibinfo {volume} {2}},\ \bibinfo {pages} {1}
  (\bibinfo {year} {1980}{\natexlab{a}})}\BibitemShut {NoStop}%
\bibitem [{\citenamefont {Kirczenow}(1982)}]{kirczenow1982}%
  \BibitemOpen
  \bibfield  {author} {\bibinfo {author} {\bibfnamefont {G.}~\bibnamefont
  {Kirczenow}},\ }\href {\doibase 10.1103/PhysRevLett.49.1853} {\bibfield
  {journal} {\bibinfo  {journal} {Physical Review Letters}\ }\textbf {\bibinfo
  {volume} {49}},\ \bibinfo {pages} {1853} (\bibinfo {year}
  {1982})}\BibitemShut {NoStop}%
\bibitem [{\citenamefont {Krishnan}\ \emph {et~al.}(2013)\citenamefont
  {Krishnan}, \citenamefont {Brenet}, \citenamefont {Machado-Charry},
  \citenamefont {Caliste}, \citenamefont {Genovese}, \citenamefont {Deutsch},\
  and\ \citenamefont {Pochet}}]{krishnan2013}%
  \BibitemOpen
  \bibfield  {author} {\bibinfo {author} {\bibfnamefont {S.}~\bibnamefont
  {Krishnan}}, \bibinfo {author} {\bibfnamefont {G.}~\bibnamefont {Brenet}},
  \bibinfo {author} {\bibfnamefont {E.}~\bibnamefont {Machado-Charry}},
  \bibinfo {author} {\bibfnamefont {D.}~\bibnamefont {Caliste}}, \bibinfo
  {author} {\bibfnamefont {L.}~\bibnamefont {Genovese}}, \bibinfo {author}
  {\bibfnamefont {T.}~\bibnamefont {Deutsch}}, \ and\ \bibinfo {author}
  {\bibfnamefont {P.}~\bibnamefont {Pochet}},\ }\href {\doibase
  10.1063/1.4850877} {\bibfield  {journal} {\bibinfo  {journal} {Applied
  Physics Letters}\ }\textbf {\bibinfo {volume} {103}},\ \bibinfo {pages}
  {251904} (\bibinfo {year} {2013})}\BibitemShut {NoStop}%
\bibitem [{\citenamefont {Clarke}\ \emph {et~al.}(1979)\citenamefont {Clarke},
  \citenamefont {Caswell},\ and\ \citenamefont {Solin}}]{clarke1979}%
  \BibitemOpen
  \bibfield  {author} {\bibinfo {author} {\bibfnamefont {R.}~\bibnamefont
  {Clarke}}, \bibinfo {author} {\bibfnamefont {N.}~\bibnamefont {Caswell}}, \
  and\ \bibinfo {author} {\bibfnamefont {S.~A.}\ \bibnamefont {Solin}},\
  }\href@noop {} {\bibfield  {journal} {\bibinfo  {journal} {Physical Review
  Letters}\ }\textbf {\bibinfo {volume} {42}},\ \bibinfo {pages} {61} (\bibinfo
  {year} {1979})}\BibitemShut {NoStop}%
\bibitem [{\citenamefont {Dimiev}\ \emph {et~al.}(2013)\citenamefont {Dimiev},
  \citenamefont {Ceriotti}, \citenamefont {Behabtu}, \citenamefont {Zakhidov},
  \citenamefont {Pasquali}, \citenamefont {Saito},\ and\ \citenamefont
  {Tour}}]{dimiev2013}%
  \BibitemOpen
  \bibfield  {author} {\bibinfo {author} {\bibfnamefont {A.~M.}\ \bibnamefont
  {Dimiev}}, \bibinfo {author} {\bibfnamefont {G.}~\bibnamefont {Ceriotti}},
  \bibinfo {author} {\bibfnamefont {N.}~\bibnamefont {Behabtu}}, \bibinfo
  {author} {\bibfnamefont {D.}~\bibnamefont {Zakhidov}}, \bibinfo {author}
  {\bibfnamefont {M.}~\bibnamefont {Pasquali}}, \bibinfo {author}
  {\bibfnamefont {R.}~\bibnamefont {Saito}}, \ and\ \bibinfo {author}
  {\bibfnamefont {J.~M.}\ \bibnamefont {Tour}},\ }\href {\doibase
  10.1021/nn400207e} {\bibfield  {journal} {\bibinfo  {journal} {ACS Nano}\
  }\textbf {\bibinfo {volume} {7}},\ \bibinfo {pages} {2773} (\bibinfo {year}
  {2013})}\BibitemShut {NoStop}%
\bibitem [{\citenamefont {Padhi}\ \emph {et~al.}(1997)\citenamefont {Padhi},
  \citenamefont {Nanjundaswamy},\ and\ \citenamefont
  {d~Goodenough}}]{padhi1997}%
  \BibitemOpen
  \bibfield  {author} {\bibinfo {author} {\bibfnamefont {A.~K.}\ \bibnamefont
  {Padhi}}, \bibinfo {author} {\bibfnamefont {K.~S.}\ \bibnamefont
  {Nanjundaswamy}}, \ and\ \bibinfo {author} {\bibfnamefont {J.~B.}\
  \bibnamefont {d~Goodenough}},\ }\href@noop {} {\bibfield  {journal} {\bibinfo
   {journal} {Journal of The Electrochemical Society}\ }\textbf {\bibinfo
  {volume} {144}},\ \bibinfo {pages} {1188} (\bibinfo {year}
  {1997})}\BibitemShut {NoStop}%
\bibitem [{\citenamefont {Ariyoshi}\ \emph {et~al.}(2004)\citenamefont
  {Ariyoshi}, \citenamefont {Iwakoshi}, \citenamefont {Nakayama},\ and\
  \citenamefont {Ohzuku}}]{ariyoshi2004}%
  \BibitemOpen
  \bibfield  {author} {\bibinfo {author} {\bibfnamefont {K.}~\bibnamefont
  {Ariyoshi}}, \bibinfo {author} {\bibfnamefont {Y.}~\bibnamefont {Iwakoshi}},
  \bibinfo {author} {\bibfnamefont {N.}~\bibnamefont {Nakayama}}, \ and\
  \bibinfo {author} {\bibfnamefont {T.}~\bibnamefont {Ohzuku}},\ }\href
  {\doibase 10.1149/1.1639162} {\bibfield  {journal} {\bibinfo  {journal}
  {Journal of The Electrochemical Society}\ }\textbf {\bibinfo {volume}
  {151}},\ \bibinfo {pages} {A296} (\bibinfo {year} {2004})}\BibitemShut
  {NoStop}%
\bibitem [{\citenamefont {Ohzuku}\ \emph {et~al.}(1995)\citenamefont {Ohzuku},
  \citenamefont {Ueda},\ and\ \citenamefont {Yamamoto}}]{ohzuku1995}%
  \BibitemOpen
  \bibfield  {author} {\bibinfo {author} {\bibfnamefont {T.}~\bibnamefont
  {Ohzuku}}, \bibinfo {author} {\bibfnamefont {A.}~\bibnamefont {Ueda}}, \ and\
  \bibinfo {author} {\bibfnamefont {N.}~\bibnamefont {Yamamoto}},\ }\href@noop
  {} {\bibfield  {journal} {\bibinfo  {journal} {Journal of the Electrochemical
  Society}\ }\textbf {\bibinfo {volume} {142}},\ \bibinfo {pages} {1431}
  (\bibinfo {year} {1995})}\BibitemShut {NoStop}%
\bibitem [{\citenamefont {Lim}\ \emph {et~al.}(2016)\citenamefont {Lim},
  \citenamefont {Li}, \citenamefont {Alsem}, \citenamefont {So}, \citenamefont
  {Lee}, \citenamefont {Bai}, \citenamefont {Cogswell}, \citenamefont {Liu},
  \citenamefont {Jin}, \citenamefont {Yu}, \citenamefont {Salmon},
  \citenamefont {Shapiro}, \citenamefont {Bazant}, \citenamefont {Tyliszczak},\
  and\ \citenamefont {Chueh}}]{lim2016origin}%
  \BibitemOpen
  \bibfield  {author} {\bibinfo {author} {\bibfnamefont {J.}~\bibnamefont
  {Lim}}, \bibinfo {author} {\bibfnamefont {Y.}~\bibnamefont {Li}}, \bibinfo
  {author} {\bibfnamefont {D.~H.}\ \bibnamefont {Alsem}}, \bibinfo {author}
  {\bibfnamefont {H.}~\bibnamefont {So}}, \bibinfo {author} {\bibfnamefont
  {S.~C.}\ \bibnamefont {Lee}}, \bibinfo {author} {\bibfnamefont
  {P.}~\bibnamefont {Bai}}, \bibinfo {author} {\bibfnamefont {D.~A.}\
  \bibnamefont {Cogswell}}, \bibinfo {author} {\bibfnamefont {X.}~\bibnamefont
  {Liu}}, \bibinfo {author} {\bibfnamefont {N.}~\bibnamefont {Jin}}, \bibinfo
  {author} {\bibfnamefont {Y.-s.}\ \bibnamefont {Yu}}, \bibinfo {author}
  {\bibfnamefont {N.~J.}\ \bibnamefont {Salmon}}, \bibinfo {author}
  {\bibfnamefont {D.~A.}\ \bibnamefont {Shapiro}}, \bibinfo {author}
  {\bibfnamefont {M.~Z.}\ \bibnamefont {Bazant}}, \bibinfo {author}
  {\bibfnamefont {T.}~\bibnamefont {Tyliszczak}}, \ and\ \bibinfo {author}
  {\bibfnamefont {W.~C.}\ \bibnamefont {Chueh}},\ }\href {\doibase
  10.1126/science.aaf4914} {\bibfield  {journal} {\bibinfo  {journal}
  {Science}\ }\textbf {\bibinfo {volume} {353}},\ \bibinfo {pages} {566}
  (\bibinfo {year} {2016})}\BibitemShut {NoStop}%
\bibitem [{\citenamefont {Tang}\ \emph {et~al.}(2010)\citenamefont {Tang},
  \citenamefont {Carter},\ and\ \citenamefont
  {Chiang}}]{tang2010electrochemically}%
  \BibitemOpen
  \bibfield  {author} {\bibinfo {author} {\bibfnamefont {M.}~\bibnamefont
  {Tang}}, \bibinfo {author} {\bibfnamefont {W.~C.}\ \bibnamefont {Carter}}, \
  and\ \bibinfo {author} {\bibfnamefont {Y.-M.}\ \bibnamefont {Chiang}},\
  }\href {\doibase 10.1146/annurev-matsci-070909-104435} {\bibfield  {journal}
  {\bibinfo  {journal} {Annual Review of Materials Research}\ }\textbf
  {\bibinfo {volume} {40}},\ \bibinfo {pages} {501} (\bibinfo {year}
  {2010})}\BibitemShut {NoStop}%
\bibitem [{\citenamefont {Bai}\ \emph {et~al.}(2011)\citenamefont {Bai},
  \citenamefont {Cogswell},\ and\ \citenamefont {Bazant}}]{bai2011}%
  \BibitemOpen
  \bibfield  {author} {\bibinfo {author} {\bibfnamefont {P.}~\bibnamefont
  {Bai}}, \bibinfo {author} {\bibfnamefont {D.~A.}\ \bibnamefont {Cogswell}}, \
  and\ \bibinfo {author} {\bibfnamefont {M.~Z.}\ \bibnamefont {Bazant}},\
  }\href@noop {} {\bibfield  {journal} {\bibinfo  {journal} {Nano letters}\
  }\textbf {\bibinfo {volume} {11}},\ \bibinfo {pages} {4890} (\bibinfo {year}
  {2011})}\BibitemShut {NoStop}%
\bibitem [{\citenamefont {Cogswell}\ and\ \citenamefont
  {Bazant}(2012)}]{cogswell2012}%
  \BibitemOpen
  \bibfield  {author} {\bibinfo {author} {\bibfnamefont {D.~A.}\ \bibnamefont
  {Cogswell}}\ and\ \bibinfo {author} {\bibfnamefont {M.~Z.}\ \bibnamefont
  {Bazant}},\ }\href@noop {} {\bibfield  {journal} {\bibinfo  {journal} {ACS
  nano}\ }\textbf {\bibinfo {volume} {6}},\ \bibinfo {pages} {2215} (\bibinfo
  {year} {2012})}\BibitemShut {NoStop}%
\bibitem [{\citenamefont {Cogswell}\ and\ \citenamefont
  {Bazant}(2013)}]{cogswell2013}%
  \BibitemOpen
  \bibfield  {author} {\bibinfo {author} {\bibfnamefont {D.~A.}\ \bibnamefont
  {Cogswell}}\ and\ \bibinfo {author} {\bibfnamefont {M.~Z.}\ \bibnamefont
  {Bazant}},\ }\href {\doibase 10.1021/nl400497t} {\bibfield  {journal}
  {\bibinfo  {journal} {Nano Letters}\ }\textbf {\bibinfo {volume} {13}},\
  \bibinfo {pages} {3036} (\bibinfo {year} {2013})}\BibitemShut {NoStop}%
\bibitem [{\citenamefont {Zeng}\ and\ \citenamefont {Bazant}(2014)}]{zeng2014}%
  \BibitemOpen
  \bibfield  {author} {\bibinfo {author} {\bibfnamefont {Y.}~\bibnamefont
  {Zeng}}\ and\ \bibinfo {author} {\bibfnamefont {M.~Z.}\ \bibnamefont
  {Bazant}},\ }\href {\doibase 10.1137/130937548} {\bibfield  {journal}
  {\bibinfo  {journal} {SIAM Journal on Applied Mathematics}\ }\textbf
  {\bibinfo {volume} {74}},\ \bibinfo {pages} {980} (\bibinfo {year}
  {2014})}\BibitemShut {NoStop}%
\bibitem [{\citenamefont {Welland}\ \emph {et~al.}(2015)\citenamefont
  {Welland}, \citenamefont {Karpeyev}, \citenamefont {O'Connor},\ and\
  \citenamefont {Heinonen}}]{welland2015miscibility}%
  \BibitemOpen
  \bibfield  {author} {\bibinfo {author} {\bibfnamefont {M.~J.}\ \bibnamefont
  {Welland}}, \bibinfo {author} {\bibfnamefont {D.}~\bibnamefont {Karpeyev}},
  \bibinfo {author} {\bibfnamefont {D.~T.}\ \bibnamefont {O'Connor}}, \ and\
  \bibinfo {author} {\bibfnamefont {O.}~\bibnamefont {Heinonen}},\ }\href
  {\doibase 10.1021/acsnano.5b02555} {\bibfield  {journal} {\bibinfo  {journal}
  {ACS Nano}\ }\textbf {\bibinfo {volume} {9}},\ \bibinfo {pages} {9757}
  (\bibinfo {year} {2015})}\BibitemShut {NoStop}%
\bibitem [{\citenamefont {Dreyer}\ \emph {et~al.}(2010)\citenamefont {Dreyer},
  \citenamefont {Jamnik}, \citenamefont {Guhlke}, \citenamefont {Huth},
  \citenamefont {Mo{\v s}kon},\ and\ \citenamefont {Gaber{\v s}{\v
  c}ek}}]{dreyer2010}%
  \BibitemOpen
  \bibfield  {author} {\bibinfo {author} {\bibfnamefont {W.}~\bibnamefont
  {Dreyer}}, \bibinfo {author} {\bibfnamefont {J.}~\bibnamefont {Jamnik}},
  \bibinfo {author} {\bibfnamefont {C.}~\bibnamefont {Guhlke}}, \bibinfo
  {author} {\bibfnamefont {R.}~\bibnamefont {Huth}}, \bibinfo {author}
  {\bibfnamefont {J.}~\bibnamefont {Mo{\v s}kon}}, \ and\ \bibinfo {author}
  {\bibfnamefont {M.}~\bibnamefont {Gaber{\v s}{\v c}ek}},\ }\href@noop {}
  {\bibfield  {journal} {\bibinfo  {journal} {Nature Materials}\ }\textbf
  {\bibinfo {volume} {9}},\ \bibinfo {pages} {448} (\bibinfo {year}
  {2010})}\BibitemShut {NoStop}%
\bibitem [{\citenamefont {Chueh}\ \emph {et~al.}(2013)\citenamefont {Chueh},
  \citenamefont {Gabaly}, \citenamefont {Sugar}, \citenamefont {Bartelt},
  \citenamefont {McDaniel}, \citenamefont {Fenton}, \citenamefont {Zavadil},
  \citenamefont {Tyliszczak}, \citenamefont {Lai},\ and\ \citenamefont
  {McCarty}}]{chueh2013}%
  \BibitemOpen
  \bibfield  {author} {\bibinfo {author} {\bibfnamefont {W.~C.}\ \bibnamefont
  {Chueh}}, \bibinfo {author} {\bibfnamefont {F.~E.}\ \bibnamefont {Gabaly}},
  \bibinfo {author} {\bibfnamefont {J.~D.}\ \bibnamefont {Sugar}}, \bibinfo
  {author} {\bibfnamefont {N.~C.}\ \bibnamefont {Bartelt}}, \bibinfo {author}
  {\bibfnamefont {A.~H.}\ \bibnamefont {McDaniel}}, \bibinfo {author}
  {\bibfnamefont {K.~R.}\ \bibnamefont {Fenton}}, \bibinfo {author}
  {\bibfnamefont {K.~R.}\ \bibnamefont {Zavadil}}, \bibinfo {author}
  {\bibfnamefont {T.}~\bibnamefont {Tyliszczak}}, \bibinfo {author}
  {\bibfnamefont {W.}~\bibnamefont {Lai}}, \ and\ \bibinfo {author}
  {\bibfnamefont {K.~F.}\ \bibnamefont {McCarty}},\ }\href@noop {} {\bibfield
  {journal} {\bibinfo  {journal} {Nano Letters}\ } (\bibinfo {year}
  {2013})}\BibitemShut {NoStop}%
\bibitem [{\citenamefont {Li}\ \emph {et~al.}(2014)\citenamefont {Li},
  \citenamefont {El~Gabaly}, \citenamefont {Ferguson}, \citenamefont {Smith},
  \citenamefont {Bartelt}, \citenamefont {Sugar}, \citenamefont {Fenton},
  \citenamefont {Cogswell}, \citenamefont {Kilcoyne}, \citenamefont
  {Tyliszczak}, \citenamefont {Bazant},\ and\ \citenamefont {Chueh}}]{li2014}%
  \BibitemOpen
  \bibfield  {author} {\bibinfo {author} {\bibfnamefont {Y.}~\bibnamefont
  {Li}}, \bibinfo {author} {\bibfnamefont {F.}~\bibnamefont {El~Gabaly}},
  \bibinfo {author} {\bibfnamefont {T.~R.}\ \bibnamefont {Ferguson}}, \bibinfo
  {author} {\bibfnamefont {R.~B.}\ \bibnamefont {Smith}}, \bibinfo {author}
  {\bibfnamefont {N.~C.}\ \bibnamefont {Bartelt}}, \bibinfo {author}
  {\bibfnamefont {J.~D.}\ \bibnamefont {Sugar}}, \bibinfo {author}
  {\bibfnamefont {K.~R.}\ \bibnamefont {Fenton}}, \bibinfo {author}
  {\bibfnamefont {D.~A.}\ \bibnamefont {Cogswell}}, \bibinfo {author}
  {\bibfnamefont {A.~L.~D.}\ \bibnamefont {Kilcoyne}}, \bibinfo {author}
  {\bibfnamefont {T.}~\bibnamefont {Tyliszczak}}, \bibinfo {author}
  {\bibfnamefont {M.~Z.}\ \bibnamefont {Bazant}}, \ and\ \bibinfo {author}
  {\bibfnamefont {W.~C.}\ \bibnamefont {Chueh}},\ }\href {\doibase
  10.1038/nmat4084} {\bibfield  {journal} {\bibinfo  {journal} {Nature
  Materials}\ }\textbf {\bibinfo {volume} {13}},\ \bibinfo {pages} {1149}
  (\bibinfo {year} {2014})}\BibitemShut {NoStop}%
\bibitem [{\citenamefont {Ferguson}\ and\ \citenamefont
  {Bazant}(2012)}]{ferguson2012}%
  \BibitemOpen
  \bibfield  {author} {\bibinfo {author} {\bibfnamefont {T.~R.}\ \bibnamefont
  {Ferguson}}\ and\ \bibinfo {author} {\bibfnamefont {M.~Z.}\ \bibnamefont
  {Bazant}},\ }\href {\doibase 10.1149/2.048212jes} {\bibfield  {journal}
  {\bibinfo  {journal} {Journal of The Electrochemical Society}\ }\textbf
  {\bibinfo {volume} {159}},\ \bibinfo {pages} {A1967} (\bibinfo {year}
  {2012})}\BibitemShut {NoStop}%
\bibitem [{\citenamefont {Ferguson}\ and\ \citenamefont
  {Bazant}(2014)}]{ferguson2014}%
  \BibitemOpen
  \bibfield  {author} {\bibinfo {author} {\bibfnamefont {T.~R.}\ \bibnamefont
  {Ferguson}}\ and\ \bibinfo {author} {\bibfnamefont {M.~Z.}\ \bibnamefont
  {Bazant}},\ }\href {\doibase 10.1016/j.electacta.2014.08.083} {\bibfield
  {journal} {\bibinfo  {journal} {Electrochimica Acta}\ }\textbf {\bibinfo
  {volume} {146}},\ \bibinfo {pages} {89} (\bibinfo {year} {2014})}\BibitemShut
  {NoStop}%
\bibitem [{\citenamefont {Orvananos}\ \emph {et~al.}(2014)\citenamefont
  {Orvananos}, \citenamefont {Ferguson}, \citenamefont {Yu}, \citenamefont
  {Bazant},\ and\ \citenamefont {Thornton}}]{orvananos2014}%
  \BibitemOpen
  \bibfield  {author} {\bibinfo {author} {\bibfnamefont {B.}~\bibnamefont
  {Orvananos}}, \bibinfo {author} {\bibfnamefont {T.~R.}\ \bibnamefont
  {Ferguson}}, \bibinfo {author} {\bibfnamefont {H.-C.}\ \bibnamefont {Yu}},
  \bibinfo {author} {\bibfnamefont {M.~Z.}\ \bibnamefont {Bazant}}, \ and\
  \bibinfo {author} {\bibfnamefont {K.}~\bibnamefont {Thornton}},\ }\href@noop
  {} {\bibfield  {journal} {\bibinfo  {journal} {Journal of The Electrochemical
  Society}\ }\textbf {\bibinfo {volume} {161}},\ \bibinfo {pages} {A535}
  (\bibinfo {year} {2014})}\BibitemShut {NoStop}%
\bibitem [{\citenamefont {Orvananos}\ \emph {et~al.}(2015)\citenamefont
  {Orvananos}, \citenamefont {Yu}, \citenamefont {Abdellahi}, \citenamefont
  {Malik}, \citenamefont {Grey}, \citenamefont {Ceder},\ and\ \citenamefont
  {Thornton}}]{orvananos2015}%
  \BibitemOpen
  \bibfield  {author} {\bibinfo {author} {\bibfnamefont {B.}~\bibnamefont
  {Orvananos}}, \bibinfo {author} {\bibfnamefont {H.-C.}\ \bibnamefont {Yu}},
  \bibinfo {author} {\bibfnamefont {A.}~\bibnamefont {Abdellahi}}, \bibinfo
  {author} {\bibfnamefont {R.}~\bibnamefont {Malik}}, \bibinfo {author}
  {\bibfnamefont {C.~P.}\ \bibnamefont {Grey}}, \bibinfo {author}
  {\bibfnamefont {G.}~\bibnamefont {Ceder}}, \ and\ \bibinfo {author}
  {\bibfnamefont {K.}~\bibnamefont {Thornton}},\ }\href@noop {} {\bibfield
  {journal} {\bibinfo  {journal} {Journal of The Electrochemical Society}\
  }\textbf {\bibinfo {volume} {162}},\ \bibinfo {pages} {A965} (\bibinfo {year}
  {2015})}\BibitemShut {NoStop}%
\bibitem [{\citenamefont {Fuller}\ \emph {et~al.}(1994)\citenamefont {Fuller},
  \citenamefont {Doyle},\ and\ \citenamefont {Newman}}]{fuller1994}%
  \BibitemOpen
  \bibfield  {author} {\bibinfo {author} {\bibfnamefont {T.~F.}\ \bibnamefont
  {Fuller}}, \bibinfo {author} {\bibfnamefont {M.}~\bibnamefont {Doyle}}, \
  and\ \bibinfo {author} {\bibfnamefont {J.}~\bibnamefont {Newman}},\ }\href
  {\doibase 10.1149/1.2054684} {\bibfield  {journal} {\bibinfo  {journal}
  {Journal of The Electrochemical Society}\ }\textbf {\bibinfo {volume}
  {141}},\ \bibinfo {pages} {1} (\bibinfo {year} {1994})}\BibitemShut {NoStop}%
\bibitem [{\citenamefont {Verbrugge}\ and\ \citenamefont
  {Koch}(2003)}]{verbrugge2003}%
  \BibitemOpen
  \bibfield  {author} {\bibinfo {author} {\bibfnamefont {M.~W.}\ \bibnamefont
  {Verbrugge}}\ and\ \bibinfo {author} {\bibfnamefont {B.~J.}\ \bibnamefont
  {Koch}},\ }\href {\doibase 10.1149/1.1553788} {\bibfield  {journal} {\bibinfo
   {journal} {Journal of The Electrochemical Society}\ }\textbf {\bibinfo
  {volume} {150}},\ \bibinfo {pages} {A374} (\bibinfo {year}
  {2003})}\BibitemShut {NoStop}%
\bibitem [{\citenamefont {Bernardi}\ and\ \citenamefont
  {Go}(2011)}]{bernardi2011}%
  \BibitemOpen
  \bibfield  {author} {\bibinfo {author} {\bibfnamefont {D.~M.}\ \bibnamefont
  {Bernardi}}\ and\ \bibinfo {author} {\bibfnamefont {J.-Y.}\ \bibnamefont
  {Go}},\ }\href {\doibase 10.1016/j.jpowsour.2010.06.107} {\bibfield
  {journal} {\bibinfo  {journal} {Journal of Power Sources}\ }\textbf {\bibinfo
  {volume} {196}},\ \bibinfo {pages} {412} (\bibinfo {year}
  {2011})}\BibitemShut {NoStop}%
\bibitem [{\citenamefont {Safari}\ and\ \citenamefont
  {Delacourt}(2011)}]{safari2011a}%
  \BibitemOpen
  \bibfield  {author} {\bibinfo {author} {\bibfnamefont {M.}~\bibnamefont
  {Safari}}\ and\ \bibinfo {author} {\bibfnamefont {C.}~\bibnamefont
  {Delacourt}},\ }\href@noop {} {\bibfield  {journal} {\bibinfo  {journal}
  {Journal of The Electrochemical Society}\ }\textbf {\bibinfo {volume}
  {158}},\ \bibinfo {pages} {A562} (\bibinfo {year} {2011})}\BibitemShut
  {NoStop}%
\bibitem [{\citenamefont {Srinivasan}\ and\ \citenamefont
  {Newman}(2004{\natexlab{a}})}]{srinivasan2004design}%
  \BibitemOpen
  \bibfield  {author} {\bibinfo {author} {\bibfnamefont {V.}~\bibnamefont
  {Srinivasan}}\ and\ \bibinfo {author} {\bibfnamefont {J.}~\bibnamefont
  {Newman}},\ }\href {\doibase 10.1149/1.1785013} {\bibfield  {journal}
  {\bibinfo  {journal} {Journal of The Electrochemical Society}\ }\textbf
  {\bibinfo {volume} {151}},\ \bibinfo {pages} {A1530} (\bibinfo {year}
  {2004}{\natexlab{a}})}\BibitemShut {NoStop}%
\bibitem [{\citenamefont {Baker}\ and\ \citenamefont
  {Verbrugge}(2012)}]{baker2012}%
  \BibitemOpen
  \bibfield  {author} {\bibinfo {author} {\bibfnamefont {D.~R.}\ \bibnamefont
  {Baker}}\ and\ \bibinfo {author} {\bibfnamefont {M.~W.}\ \bibnamefont
  {Verbrugge}},\ }\href {\doibase 10.1149/2.002208jes} {\bibfield  {journal}
  {\bibinfo  {journal} {Journal of The Electrochemical Society}\ }\textbf
  {\bibinfo {volume} {159}},\ \bibinfo {pages} {A1341} (\bibinfo {year}
  {2012})}\BibitemShut {NoStop}%
\bibitem [{\citenamefont {Funabiki}\ \emph {et~al.}(1999)\citenamefont
  {Funabiki}, \citenamefont {Inaba}, \citenamefont {Abe},\ and\ \citenamefont
  {Ogumi}}]{funabiki1999stage}%
  \BibitemOpen
  \bibfield  {author} {\bibinfo {author} {\bibfnamefont {A.}~\bibnamefont
  {Funabiki}}, \bibinfo {author} {\bibfnamefont {M.}~\bibnamefont {Inaba}},
  \bibinfo {author} {\bibfnamefont {T.}~\bibnamefont {Abe}}, \ and\ \bibinfo
  {author} {\bibfnamefont {Z.}~\bibnamefont {Ogumi}},\ }\href@noop {}
  {\bibfield  {journal} {\bibinfo  {journal} {Journal of The Electrochemical
  Society}\ }\textbf {\bibinfo {volume} {146}},\ \bibinfo {pages} {2443}
  (\bibinfo {year} {1999})}\BibitemShut {NoStop}%
\bibitem [{\citenamefont {Srinivasan}\ and\ \citenamefont
  {Newman}(2004{\natexlab{b}})}]{srinivasan2004}%
  \BibitemOpen
  \bibfield  {author} {\bibinfo {author} {\bibfnamefont {V.}~\bibnamefont
  {Srinivasan}}\ and\ \bibinfo {author} {\bibfnamefont {J.}~\bibnamefont
  {Newman}},\ }\href {\doibase 10.1149/1.1785012} {\bibfield  {journal}
  {\bibinfo  {journal} {Journal of The Electrochemical Society}\ }\textbf
  {\bibinfo {volume} {151}},\ \bibinfo {pages} {A1517} (\bibinfo {year}
  {2004}{\natexlab{b}})}\BibitemShut {NoStop}%
\bibitem [{\citenamefont {He\ss{}}\ and\ \citenamefont
  {Nov{\'a}k}(2013)}]{hess2013}%
  \BibitemOpen
  \bibfield  {author} {\bibinfo {author} {\bibfnamefont {M.}~\bibnamefont
  {He\ss{}}}\ and\ \bibinfo {author} {\bibfnamefont {P.}~\bibnamefont
  {Nov{\'a}k}},\ }\href {\doibase 10.1016/j.electacta.2013.05.056} {\bibfield
  {journal} {\bibinfo  {journal} {Electrochimica Acta}\ }\textbf {\bibinfo
  {volume} {106}},\ \bibinfo {pages} {149} (\bibinfo {year}
  {2013})}\BibitemShut {NoStop}%
\bibitem [{\citenamefont {Gallagher}\ \emph {et~al.}(2012)\citenamefont
  {Gallagher}, \citenamefont {Dees}, \citenamefont {Jansen}, \citenamefont
  {Abraham},\ and\ \citenamefont {Kang}}]{gallagher2012}%
  \BibitemOpen
  \bibfield  {author} {\bibinfo {author} {\bibfnamefont {K.~G.}\ \bibnamefont
  {Gallagher}}, \bibinfo {author} {\bibfnamefont {D.~W.}\ \bibnamefont {Dees}},
  \bibinfo {author} {\bibfnamefont {A.~N.}\ \bibnamefont {Jansen}}, \bibinfo
  {author} {\bibfnamefont {D.~P.}\ \bibnamefont {Abraham}}, \ and\ \bibinfo
  {author} {\bibfnamefont {S.-H.}\ \bibnamefont {Kang}},\ }\href@noop {}
  {\bibfield  {journal} {\bibinfo  {journal} {Journal of The Electrochemical
  Society}\ }\textbf {\bibinfo {volume} {159}},\ \bibinfo {pages} {A2029}
  (\bibinfo {year} {2012})}\BibitemShut {NoStop}%
\bibitem [{\citenamefont {Rowlinson}(1979)}]{rowlinson1979translation}%
  \BibitemOpen
  \bibfield  {author} {\bibinfo {author} {\bibfnamefont {J.~S.}\ \bibnamefont
  {Rowlinson}},\ }\href@noop {} {\bibfield  {journal} {\bibinfo  {journal}
  {Journal of Statistical Physics}\ }\textbf {\bibinfo {volume} {20}},\
  \bibinfo {pages} {197} (\bibinfo {year} {1979})}\BibitemShut {NoStop}%
\bibitem [{\citenamefont {Cahn}\ and\ \citenamefont
  {Hilliard}(1958)}]{cahn1958}%
  \BibitemOpen
  \bibfield  {author} {\bibinfo {author} {\bibfnamefont {J.~W.}\ \bibnamefont
  {Cahn}}\ and\ \bibinfo {author} {\bibfnamefont {J.~E.}\ \bibnamefont
  {Hilliard}},\ }\href@noop {} {\bibfield  {journal} {\bibinfo  {journal} {The
  Journal of Chemical Physics}\ }\textbf {\bibinfo {volume} {28}},\ \bibinfo
  {pages} {258} (\bibinfo {year} {1958})}\BibitemShut {NoStop}%
\bibitem [{\citenamefont {Cahn}(1961)}]{cahn1961}%
  \BibitemOpen
  \bibfield  {author} {\bibinfo {author} {\bibfnamefont {J.~W.}\ \bibnamefont
  {Cahn}},\ }\href@noop {} {\bibfield  {journal} {\bibinfo  {journal} {Acta
  metallurgica}\ }\textbf {\bibinfo {volume} {9}},\ \bibinfo {pages} {795}
  (\bibinfo {year} {1961})}\BibitemShut {NoStop}%
\bibitem [{\citenamefont {Groot}\ and\ \citenamefont
  {Mazur}(1962)}]{groot1962}%
  \BibitemOpen
  \bibfield  {author} {\bibinfo {author} {\bibfnamefont {S.~R.~D.}\
  \bibnamefont {Groot}}\ and\ \bibinfo {author} {\bibfnamefont
  {P.}~\bibnamefont {Mazur}},\ }\href@noop {} {\emph {\bibinfo {title}
  {Non-Equilibrium {{Thermodynamics}}}}}\ (\bibinfo  {publisher} {{Interscience
  Publishers, Inc.}},\ \bibinfo {address} {New York},\ \bibinfo {year}
  {1962})\BibitemShut {NoStop}%
\bibitem [{\citenamefont {Garc\i\'a}\ \emph {et~al.}(2004)\citenamefont
  {Garc\i\'a}, \citenamefont {Bishop},\ and\ \citenamefont
  {Carter}}]{garcia2004}%
  \BibitemOpen
  \bibfield  {author} {\bibinfo {author} {\bibfnamefont {R.}~\bibnamefont
  {Garc\i\'a}}, \bibinfo {author} {\bibfnamefont {C.~M.}\ \bibnamefont
  {Bishop}}, \ and\ \bibinfo {author} {\bibfnamefont {W.}~\bibnamefont
  {Carter}},\ }\href {\doibase 10.1016/j.actamat.2003.08.020} {\bibfield
  {journal} {\bibinfo  {journal} {Acta Materialia}\ }\textbf {\bibinfo {volume}
  {52}},\ \bibinfo {pages} {11} (\bibinfo {year} {2004})}\BibitemShut {NoStop}%
\bibitem [{\citenamefont {Hawrylak}\ and\ \citenamefont
  {Subbaswamy}(1984)}]{hawrylak1984}%
  \BibitemOpen
  \bibfield  {author} {\bibinfo {author} {\bibfnamefont {P.}~\bibnamefont
  {Hawrylak}}\ and\ \bibinfo {author} {\bibfnamefont {K.~R.}\ \bibnamefont
  {Subbaswamy}},\ }\href@noop {} {\bibfield  {journal} {\bibinfo  {journal}
  {Physical Review Letters}\ }\textbf {\bibinfo {volume} {53}},\ \bibinfo
  {pages} {2098} (\bibinfo {year} {1984})}\BibitemShut {NoStop}%
\bibitem [{\citenamefont {Guo}\ \emph {et~al.}(2016)\citenamefont {Guo},
  \citenamefont {Smith}, \citenamefont {Yu}, \citenamefont {Efetov},
  \citenamefont {Wang}, \citenamefont {Kim}, \citenamefont {Bazant},\ and\
  \citenamefont {Brus}}]{guo2016}%
  \BibitemOpen
  \bibfield  {author} {\bibinfo {author} {\bibfnamefont {Y.}~\bibnamefont
  {Guo}}, \bibinfo {author} {\bibfnamefont {R.~B.}\ \bibnamefont {Smith}},
  \bibinfo {author} {\bibfnamefont {Z.}~\bibnamefont {Yu}}, \bibinfo {author}
  {\bibfnamefont {D.~K.}\ \bibnamefont {Efetov}}, \bibinfo {author}
  {\bibfnamefont {J.}~\bibnamefont {Wang}}, \bibinfo {author} {\bibfnamefont
  {P.}~\bibnamefont {Kim}}, \bibinfo {author} {\bibfnamefont {M.~Z.}\
  \bibnamefont {Bazant}}, \ and\ \bibinfo {author} {\bibfnamefont {L.~E.}\
  \bibnamefont {Brus}},\ }\href {\doibase 10.1021/acs.jpclett.6b00625}
  {\bibfield  {journal} {\bibinfo  {journal} {The Journal of Physical Chemistry
  Letters}\ }\textbf {\bibinfo {volume} {7}},\ \bibinfo {pages} {2151}
  (\bibinfo {year} {2016})}\BibitemShut {NoStop}%
\bibitem [{\citenamefont {Thomas-Alyea}\ \emph {et~al.}(2016)\citenamefont
  {Thomas-Alyea}, \citenamefont {Jung}, \citenamefont {Smith},\ and\
  \citenamefont {Bazant}}]{thomas-alyea2016}%
  \BibitemOpen
  \bibfield  {author} {\bibinfo {author} {\bibfnamefont {K.~E.}\ \bibnamefont
  {Thomas-Alyea}}, \bibinfo {author} {\bibfnamefont {C.}~\bibnamefont {Jung}},
  \bibinfo {author} {\bibfnamefont {R.~B.}\ \bibnamefont {Smith}}, \ and\
  \bibinfo {author} {\bibfnamefont {M.~Z.}\ \bibnamefont {Bazant}},\
  }\href@noop {} {\enquote {\bibinfo {title} {In {{Situ Observation}} and
  {{Mathematical Modeling}} of {{Lithium Distribution}} within {{Graphite}} (in
  preparation)},}\ } (\bibinfo {year} {2016})\BibitemShut {NoStop}%
\bibitem [{\citenamefont {Derosa}\ and\ \citenamefont
  {Balbuena}(1999)}]{derosa1999lattice}%
  \BibitemOpen
  \bibfield  {author} {\bibinfo {author} {\bibfnamefont {P.~A.}\ \bibnamefont
  {Derosa}}\ and\ \bibinfo {author} {\bibfnamefont {P.~B.}\ \bibnamefont
  {Balbuena}},\ }\href@noop {} {\bibfield  {journal} {\bibinfo  {journal}
  {Journal of The Electrochemical Society}\ }\textbf {\bibinfo {volume}
  {146}},\ \bibinfo {pages} {3630} (\bibinfo {year} {1999})}\BibitemShut
  {NoStop}%
\bibitem [{\citenamefont {Ledovskikh}\ and\ \citenamefont
  {Wagemaker}(2016)}]{ledovskikh2016}%
  \BibitemOpen
  \bibfield  {author} {\bibinfo {author} {\bibfnamefont {A.~V.}\ \bibnamefont
  {Ledovskikh}}\ and\ \bibinfo {author} {\bibfnamefont {M.}~\bibnamefont
  {Wagemaker}},\ }\href {\doibase 10.1021/acs.jpcc.6b00914} {\bibfield
  {journal} {\bibinfo  {journal} {The Journal of Physical Chemistry C}\ }
  (\bibinfo {year} {2016}),\ 10.1021/acs.jpcc.6b00914}\BibitemShut {NoStop}%
\bibitem [{\citenamefont {Schiffer}\ \emph {et~al.}(2016)\citenamefont
  {Schiffer}, \citenamefont {Cannarella},\ and\ \citenamefont
  {Arnold}}]{schiffer2016strain}%
  \BibitemOpen
  \bibfield  {author} {\bibinfo {author} {\bibfnamefont {Z.~J.}\ \bibnamefont
  {Schiffer}}, \bibinfo {author} {\bibfnamefont {J.}~\bibnamefont
  {Cannarella}}, \ and\ \bibinfo {author} {\bibfnamefont {C.~B.}\ \bibnamefont
  {Arnold}},\ }\href@noop {} {\bibfield  {journal} {\bibinfo  {journal}
  {Journal of The Electrochemical Society}\ }\textbf {\bibinfo {volume}
  {163}},\ \bibinfo {pages} {A427} (\bibinfo {year} {2016})}\BibitemShut
  {NoStop}%
\bibitem [{\citenamefont {Safran}\ and\ \citenamefont
  {Hamann}(1979)}]{safran1979}%
  \BibitemOpen
  \bibfield  {author} {\bibinfo {author} {\bibfnamefont {S.~A.}\ \bibnamefont
  {Safran}}\ and\ \bibinfo {author} {\bibfnamefont {D.~R.}\ \bibnamefont
  {Hamann}},\ }\href@noop {} {\bibfield  {journal} {\bibinfo  {journal}
  {Physical Review Letters}\ }\textbf {\bibinfo {volume} {42}},\ \bibinfo
  {pages} {1410} (\bibinfo {year} {1979})}\BibitemShut {NoStop}%
\bibitem [{\citenamefont {Safran}(1980{\natexlab{b}})}]{safran1980b}%
  \BibitemOpen
  \bibfield  {author} {\bibinfo {author} {\bibfnamefont {S.~A.}\ \bibnamefont
  {Safran}},\ }\href@noop {} {\bibfield  {journal} {\bibinfo  {journal}
  {Physical Review Letters}\ }\textbf {\bibinfo {volume} {44}},\ \bibinfo
  {pages} {937} (\bibinfo {year} {1980}{\natexlab{b}})}\BibitemShut {NoStop}%
\bibitem [{\citenamefont {Kirczenow}(1984)}]{kirczenow1984}%
  \BibitemOpen
  \bibfield  {author} {\bibinfo {author} {\bibfnamefont {G.}~\bibnamefont
  {Kirczenow}},\ }\href@noop {} {\bibfield  {journal} {\bibinfo  {journal}
  {Physical review letters}\ }\textbf {\bibinfo {volume} {52}},\ \bibinfo
  {pages} {437} (\bibinfo {year} {1984})}\BibitemShut {NoStop}%
\bibitem [{\citenamefont {Kirczenow}(1990)}]{kirczenow1990}%
  \BibitemOpen
  \bibfield  {author} {\bibinfo {author} {\bibfnamefont {G.}~\bibnamefont
  {Kirczenow}},\ }in\ \href@noop {} {\emph {\bibinfo {booktitle} {Graphite
  {{Intercalation Compounds I}}}}}\ (\bibinfo  {publisher} {{Springer}},\
  \bibinfo {year} {1990})\ pp.\ \bibinfo {pages} {59--100}\BibitemShut
  {NoStop}%
\bibitem [{\citenamefont {Cahn}(1962)}]{cahn1962}%
  \BibitemOpen
  \bibfield  {author} {\bibinfo {author} {\bibfnamefont {J.~W.}\ \bibnamefont
  {Cahn}},\ }\href@noop {} {\bibfield  {journal} {\bibinfo  {journal} {Acta
  Metallurgica}\ }\textbf {\bibinfo {volume} {10}},\ \bibinfo {pages} {179}
  (\bibinfo {year} {1962})}\BibitemShut {NoStop}%
\bibitem [{\citenamefont {Aziz}\ \emph {et~al.}(1991)\citenamefont {Aziz},
  \citenamefont {Sabin},\ and\ \citenamefont {Lu}}]{aziz1991}%
  \BibitemOpen
  \bibfield  {author} {\bibinfo {author} {\bibfnamefont {M.~J.}\ \bibnamefont
  {Aziz}}, \bibinfo {author} {\bibfnamefont {P.~C.}\ \bibnamefont {Sabin}}, \
  and\ \bibinfo {author} {\bibfnamefont {G.-Q.}\ \bibnamefont {Lu}},\
  }\href@noop {} {\bibfield  {journal} {\bibinfo  {journal} {Physical Review
  B}\ }\textbf {\bibinfo {volume} {44}},\ \bibinfo {pages} {9812} (\bibinfo
  {year} {1991})}\BibitemShut {NoStop}%
\bibitem [{\citenamefont {Meunier}\ \emph {et~al.}(2002)\citenamefont
  {Meunier}, \citenamefont {Kephart}, \citenamefont {Roland},\ and\
  \citenamefont {Bernholc}}]{meunier2002}%
  \BibitemOpen
  \bibfield  {author} {\bibinfo {author} {\bibfnamefont {V.}~\bibnamefont
  {Meunier}}, \bibinfo {author} {\bibfnamefont {J.}~\bibnamefont {Kephart}},
  \bibinfo {author} {\bibfnamefont {C.}~\bibnamefont {Roland}}, \ and\ \bibinfo
  {author} {\bibfnamefont {J.}~\bibnamefont {Bernholc}},\ }\href {\doibase
  10.1103/PhysRevLett.88.075506} {\bibfield  {journal} {\bibinfo  {journal}
  {Physical Review Letters}\ }\textbf {\bibinfo {volume} {88}} (\bibinfo {year}
  {2002}),\ 10.1103/PhysRevLett.88.075506}\BibitemShut {NoStop}%
\bibitem [{\citenamefont {Yao}\ \emph {et~al.}(2012)\citenamefont {Yao},
  \citenamefont {G{\"u}ne{\c s}}, \citenamefont {Ta}, \citenamefont {Lee},
  \citenamefont {Chae}, \citenamefont {Sheem}, \citenamefont {Cojocaru},
  \citenamefont {Xie},\ and\ \citenamefont {Lee}}]{yao2012diffusion}%
  \BibitemOpen
  \bibfield  {author} {\bibinfo {author} {\bibfnamefont {F.}~\bibnamefont
  {Yao}}, \bibinfo {author} {\bibfnamefont {F.}~\bibnamefont {G{\"u}ne{\c s}}},
  \bibinfo {author} {\bibfnamefont {H.~Q.}\ \bibnamefont {Ta}}, \bibinfo
  {author} {\bibfnamefont {S.~M.}\ \bibnamefont {Lee}}, \bibinfo {author}
  {\bibfnamefont {S.~J.}\ \bibnamefont {Chae}}, \bibinfo {author}
  {\bibfnamefont {K.~Y.}\ \bibnamefont {Sheem}}, \bibinfo {author}
  {\bibfnamefont {C.~S.}\ \bibnamefont {Cojocaru}}, \bibinfo {author}
  {\bibfnamefont {S.~S.}\ \bibnamefont {Xie}}, \ and\ \bibinfo {author}
  {\bibfnamefont {Y.~H.}\ \bibnamefont {Lee}},\ }\href {\doibase
  10.1021/ja301586m} {\bibfield  {journal} {\bibinfo  {journal} {Journal of the
  American Chemical Society}\ }\textbf {\bibinfo {volume} {134}},\ \bibinfo
  {pages} {8646} (\bibinfo {year} {2012})}\BibitemShut {NoStop}%
\bibitem [{\citenamefont {Cahn}(1977)}]{cahn1977}%
  \BibitemOpen
  \bibfield  {author} {\bibinfo {author} {\bibfnamefont {J.~W.}\ \bibnamefont
  {Cahn}},\ }\href@noop {} {\bibfield  {journal} {\bibinfo  {journal} {Journal
  of Chemical Physics}\ }\textbf {\bibinfo {volume} {66}},\ \bibinfo {pages}
  {3667} (\bibinfo {year} {1977})}\BibitemShut {NoStop}%
\bibitem [{\citenamefont {Bard}\ and\ \citenamefont
  {Faulkner}(2001)}]{bard2001}%
  \BibitemOpen
  \bibfield  {author} {\bibinfo {author} {\bibfnamefont {A.~J.}\ \bibnamefont
  {Bard}}\ and\ \bibinfo {author} {\bibfnamefont {L.~R.}\ \bibnamefont
  {Faulkner}},\ }\href@noop {} {\emph {\bibinfo {title} {Electrochemical
  {{Methods}}}}}\ (\bibinfo  {publisher} {{J. Wiley \& Sons, Inc.}},\ \bibinfo
  {address} {New York, NY},\ \bibinfo {year} {2001})\BibitemShut {NoStop}%
\bibitem [{\citenamefont {Chidsey}(1991)}]{chidsey1991}%
  \BibitemOpen
  \bibfield  {author} {\bibinfo {author} {\bibfnamefont {C.~E.}\ \bibnamefont
  {Chidsey}},\ }\href {\doibase 10.1126/science.251.4996.919} {\bibfield
  {journal} {\bibinfo  {journal} {Science}\ }\textbf {\bibinfo {volume}
  {251}},\ \bibinfo {pages} {919} (\bibinfo {year} {1991})}\BibitemShut
  {NoStop}%
\bibitem [{\citenamefont {Bai}\ and\ \citenamefont {Bazant}(2014)}]{bai2014}%
  \BibitemOpen
  \bibfield  {author} {\bibinfo {author} {\bibfnamefont {P.}~\bibnamefont
  {Bai}}\ and\ \bibinfo {author} {\bibfnamefont {M.~Z.}\ \bibnamefont
  {Bazant}},\ }\href {\doibase 10.1038/ncomms4585} {\bibfield  {journal}
  {\bibinfo  {journal} {Nature Communications}\ }\textbf {\bibinfo {volume}
  {5}} (\bibinfo {year} {2014}),\ 10.1038/ncomms4585}\BibitemShut {NoStop}%
\bibitem [{\citenamefont {Laborda}\ \emph {et~al.}(2013)\citenamefont
  {Laborda}, \citenamefont {Henstridge}, \citenamefont {Batchelor-McAuley},\
  and\ \citenamefont {Compton}}]{laborda2013}%
  \BibitemOpen
  \bibfield  {author} {\bibinfo {author} {\bibfnamefont {E.}~\bibnamefont
  {Laborda}}, \bibinfo {author} {\bibfnamefont {M.~C.}\ \bibnamefont
  {Henstridge}}, \bibinfo {author} {\bibfnamefont {C.}~\bibnamefont
  {Batchelor-McAuley}}, \ and\ \bibinfo {author} {\bibfnamefont {R.~G.}\
  \bibnamefont {Compton}},\ }\href {\doibase 10.1039/c3cs35487c} {\bibfield
  {journal} {\bibinfo  {journal} {Chemical Society Reviews}\ }\textbf {\bibinfo
  {volume} {42}},\ \bibinfo {pages} {4894} (\bibinfo {year}
  {2013})}\BibitemShut {NoStop}%
\bibitem [{\citenamefont {Levi}\ and\ \citenamefont
  {Aurbach}(1997{\natexlab{a}})}]{levi1997simultaneous}%
  \BibitemOpen
  \bibfield  {author} {\bibinfo {author} {\bibfnamefont {M.~D.}\ \bibnamefont
  {Levi}}\ and\ \bibinfo {author} {\bibfnamefont {D.}~\bibnamefont {Aurbach}},\
  }\href@noop {} {\bibfield  {journal} {\bibinfo  {journal} {The Journal of
  Physical Chemistry B}\ }\textbf {\bibinfo {volume} {101}},\ \bibinfo {pages}
  {4630} (\bibinfo {year} {1997}{\natexlab{a}})}\BibitemShut {NoStop}%
\bibitem [{\citenamefont {Wissler}(2006)}]{wissler2006}%
  \BibitemOpen
  \bibfield  {author} {\bibinfo {author} {\bibfnamefont {M.}~\bibnamefont
  {Wissler}},\ }\href {\doibase 10.1016/j.jpowsour.2006.02.064} {\bibfield
  {journal} {\bibinfo  {journal} {Journal of Power Sources}\ }\textbf {\bibinfo
  {volume} {156}},\ \bibinfo {pages} {142} (\bibinfo {year}
  {2006})}\BibitemShut {NoStop}%
\bibitem [{\citenamefont {Kganyago}\ and\ \citenamefont
  {Ngoepe}(2003)}]{kganyago2003}%
  \BibitemOpen
  \bibfield  {author} {\bibinfo {author} {\bibfnamefont {K.}~\bibnamefont
  {Kganyago}}\ and\ \bibinfo {author} {\bibfnamefont {P.}~\bibnamefont
  {Ngoepe}},\ }\href {\doibase 10.1103/PhysRevB.68.205111} {\bibfield
  {journal} {\bibinfo  {journal} {Physical Review B}\ }\textbf {\bibinfo
  {volume} {68}} (\bibinfo {year} {2003}),\
  10.1103/PhysRevB.68.205111}\BibitemShut {NoStop}%
\bibitem [{\citenamefont {Persson}\ \emph {et~al.}(2010)\citenamefont
  {Persson}, \citenamefont {Hinuma}, \citenamefont {Meng}, \citenamefont {{Van
  der Ven}},\ and\ \citenamefont {Ceder}}]{persson2010}%
  \BibitemOpen
  \bibfield  {author} {\bibinfo {author} {\bibfnamefont {K.}~\bibnamefont
  {Persson}}, \bibinfo {author} {\bibfnamefont {Y.}~\bibnamefont {Hinuma}},
  \bibinfo {author} {\bibfnamefont {Y.~S.}\ \bibnamefont {Meng}}, \bibinfo
  {author} {\bibfnamefont {A.}~\bibnamefont {{Van der Ven}}}, \ and\ \bibinfo
  {author} {\bibfnamefont {G.}~\bibnamefont {Ceder}},\ }\href {\doibase
  10.1103/PhysRevB.82.125416} {\bibfield  {journal} {\bibinfo  {journal}
  {Physical Review B}\ }\textbf {\bibinfo {volume} {82}} (\bibinfo {year}
  {2010}),\ 10.1103/PhysRevB.82.125416}\BibitemShut {NoStop}%
\bibitem [{\citenamefont {Nikoli{\'c}}(2016)}]{nikolic2016}%
  \BibitemOpen
  \bibfield  {author} {\bibinfo {author} {\bibfnamefont {D.~D.}\ \bibnamefont
  {Nikoli{\'c}}},\ }\href {\doibase 10.7717/peerj-cs.54} {\bibfield  {journal}
  {\bibinfo  {journal} {PeerJ Computer Science}\ }\textbf {\bibinfo {volume}
  {2}},\ \bibinfo {pages} {e54} (\bibinfo {year} {2016})}\BibitemShut {NoStop}%
\bibitem [{\citenamefont {Hindmarsh}\ \emph {et~al.}(2005)\citenamefont
  {Hindmarsh}, \citenamefont {Brown}, \citenamefont {Grant}, \citenamefont
  {Lee}, \citenamefont {Serban}, \citenamefont {Shumaker},\ and\ \citenamefont
  {Woodward}}]{hindmarsh2005}%
  \BibitemOpen
  \bibfield  {author} {\bibinfo {author} {\bibfnamefont {A.~C.}\ \bibnamefont
  {Hindmarsh}}, \bibinfo {author} {\bibfnamefont {P.~N.}\ \bibnamefont
  {Brown}}, \bibinfo {author} {\bibfnamefont {K.~E.}\ \bibnamefont {Grant}},
  \bibinfo {author} {\bibfnamefont {S.~L.}\ \bibnamefont {Lee}}, \bibinfo
  {author} {\bibfnamefont {R.}~\bibnamefont {Serban}}, \bibinfo {author}
  {\bibfnamefont {D.~E.}\ \bibnamefont {Shumaker}}, \ and\ \bibinfo {author}
  {\bibfnamefont {C.~S.}\ \bibnamefont {Woodward}},\ }\href {\doibase
  10.1145/1089014.1089020} {\bibfield  {journal} {\bibinfo  {journal} {ACM
  Transactions on Mathematical Software}\ }\textbf {\bibinfo {volume} {31}},\
  \bibinfo {pages} {363} (\bibinfo {year} {2005})}\BibitemShut {NoStop}%
\bibitem [{\citenamefont {Griewank}\ \emph {et~al.}(1996)\citenamefont
  {Griewank}, \citenamefont {Juedes},\ and\ \citenamefont
  {Utke}}]{griewank1996}%
  \BibitemOpen
  \bibfield  {author} {\bibinfo {author} {\bibfnamefont {A.}~\bibnamefont
  {Griewank}}, \bibinfo {author} {\bibfnamefont {D.}~\bibnamefont {Juedes}}, \
  and\ \bibinfo {author} {\bibfnamefont {J.}~\bibnamefont {Utke}},\ }\href
  {\doibase 10.1145/229473.229474} {\bibfield  {journal} {\bibinfo  {journal}
  {ACM Transactions on Mathematical Software}\ }\textbf {\bibinfo {volume}
  {22}},\ \bibinfo {pages} {131} (\bibinfo {year} {1996})}\BibitemShut
  {NoStop}%
\bibitem [{\citenamefont {Song}\ \emph {et~al.}(1996)\citenamefont {Song},
  \citenamefont {Kinoshita},\ and\ \citenamefont
  {Tran}}]{song1996microstructural}%
  \BibitemOpen
  \bibfield  {author} {\bibinfo {author} {\bibfnamefont {X.~Y.}\ \bibnamefont
  {Song}}, \bibinfo {author} {\bibfnamefont {K.}~\bibnamefont {Kinoshita}}, \
  and\ \bibinfo {author} {\bibfnamefont {T.~D.}\ \bibnamefont {Tran}},\
  }\href@noop {} {\bibfield  {journal} {\bibinfo  {journal} {Journal of the
  Electrochemical Society}\ }\textbf {\bibinfo {volume} {143}},\ \bibinfo
  {pages} {L120} (\bibinfo {year} {1996})}\BibitemShut {NoStop}%
\bibitem [{\citenamefont {Fischer}\ \emph {et~al.}(1983)\citenamefont
  {Fischer}, \citenamefont {Fuerst},\ and\ \citenamefont {Woo}}]{fischer1983}%
  \BibitemOpen
  \bibfield  {author} {\bibinfo {author} {\bibfnamefont {J.~E.}\ \bibnamefont
  {Fischer}}, \bibinfo {author} {\bibfnamefont {C.~D.}\ \bibnamefont {Fuerst}},
  \ and\ \bibinfo {author} {\bibfnamefont {K.~C.}\ \bibnamefont {Woo}},\ }\href
  {\doibase 10.1016/0379-6779(83)90076-0} {\bibfield  {journal} {\bibinfo
  {journal} {Synthetic Metals}\ }\textbf {\bibinfo {volume} {7}},\ \bibinfo
  {pages} {1} (\bibinfo {year} {1983})}\BibitemShut {NoStop}%
\bibitem [{\citenamefont {Fischer}\ and\ \citenamefont
  {Kim}(1988)}]{fischer1988}%
  \BibitemOpen
  \bibfield  {author} {\bibinfo {author} {\bibfnamefont {J.~E.}\ \bibnamefont
  {Fischer}}\ and\ \bibinfo {author} {\bibfnamefont {H.~J.}\ \bibnamefont
  {Kim}},\ }\href@noop {} {\bibfield  {journal} {\bibinfo  {journal} {Synthetic
  Metals}\ }\textbf {\bibinfo {volume} {23}},\ \bibinfo {pages} {121} (\bibinfo
  {year} {1988})}\BibitemShut {NoStop}%
\bibitem [{\citenamefont {Woo}\ \emph {et~al.}(1983)\citenamefont {Woo},
  \citenamefont {Mertwoy}, \citenamefont {Fischer}, \citenamefont
  {Kamitakahara},\ and\ \citenamefont {Robinson}}]{woo1983}%
  \BibitemOpen
  \bibfield  {author} {\bibinfo {author} {\bibfnamefont {K.~C.}\ \bibnamefont
  {Woo}}, \bibinfo {author} {\bibfnamefont {H.}~\bibnamefont {Mertwoy}},
  \bibinfo {author} {\bibfnamefont {J.~E.}\ \bibnamefont {Fischer}}, \bibinfo
  {author} {\bibfnamefont {W.~A.}\ \bibnamefont {Kamitakahara}}, \ and\
  \bibinfo {author} {\bibfnamefont {D.~S.}\ \bibnamefont {Robinson}},\
  }\href@noop {} {\bibfield  {journal} {\bibinfo  {journal} {Physical Review
  B}\ }\textbf {\bibinfo {volume} {27}},\ \bibinfo {pages} {7831} (\bibinfo
  {year} {1983})}\BibitemShut {NoStop}%
\bibitem [{\citenamefont {Reynier}\ \emph {et~al.}(2004)\citenamefont
  {Reynier}, \citenamefont {Yazami},\ and\ \citenamefont
  {Fultz}}]{reynier2004}%
  \BibitemOpen
  \bibfield  {author} {\bibinfo {author} {\bibfnamefont {Y.~F.}\ \bibnamefont
  {Reynier}}, \bibinfo {author} {\bibfnamefont {R.}~\bibnamefont {Yazami}}, \
  and\ \bibinfo {author} {\bibfnamefont {B.}~\bibnamefont {Fultz}},\ }\href
  {\doibase 10.1149/1.1646152} {\bibfield  {journal} {\bibinfo  {journal}
  {Journal of The Electrochemical Society}\ }\textbf {\bibinfo {volume}
  {151}},\ \bibinfo {pages} {A422} (\bibinfo {year} {2004})}\BibitemShut
  {NoStop}%
\bibitem [{\citenamefont {Millman}\ and\ \citenamefont
  {Kirczenow}(1982)}]{millman1982origin}%
  \BibitemOpen
  \bibfield  {author} {\bibinfo {author} {\bibfnamefont {S.~E.}\ \bibnamefont
  {Millman}}\ and\ \bibinfo {author} {\bibfnamefont {G.}~\bibnamefont
  {Kirczenow}},\ }\href@noop {} {\bibfield  {journal} {\bibinfo  {journal}
  {Physical Review B}\ }\textbf {\bibinfo {volume} {26}},\ \bibinfo {pages}
  {2310} (\bibinfo {year} {1982})}\BibitemShut {NoStop}%
\bibitem [{\citenamefont {Millman}\ and\ \citenamefont
  {Kirczenow}(1983)}]{millman1983study}%
  \BibitemOpen
  \bibfield  {author} {\bibinfo {author} {\bibfnamefont {S.~E.}\ \bibnamefont
  {Millman}}\ and\ \bibinfo {author} {\bibfnamefont {G.}~\bibnamefont
  {Kirczenow}},\ }\href@noop {} {\bibfield  {journal} {\bibinfo  {journal}
  {Physical Review B}\ }\textbf {\bibinfo {volume} {28}},\ \bibinfo {pages}
  {3482} (\bibinfo {year} {1983})}\BibitemShut {NoStop}%
\bibitem [{\citenamefont {Balluffi}\ \emph {et~al.}(1954)\citenamefont
  {Balluffi}, \citenamefont {Allen},\ and\ \citenamefont
  {Carter}}]{balluffi1954}%
  \BibitemOpen
  \bibfield  {author} {\bibinfo {author} {\bibfnamefont {R.~W.}\ \bibnamefont
  {Balluffi}}, \bibinfo {author} {\bibfnamefont {S.~M.}\ \bibnamefont {Allen}},
  \ and\ \bibinfo {author} {\bibfnamefont {W.~C.}\ \bibnamefont {Carter}},\
  }\href@noop {} {\emph {\bibinfo {title} {Kinetics of {{Materials}}}}}\
  (\bibinfo  {publisher} {{John Wiley and Sons}},\ \bibinfo {address} {New
  York},\ \bibinfo {year} {1954})\BibitemShut {NoStop}%
\bibitem [{\citenamefont {Yu}\ \emph {et~al.}(1999)\citenamefont {Yu},
  \citenamefont {Popov}, \citenamefont {Ritter},\ and\ \citenamefont
  {White}}]{yu1999determination}%
  \BibitemOpen
  \bibfield  {author} {\bibinfo {author} {\bibfnamefont {P.}~\bibnamefont
  {Yu}}, \bibinfo {author} {\bibfnamefont {B.~N.}\ \bibnamefont {Popov}},
  \bibinfo {author} {\bibfnamefont {J.~A.}\ \bibnamefont {Ritter}}, \ and\
  \bibinfo {author} {\bibfnamefont {R.~E.}\ \bibnamefont {White}},\ }\href@noop
  {} {\bibfield  {journal} {\bibinfo  {journal} {Journal of The Electrochemical
  Society}\ }\textbf {\bibinfo {volume} {146}},\ \bibinfo {pages} {8} (\bibinfo
  {year} {1999})}\BibitemShut {NoStop}%
\bibitem [{\citenamefont {Aurbach}\ \emph {et~al.}(1998)\citenamefont
  {Aurbach}, \citenamefont {Levi}, \citenamefont {Levi}, \citenamefont
  {Teller}, \citenamefont {Markovsky}, \citenamefont {Salitra}, \citenamefont
  {Heider},\ and\ \citenamefont {Heider}}]{aurbach1998}%
  \BibitemOpen
  \bibfield  {author} {\bibinfo {author} {\bibfnamefont {D.}~\bibnamefont
  {Aurbach}}, \bibinfo {author} {\bibfnamefont {M.~D.}\ \bibnamefont {Levi}},
  \bibinfo {author} {\bibfnamefont {E.}~\bibnamefont {Levi}}, \bibinfo {author}
  {\bibfnamefont {H.}~\bibnamefont {Teller}}, \bibinfo {author} {\bibfnamefont
  {B.}~\bibnamefont {Markovsky}}, \bibinfo {author} {\bibfnamefont
  {G.}~\bibnamefont {Salitra}}, \bibinfo {author} {\bibfnamefont
  {U.}~\bibnamefont {Heider}}, \ and\ \bibinfo {author} {\bibfnamefont
  {L.}~\bibnamefont {Heider}},\ }\href@noop {} {\bibfield  {journal} {\bibinfo
  {journal} {Journal of The Electrochemical Society}\ }\textbf {\bibinfo
  {volume} {145}},\ \bibinfo {pages} {3024} (\bibinfo {year}
  {1998})}\BibitemShut {NoStop}%
\bibitem [{\citenamefont {Funabiki}\ \emph {et~al.}(1998)\citenamefont
  {Funabiki}, \citenamefont {Inaba}, \citenamefont {Ogumi}, \citenamefont
  {Yuasa}, \citenamefont {Otsuji},\ and\ \citenamefont
  {Tasaka}}]{funabiki1998impedance}%
  \BibitemOpen
  \bibfield  {author} {\bibinfo {author} {\bibfnamefont {A.}~\bibnamefont
  {Funabiki}}, \bibinfo {author} {\bibfnamefont {M.}~\bibnamefont {Inaba}},
  \bibinfo {author} {\bibfnamefont {Z.}~\bibnamefont {Ogumi}}, \bibinfo
  {author} {\bibfnamefont {S.-i.}\ \bibnamefont {Yuasa}}, \bibinfo {author}
  {\bibfnamefont {J.}~\bibnamefont {Otsuji}}, \ and\ \bibinfo {author}
  {\bibfnamefont {A.}~\bibnamefont {Tasaka}},\ }\href@noop {} {\bibfield
  {journal} {\bibinfo  {journal} {Journal of the Electrochemical Society}\
  }\textbf {\bibinfo {volume} {145}},\ \bibinfo {pages} {172} (\bibinfo {year}
  {1998})}\BibitemShut {NoStop}%
\bibitem [{\citenamefont {Levi}\ \emph {et~al.}(1997)\citenamefont {Levi},
  \citenamefont {Levi},\ and\ \citenamefont {Aurbach}}]{levi1997mechanism2}%
  \BibitemOpen
  \bibfield  {author} {\bibinfo {author} {\bibfnamefont {M.}~\bibnamefont
  {Levi}}, \bibinfo {author} {\bibfnamefont {E.}~\bibnamefont {Levi}}, \ and\
  \bibinfo {author} {\bibfnamefont {D.}~\bibnamefont {Aurbach}},\ }\href
  {\doibase 10.1016/S0022-0728(96)04833-4} {\bibfield  {journal} {\bibinfo
  {journal} {Journal of Electroanalytical Chemistry}\ }\textbf {\bibinfo
  {volume} {421}},\ \bibinfo {pages} {89} (\bibinfo {year} {1997})}\BibitemShut
  {NoStop}%
\bibitem [{\citenamefont {Levi}\ and\ \citenamefont
  {Aurbach}(1997{\natexlab{b}})}]{levi1997diffusion}%
  \BibitemOpen
  \bibfield  {author} {\bibinfo {author} {\bibfnamefont {M.~D.}\ \bibnamefont
  {Levi}}\ and\ \bibinfo {author} {\bibfnamefont {D.}~\bibnamefont {Aurbach}},\
  }\href@noop {} {\bibfield  {journal} {\bibinfo  {journal} {The Journal of
  Physical Chemistry B}\ }\textbf {\bibinfo {volume} {101}},\ \bibinfo {pages}
  {4641} (\bibinfo {year} {1997}{\natexlab{b}})}\BibitemShut {NoStop}%
\bibitem [{\citenamefont {Levi}\ \emph {et~al.}(2005)\citenamefont {Levi},
  \citenamefont {Markevich},\ and\ \citenamefont
  {Aurbach}}]{levi2005comparison}%
  \BibitemOpen
  \bibfield  {author} {\bibinfo {author} {\bibfnamefont {M.}~\bibnamefont
  {Levi}}, \bibinfo {author} {\bibfnamefont {E.}~\bibnamefont {Markevich}}, \
  and\ \bibinfo {author} {\bibfnamefont {D.}~\bibnamefont {Aurbach}},\ }\href
  {\doibase 10.1016/j.electacta.2005.04.007} {\bibfield  {journal} {\bibinfo
  {journal} {Electrochimica Acta}\ }\textbf {\bibinfo {volume} {51}},\ \bibinfo
  {pages} {98} (\bibinfo {year} {2005})}\BibitemShut {NoStop}%
\bibitem [{\citenamefont {Newman}\ and\ \citenamefont
  {Thomas-Alyea}(2004)}]{newman2004}%
  \BibitemOpen
  \bibfield  {author} {\bibinfo {author} {\bibfnamefont {J.}~\bibnamefont
  {Newman}}\ and\ \bibinfo {author} {\bibfnamefont {K.~E.}\ \bibnamefont
  {Thomas-Alyea}},\ }\href@noop {} {\emph {\bibinfo {title} {Electrochemical
  {{Systems}}}}},\ \bibinfo {edition} {3rd}\ ed.\ (\bibinfo  {publisher} {{John
  Wiley and Sons}},\ \bibinfo {address} {Hoboken, New Jersey},\ \bibinfo {year}
  {2004})\BibitemShut {NoStop}%
\bibitem [{\citenamefont {Doyle}\ \emph {et~al.}(1993)\citenamefont {Doyle},
  \citenamefont {Fuller},\ and\ \citenamefont {Newman}}]{doyle1993}%
  \BibitemOpen
  \bibfield  {author} {\bibinfo {author} {\bibfnamefont {M.}~\bibnamefont
  {Doyle}}, \bibinfo {author} {\bibfnamefont {T.~F.}\ \bibnamefont {Fuller}}, \
  and\ \bibinfo {author} {\bibfnamefont {J.}~\bibnamefont {Newman}},\
  }\href@noop {} {\bibfield  {journal} {\bibinfo  {journal} {Journal of the
  Electrochemical Society}\ }\textbf {\bibinfo {volume} {140}},\ \bibinfo
  {pages} {1526} (\bibinfo {year} {1993})}\BibitemShut {NoStop}%
\bibitem [{\citenamefont {Chung}\ \emph {et~al.}(2014)\citenamefont {Chung},
  \citenamefont {Shearing}, \citenamefont {Brandon}, \citenamefont {Harris},\
  and\ \citenamefont {Garc{\'\i}a}}]{chung2014particle}%
  \BibitemOpen
  \bibfield  {author} {\bibinfo {author} {\bibfnamefont {D.-W.}\ \bibnamefont
  {Chung}}, \bibinfo {author} {\bibfnamefont {P.~R.}\ \bibnamefont {Shearing}},
  \bibinfo {author} {\bibfnamefont {N.~P.}\ \bibnamefont {Brandon}}, \bibinfo
  {author} {\bibfnamefont {S.~J.}\ \bibnamefont {Harris}}, \ and\ \bibinfo
  {author} {\bibfnamefont {R.~E.}\ \bibnamefont {Garc{\'\i}a}},\ }\href@noop {}
  {\bibfield  {journal} {\bibinfo  {journal} {Journal of The Electrochemical
  Society}\ }\textbf {\bibinfo {volume} {161}},\ \bibinfo {pages} {A422}
  (\bibinfo {year} {2014})}\BibitemShut {NoStop}%
\bibitem [{\citenamefont {Stephenson}\ \emph {et~al.}(2007)\citenamefont
  {Stephenson}, \citenamefont {Hartman}, \citenamefont {Harb},\ and\
  \citenamefont {Wheeler}}]{stephenson2007modeling}%
  \BibitemOpen
  \bibfield  {author} {\bibinfo {author} {\bibfnamefont {D.~E.}\ \bibnamefont
  {Stephenson}}, \bibinfo {author} {\bibfnamefont {E.~M.}\ \bibnamefont
  {Hartman}}, \bibinfo {author} {\bibfnamefont {J.~N.}\ \bibnamefont {Harb}}, \
  and\ \bibinfo {author} {\bibfnamefont {D.~R.}\ \bibnamefont {Wheeler}},\
  }\href {\doibase 10.1149/1.2783772} {\bibfield  {journal} {\bibinfo
  {journal} {Journal of The Electrochemical Society}\ }\textbf {\bibinfo
  {volume} {154}},\ \bibinfo {pages} {A1146} (\bibinfo {year}
  {2007})}\BibitemShut {NoStop}%
\bibitem [{\citenamefont {Garc\i\'a}\ \emph {et~al.}(2005)\citenamefont
  {Garc\i\'a}, \citenamefont {Chiang}, \citenamefont {Craig~Carter},
  \citenamefont {Limthongkul},\ and\ \citenamefont {Bishop}}]{garcia2005}%
  \BibitemOpen
  \bibfield  {author} {\bibinfo {author} {\bibfnamefont {R.~E.}\ \bibnamefont
  {Garc\i\'a}}, \bibinfo {author} {\bibfnamefont {Y.-M.}\ \bibnamefont
  {Chiang}}, \bibinfo {author} {\bibfnamefont {W.}~\bibnamefont
  {Craig~Carter}}, \bibinfo {author} {\bibfnamefont {P.}~\bibnamefont
  {Limthongkul}}, \ and\ \bibinfo {author} {\bibfnamefont {C.~M.}\ \bibnamefont
  {Bishop}},\ }\href {\doibase 10.1149/1.1836132} {\bibfield  {journal}
  {\bibinfo  {journal} {Journal of The Electrochemical Society}\ }\textbf
  {\bibinfo {volume} {152}},\ \bibinfo {pages} {A255} (\bibinfo {year}
  {2005})}\BibitemShut {NoStop}%
\bibitem [{\citenamefont {Di~Leo}\ \emph {et~al.}(2014)\citenamefont {Di~Leo},
  \citenamefont {Rejovitzky},\ and\ \citenamefont {Anand}}]{dileo2014}%
  \BibitemOpen
  \bibfield  {author} {\bibinfo {author} {\bibfnamefont {C.~V.}\ \bibnamefont
  {Di~Leo}}, \bibinfo {author} {\bibfnamefont {E.}~\bibnamefont {Rejovitzky}},
  \ and\ \bibinfo {author} {\bibfnamefont {L.}~\bibnamefont {Anand}},\ }\href
  {\doibase 10.1016/j.jmps.2014.05.001} {\bibfield  {journal} {\bibinfo
  {journal} {Journal of the Mechanics and Physics of Solids}\ }\textbf
  {\bibinfo {volume} {70}},\ \bibinfo {pages} {1} (\bibinfo {year}
  {2014})}\BibitemShut {NoStop}%
\bibitem [{\citenamefont {Christensen}\ and\ \citenamefont
  {Newman}(2006)}]{christensen2006}%
  \BibitemOpen
  \bibfield  {author} {\bibinfo {author} {\bibfnamefont {J.}~\bibnamefont
  {Christensen}}\ and\ \bibinfo {author} {\bibfnamefont {J.}~\bibnamefont
  {Newman}},\ }\href {\doibase 10.1007/s10008-006-0095-1} {\bibfield  {journal}
  {\bibinfo  {journal} {Journal of Solid State Electrochemistry}\ }\textbf
  {\bibinfo {volume} {10}},\ \bibinfo {pages} {293} (\bibinfo {year}
  {2006})}\BibitemShut {NoStop}%
\bibitem [{\citenamefont {Cheng}\ and\ \citenamefont
  {Verbrugge}(2010)}]{cheng2010}%
  \BibitemOpen
  \bibfield  {author} {\bibinfo {author} {\bibfnamefont {Y.-T.}\ \bibnamefont
  {Cheng}}\ and\ \bibinfo {author} {\bibfnamefont {M.~W.}\ \bibnamefont
  {Verbrugge}},\ }\href {\doibase 10.1149/1.3298892} {\bibfield  {journal}
  {\bibinfo  {journal} {Journal of The Electrochemical Society}\ }\textbf
  {\bibinfo {volume} {157}},\ \bibinfo {pages} {A508} (\bibinfo {year}
  {2010})}\BibitemShut {NoStop}%
\bibitem [{\citenamefont {Grantab}\ and\ \citenamefont
  {Shenoy}(2011)}]{grantab2011}%
  \BibitemOpen
  \bibfield  {author} {\bibinfo {author} {\bibfnamefont {R.}~\bibnamefont
  {Grantab}}\ and\ \bibinfo {author} {\bibfnamefont {V.~B.}\ \bibnamefont
  {Shenoy}},\ }\href@noop {} {\bibfield  {journal} {\bibinfo  {journal}
  {Journal of The Electrochemical Society}\ }\textbf {\bibinfo {volume}
  {158}},\ \bibinfo {pages} {A948} (\bibinfo {year} {2011})}\BibitemShut
  {NoStop}%
\bibitem [{\citenamefont {Woodford}\ \emph {et~al.}(2010)\citenamefont
  {Woodford}, \citenamefont {Chiang},\ and\ \citenamefont
  {Carter}}]{woodford2010}%
  \BibitemOpen
  \bibfield  {author} {\bibinfo {author} {\bibfnamefont {W.~H.}\ \bibnamefont
  {Woodford}}, \bibinfo {author} {\bibfnamefont {Y.-M.}\ \bibnamefont
  {Chiang}}, \ and\ \bibinfo {author} {\bibfnamefont {W.~C.}\ \bibnamefont
  {Carter}},\ }\href {\doibase 10.1149/1.3464773} {\bibfield  {journal}
  {\bibinfo  {journal} {Journal of The Electrochemical Society}\ }\textbf
  {\bibinfo {volume} {157}},\ \bibinfo {pages} {A1052} (\bibinfo {year}
  {2010})}\BibitemShut {NoStop}%
\bibitem [{\citenamefont {Thomas}\ and\ \citenamefont
  {Newman}(2003)}]{thomas2003heats}%
  \BibitemOpen
  \bibfield  {author} {\bibinfo {author} {\bibfnamefont {K.~E.}\ \bibnamefont
  {Thomas}}\ and\ \bibinfo {author} {\bibfnamefont {J.}~\bibnamefont
  {Newman}},\ }\href {\doibase 10.1016/S0378-7753(03)00283-0} {\bibfield
  {journal} {\bibinfo  {journal} {Journal of Power Sources}\ }\textbf {\bibinfo
  {volume} {119-121}},\ \bibinfo {pages} {844} (\bibinfo {year}
  {2003})}\BibitemShut {NoStop}%
\bibitem [{\citenamefont {Dahn}\ \emph {et~al.}(1990)\citenamefont {Dahn},
  \citenamefont {Fong},\ and\ \citenamefont {Spoon}}]{dahn1990}%
  \BibitemOpen
  \bibfield  {author} {\bibinfo {author} {\bibfnamefont {J.~R.}\ \bibnamefont
  {Dahn}}, \bibinfo {author} {\bibfnamefont {R.}~\bibnamefont {Fong}}, \ and\
  \bibinfo {author} {\bibfnamefont {M.~J.}\ \bibnamefont {Spoon}},\ }\href@noop
  {} {\bibfield  {journal} {\bibinfo  {journal} {Physical Review B}\ }\textbf
  {\bibinfo {volume} {42}},\ \bibinfo {pages} {6424} (\bibinfo {year}
  {1990})}\BibitemShut {NoStop}%
\bibitem [{\citenamefont {Dahn}\ \emph {et~al.}(1982)\citenamefont {Dahn},
  \citenamefont {Dahn},\ and\ \citenamefont {Haering}}]{dahn1982elastic}%
  \BibitemOpen
  \bibfield  {author} {\bibinfo {author} {\bibfnamefont {J.~R.}\ \bibnamefont
  {Dahn}}, \bibinfo {author} {\bibfnamefont {D.~C.}\ \bibnamefont {Dahn}}, \
  and\ \bibinfo {author} {\bibfnamefont {R.~R.}\ \bibnamefont {Haering}},\
  }\href {\doibase 10.1016/0038-1098(82)90999-1} {\bibfield  {journal}
  {\bibinfo  {journal} {Solid State Communications}\ }\textbf {\bibinfo
  {volume} {42}},\ \bibinfo {pages} {179} (\bibinfo {year} {1982})}\BibitemShut
  {NoStop}%
\bibitem [{\citenamefont {Qi}\ \emph {et~al.}(2010)\citenamefont {Qi},
  \citenamefont {Guo}, \citenamefont {Hector},\ and\ \citenamefont
  {Timmons}}]{qi2010threefold}%
  \BibitemOpen
  \bibfield  {author} {\bibinfo {author} {\bibfnamefont {Y.}~\bibnamefont
  {Qi}}, \bibinfo {author} {\bibfnamefont {H.}~\bibnamefont {Guo}}, \bibinfo
  {author} {\bibfnamefont {L.~G.}\ \bibnamefont {Hector}}, \ and\ \bibinfo
  {author} {\bibfnamefont {A.}~\bibnamefont {Timmons}},\ }\href {\doibase
  10.1149/1.3327913} {\bibfield  {journal} {\bibinfo  {journal} {Journal of The
  Electrochemical Society}\ }\textbf {\bibinfo {volume} {157}},\ \bibinfo
  {pages} {A558} (\bibinfo {year} {2010})}\BibitemShut {NoStop}%
\bibitem [{\citenamefont {Qi}\ and\ \citenamefont
  {Harris}(2010)}]{qi2010insitu}%
  \BibitemOpen
  \bibfield  {author} {\bibinfo {author} {\bibfnamefont {Y.}~\bibnamefont
  {Qi}}\ and\ \bibinfo {author} {\bibfnamefont {S.~J.}\ \bibnamefont
  {Harris}},\ }\href {\doibase 10.1149/1.3377130} {\bibfield  {journal}
  {\bibinfo  {journal} {Journal of The Electrochemical Society}\ }\textbf
  {\bibinfo {volume} {157}},\ \bibinfo {pages} {A741} (\bibinfo {year}
  {2010})}\BibitemShut {NoStop}%
\bibitem [{\citenamefont {Bower}\ \emph {et~al.}(2011)\citenamefont {Bower},
  \citenamefont {Guduru},\ and\ \citenamefont {Sethuraman}}]{bower2011finite}%
  \BibitemOpen
  \bibfield  {author} {\bibinfo {author} {\bibfnamefont {A.}~\bibnamefont
  {Bower}}, \bibinfo {author} {\bibfnamefont {P.}~\bibnamefont {Guduru}}, \
  and\ \bibinfo {author} {\bibfnamefont {V.}~\bibnamefont {Sethuraman}},\
  }\href {\doibase 10.1016/j.jmps.2011.01.003} {\bibfield  {journal} {\bibinfo
  {journal} {Journal of the Mechanics and Physics of Solids}\ }\textbf
  {\bibinfo {volume} {59}},\ \bibinfo {pages} {804} (\bibinfo {year}
  {2011})}\BibitemShut {NoStop}%
\bibitem [{\citenamefont {Sethuraman}\ \emph {et~al.}(2012)\citenamefont
  {Sethuraman}, \citenamefont {Van~Winkle}, \citenamefont {Abraham},
  \citenamefont {Bower},\ and\ \citenamefont
  {Guduru}}]{sethuraman2012realtime}%
  \BibitemOpen
  \bibfield  {author} {\bibinfo {author} {\bibfnamefont {V.}~\bibnamefont
  {Sethuraman}}, \bibinfo {author} {\bibfnamefont {N.}~\bibnamefont
  {Van~Winkle}}, \bibinfo {author} {\bibfnamefont {D.}~\bibnamefont {Abraham}},
  \bibinfo {author} {\bibfnamefont {A.}~\bibnamefont {Bower}}, \ and\ \bibinfo
  {author} {\bibfnamefont {P.}~\bibnamefont {Guduru}},\ }\href {\doibase
  10.1016/j.jpowsour.2012.01.036} {\bibfield  {journal} {\bibinfo  {journal}
  {Journal of Power Sources}\ }\textbf {\bibinfo {volume} {206}},\ \bibinfo
  {pages} {334} (\bibinfo {year} {2012})}\BibitemShut {NoStop}%
\bibitem [{\citenamefont {Qi}\ \emph {et~al.}(2014)\citenamefont {Qi},
  \citenamefont {Hector}, \citenamefont {James},\ and\ \citenamefont
  {Kim}}]{qi2014lithium}%
  \BibitemOpen
  \bibfield  {author} {\bibinfo {author} {\bibfnamefont {Y.}~\bibnamefont
  {Qi}}, \bibinfo {author} {\bibfnamefont {L.~G.}\ \bibnamefont {Hector}},
  \bibinfo {author} {\bibfnamefont {C.}~\bibnamefont {James}}, \ and\ \bibinfo
  {author} {\bibfnamefont {K.~J.}\ \bibnamefont {Kim}},\ }\href@noop {}
  {\bibfield  {journal} {\bibinfo  {journal} {Journal of The Electrochemical
  Society}\ }\textbf {\bibinfo {volume} {161}},\ \bibinfo {pages} {F3010}
  (\bibinfo {year} {2014})}\BibitemShut {NoStop}%
\bibitem [{\citenamefont {Naguib}\ \emph {et~al.}(2014)\citenamefont {Naguib},
  \citenamefont {Mochalin}, \citenamefont {Barsoum},\ and\ \citenamefont
  {Gogotsi}}]{naguib2014}%
  \BibitemOpen
  \bibfield  {author} {\bibinfo {author} {\bibfnamefont {M.}~\bibnamefont
  {Naguib}}, \bibinfo {author} {\bibfnamefont {V.~N.}\ \bibnamefont
  {Mochalin}}, \bibinfo {author} {\bibfnamefont {M.~W.}\ \bibnamefont
  {Barsoum}}, \ and\ \bibinfo {author} {\bibfnamefont {Y.}~\bibnamefont
  {Gogotsi}},\ }\href {\doibase 10.1002/adma.201304138} {\bibfield  {journal}
  {\bibinfo  {journal} {Advanced Materials}\ }\textbf {\bibinfo {volume}
  {26}},\ \bibinfo {pages} {992} (\bibinfo {year} {2014})}\BibitemShut
  {NoStop}%
\bibitem [{\citenamefont {Lin}\ \emph {et~al.}(2015)\citenamefont {Lin},
  \citenamefont {Gong}, \citenamefont {Lu}, \citenamefont {Wu}, \citenamefont
  {Wang}, \citenamefont {Guan}, \citenamefont {Angell}, \citenamefont {Chen},
  \citenamefont {Yang}, \citenamefont {Hwang},\ and\ \citenamefont
  {Dai}}]{lin2015}%
  \BibitemOpen
  \bibfield  {author} {\bibinfo {author} {\bibfnamefont {M.-C.}\ \bibnamefont
  {Lin}}, \bibinfo {author} {\bibfnamefont {M.}~\bibnamefont {Gong}}, \bibinfo
  {author} {\bibfnamefont {B.}~\bibnamefont {Lu}}, \bibinfo {author}
  {\bibfnamefont {Y.}~\bibnamefont {Wu}}, \bibinfo {author} {\bibfnamefont
  {D.-Y.}\ \bibnamefont {Wang}}, \bibinfo {author} {\bibfnamefont
  {M.}~\bibnamefont {Guan}}, \bibinfo {author} {\bibfnamefont {M.}~\bibnamefont
  {Angell}}, \bibinfo {author} {\bibfnamefont {C.}~\bibnamefont {Chen}},
  \bibinfo {author} {\bibfnamefont {J.}~\bibnamefont {Yang}}, \bibinfo {author}
  {\bibfnamefont {B.-J.}\ \bibnamefont {Hwang}}, \ and\ \bibinfo {author}
  {\bibfnamefont {H.}~\bibnamefont {Dai}},\ }\href {\doibase
  10.1038/nature14340} {\bibfield  {journal} {\bibinfo  {journal} {Nature}\ }
  (\bibinfo {year} {2015}),\ 10.1038/nature14340}\BibitemShut {NoStop}%
\bibitem [{\citenamefont {Kim}\ \emph {et~al.}(2015)\citenamefont {Kim},
  \citenamefont {Hong}, \citenamefont {Park}, \citenamefont {Kim},
  \citenamefont {Hwang},\ and\ \citenamefont {Kang}}]{kim2015sodium}%
  \BibitemOpen
  \bibfield  {author} {\bibinfo {author} {\bibfnamefont {H.}~\bibnamefont
  {Kim}}, \bibinfo {author} {\bibfnamefont {J.}~\bibnamefont {Hong}}, \bibinfo
  {author} {\bibfnamefont {Y.-U.}\ \bibnamefont {Park}}, \bibinfo {author}
  {\bibfnamefont {J.}~\bibnamefont {Kim}}, \bibinfo {author} {\bibfnamefont
  {I.}~\bibnamefont {Hwang}}, \ and\ \bibinfo {author} {\bibfnamefont
  {K.}~\bibnamefont {Kang}},\ }\href {\doibase 10.1002/adfm.201402984}
  {\bibfield  {journal} {\bibinfo  {journal} {Advanced Functional Materials}\
  }\textbf {\bibinfo {volume} {25}},\ \bibinfo {pages} {534} (\bibinfo {year}
  {2015})}\BibitemShut {NoStop}%
\bibitem [{\citenamefont {Wu}\ \emph {et~al.}(2015)\citenamefont {Wu},
  \citenamefont {Li}, \citenamefont {Xu}, \citenamefont {Twu}, \citenamefont
  {Liu},\ and\ \citenamefont {Ceder}}]{wu2015sodium}%
  \BibitemOpen
  \bibfield  {author} {\bibinfo {author} {\bibfnamefont {D.}~\bibnamefont
  {Wu}}, \bibinfo {author} {\bibfnamefont {X.}~\bibnamefont {Li}}, \bibinfo
  {author} {\bibfnamefont {B.}~\bibnamefont {Xu}}, \bibinfo {author}
  {\bibfnamefont {N.}~\bibnamefont {Twu}}, \bibinfo {author} {\bibfnamefont
  {L.}~\bibnamefont {Liu}}, \ and\ \bibinfo {author} {\bibfnamefont
  {G.}~\bibnamefont {Ceder}},\ }\href {\doibase 10.1039/C4EE03045A} {\bibfield
  {journal} {\bibinfo  {journal} {Energy Environ. Sci.}\ }\textbf {\bibinfo
  {volume} {8}},\ \bibinfo {pages} {195} (\bibinfo {year} {2015})}\BibitemShut
  {NoStop}%
\end{thebibliography}%
\end{document}